\documentclass[useAMS,onecolumn,usenatbib]{mn2e}

\newif\ifAMStwofonts

\usepackage{graphicx}	
\usepackage{amssymb}	
\usepackage{epsfig}
\usepackage{color}

\def\pmb#1{\mbox{\boldmath$#1$}}
\def\gtsim {>\kern-1.2em\lower1.1ex\hbox{$\sim$}}
\def\ltsim {<\kern-1.2em\lower1.1ex\hbox{$\sim$}}
\def\gtsim {>\kern-1.2em\lower1.1ex\hbox{$\sim$}}
\def\ltsim {<\kern-1.2em\lower1.1ex\hbox{$\sim$}}

\def\be{\begin{equation}}
\def\ee{\end{equation}}

\def\pmbmt#1{\pmb{\sf #1}}
\def\rmi{{\rm i}}

\begin{document}

\title[Pulsation-driven mean flows]{Pulsation-driven mean zonal and meridional flows in rotating massive stars}

\author[U. Lee, S. Mathis \& C. Neiner]{
Umin Lee$^{1}$\thanks{E-mail: lee@astr.tohoku.ac.jp},
St\'ephane Mathis$^{2,3}$,
and Coralie Neiner$^{3}$
\\
$^{1}$Astronomical Institute, Tohoku University, Sendai, Miyagi 980-8578, Japan\\
$^{2}$Laboratoire AIM Paris-Saclay, CEA/DSM-Universit\'e Paris Diderot-CNRS,
IRFU/SAp Centre de Saclay, 91191 Gif-sur-Yvette,\\
France\\
$^{3}$LESIA, Observatoire de Paris, PSL Research University, CNRS, Sorbonne
Universit\'es, UPMC Univ. Paris 06, Univ. Paris Diderot,\\
Sorbonne Paris Cit\'e, 5 place Jules Janssen, 92195 Meudon, France
}

\date{Accepted XXX. Received YYY; in original form ZZZ}
\pubyear{2015}

\maketitle

\begin{abstract}
Zonal and meridional axisymmetric flows can deeply impact the rotational
and chemical evolution of stars. Therefore, momentum
exchanges between waves propagating in stars, differential rotation, and
meridional circulation must be carefully evaluated. In this work,
we study axisymmetric mean flows in rapidly and initially uniformly rotating 
massive stars driven by small amplitude
non-axisymmetric {$\kappa$-driven} oscillations. We treat {them} 
as perturbations of second-order of the oscillation amplitudes
{and} derive {their} governing equations as a set of
coupled linear ordinary differential {equations.} {This
allows us} to compute {2-D zonal and meridional} mean flows driven by low
frequency $g$- and $r$-modes in slowly pulsating B (SPB) stars and $p$-modes in
$\beta$ Cephei stars. {Oscillation-driven} mean flows
{usually} have large amplitudes only in the surface layers.
{In addition,} {the kinetic energy of the induced
{2-D} zonal rotational motions {is much}
larger than {that of the} meridional
{motions}. In some cases, meridional flows have a complex
{radial and latitudinal} structure.} 
We find pulsation-driven and
rotation-driven meridional flows can have similar
amplitudes. These results show the importance of taking wave
-- mean flow interactions into account when studying the
evolution of massive stars.
\end{abstract}



\section{introduction}

It is well known that {differential rotation and related viscous turbulent
transport, structural adjustments, and applied torques at stellar surfaces drive
large-scale meridional circulation in stellar radiation zones 
(e.g., Zahn 1992; Rieutord 2006; Decressin et al. 2009; Hypolite \& Rieutord 2014).
{When} magnetic fields and internal
gravity waves are neglected,} the magnitudes of {the radial component of}
such meridional {flow} ($v_r$) {scales as}
(e.g., Kippenhahn et al. 2012)
\be
v_{r:\rm MC}\sim \frac{LR^2}{GM^2}\frac{\Omega^2}{2\pi G\bar\rho},
\ee
where $L$, $M$, $R$, $\bar\rho$ and $\Omega$ are the luminosity, mass, radius,
mean density and angular velocity of the star respectively. It is
anticipated that {such meridional circulation} in rotating
stars mix material {and transport angular momentum} in their interior during
their evolution 
(e.g., Zahn 1992; Talon et al. 1997; Maeder \& Zahn 1998; Meynet \& Maeder 2000;
Mathis \& Zahn 2004).

Stellar {non-radial} pulsations in rotating stars can
also drive such non-oscillatory, {large-scale} fluid motions, which we may
call mean flows. {Because of dissipative mechanisms, co-rotation
resonances and breaking mechanisms 
(e.g., Andrews \& McIntyre 1978a; Lindzen 1981; Goldreich \& Nicholson 1989),
{pulsations} transport
angular momentum, which induces mean zonal flows leading to differential rotation.
Indeed, angular momentum transport by internal gravity waves takes place in
rotating stars, if the wave propagation is accompanied by thermal
diffusion/critical layers/non-linear breaking processes that lead to damping
and/or excitation of the waves 
(e.g., Press 1981; Schatzman 1993, 1996; Zahn et al. 1997; Alvan et al. 2013;
Rogers et al. 2013).
For example, if prograde internal gravity waves
propagating in a rotating star damp at a {certain} place, they
will deposit their angular momentum, which accelerates zonal rotational motion of
the fluid there. On the other hand, {when} prograde waves are
excited in a place, they extract angular momentum from the flow at this
location, leading to deceleration of the rotational fluid motion. The opposite
results are obtained for retrograde waves. Therefore, extraction and deposition
of angular momentum from rotational fluid motion by waves change the zonal
velocity field in the interior of rotating stars.} Because {of the related
modification of angular velocity gradients, of the radial and latitudinal
components of the momentum equation, and of the equation of heat, waves modify
the general 3-D dynamical balance in stellar radiation zones.
{As a consequence}, they also profoundly affect large-scale
meridional flows in the radial and latitudinal directions {in
these regions} 
(e.g., Bretherton 1969; Andrews \& McIntyre 1978a; Holton 1982; Mathis et al. 2013;
Belkacem et al. 2015a).
{In} the case of
{small-amplitude linear} pulsations, magnitudes of mean flows {are} of
second-order of the amplitudes {and their characteristic timescale
{is} the growth or damping timescales of the waves. Then,
these} {pulsation-driven} mean flows will {impact} material
mixing and angular momentum transport processes in rotating stars and will
affect pulsations in return.

{These processes have been relatively well studied for the case
of low-mass stars (e.g., Mathis et al. 2013),
but poorly investigated
for massive stars. Therefore, it is} {important} to investigate the velocity
fields and the magnitudes of {both zonal and meridional} mean flows driven
by stellar pulsations in rotating stars and to investigate their
geometrical properties and amplitude in the case of massive
{stars.}

The problem of angular momentum transport and redistribution by internal gravity
waves in rotating stars has been investigated by many authors 
{for cool stars. For example, for} an explanation {of the}
uniform rotation observationally inferred for the radiative core of the Sun 
until $0.2R_{\odot}$ thanks to helioseismology
(e.g., Schou et al. 1998; Garc\'ia et al. 2007),
Schatzman (1993) suggested that angular momentum transport by 
{internal} gravity waves generated in the {external} convective envelope can
play an essential role to extract angular momentum 
[see also Zahn et al. (1997); Talon et al.
(2002); Talon \& Charbonnel (2005); Rogers et al. (2008); Mathis et al. (2008)
in the case where the Coriolis
acceleration is taken into account; 
Kumar et al. (1999); Rogers \& MacGregor (2010); Mathis \& de Brye
(2012) in the magnetized case].
More recently,
{thanks to new asteroseismic information} obtained on the weak
differential rotation between the surface and the core of low-mass subgiant and
red giant stars (e.g., Beck et al.
2012; Mosser et al. 2012; Deheuvels et al. 2012, 2014),
the potential extraction of angular
momentum by internal gravity waves and normal mixed gravito-acoustic modes have
been studied 
(Talon \& Charbonnel 2008; Fuller et al. 2014; Belkacem et al. 2015a,b).

In the case of massive stars, the transport of angular momentum by internal
gravity waves has been less studied 
(Lee \& Saio 1993; Pantillon et al. 2007; Lee et al. 2014).
Zahn (1975, 1977)
{investigated the case of} gravity waves tidally excited in massive main-sequence
stars by the orbital motion of a companion star in a binary system 
(see also Goldreich \& Nicholson 1989).
{ It {allowed} him} to
estimate the timescales of circularization of the orbit and of synchronization
of the stellar rotation with the orbital motion. However, the impact of internal
gravity waves} on large-scale meridional flows in these stars has never been
studied. 

In this context, {one important issue} is the case of Be
stars. Be stars are rapidly rotating main-sequence {late-O, B,
and early-A} stars {that} have circumstellar gaseous
{discs,} which are responsible for {the}
generation of emission lines {(see Rivinius et al. 2013 for a complete review).} 
Gaseous discs in Be stars {are
thought} to be viscous Keplerian discs (Lee et al. 1991).
Although
mechanisms for disc formation have not {been completely
identified yet,} rapid rotation {and pulsations} of {these}
stars must be {key ingredients}. { Moreover,} to sustain the
discs for several decades, a good amount of angular momentum must be
continuously supplied to them. {
Ando (1983, 1986)} employed a wave mean flow interaction theory
(e.g., Andrews \& McIntyre
1976, 1978a,b; Dunkerton 1980; Grimshaw 1984; Craik 1988; Pedlosky 1982; B\"uhler 2014)
to explain episodic mass loss phenomena observed in Be stars 
(see also Lee et al. 1991, 2014).
As suggested by Lee (2013),
if {enough} angular momentum is
supplied to the surface regions of rapidly rotating stars, we can construct a
steady system composed of a rotating star and a viscous Keplerian disc around
the star. As a mechanism to provide the surface layers with angular momentum, we
may {invoke the rotation-driven} meridional circulation
(Meynet \& Maeder 1997; Ekstr\"om et al. 2008; Granada et al. 2013)
and/or {internal} gravity waves, which may be destabilized by the opacity mechanism or
stochastically excited by convective motion in the core 
(see e.g., Rogers et al. 2013; Lee et al. 2014).

In this paper, we are {thus} interested in velocity fields of pulsation
driven mean {zonal and meridional} flows in rotating {massive} stars,
and we {investigate} whether the velocity fields of the mean
flows can be favorable to a disc formation around Be stars. {Therefore,} we
calculate pulsation-driven mean flows in the interior {of} rotating SPB 
{and} $\beta$ Cephei stars, in which pulsations are
{mainly} excited by the iron opacity bump
($\kappa$) mechanism. Assuming that the pulsation-driven
velocity fields are of second-order of the {linear} pulsation amplitudes, we
derive for the second-order perturbations a set of linear ordinary differential
equations, which have inhomogeneous terms generated by {non-linear} products
of the eigenfunctions of the pulsation mode. To derive
analytical expressions of these non-linear terms, we introduce and use the so-called Spin
Weighted Spherical Harmonics and their complex and tedious derivation is detailed in a
complete appendix. The set of linear differential equations we solve
is given in \S 2. In \S 3, we calculate and discuss the properties of second-order
velocity fields driven by unstable $g$- and $r$-modes in SPB stars and $p$-modes
in $\beta$ Cephei stars assuming moderate rotation of the stars.
In \S4, we summarize the results
we obtained, and we discuss drawbacks and
perspectives of this work.

\section{Theoretical formalism for mean flows induced by pulsations}

\subsection{Dynamical equations}

The basic equations governing fluid motions in a star {are} given by
\be
\frac{\partial\pmb{v}}{\partial t}+\pmb{v}\cdot\nabla\pmb{v}=-\frac{1}{\rho}\nabla p-\nabla\Phi,
\ee
\be
\nabla^2\Phi=4\pi G\rho,
\ee
\be
\frac{\partial\rho}{\partial t}+\nabla\cdot\left(\rho\pmb{v}\right)=0,
\ee
\be
\rho T \frac{ds}{dt}=\rho\epsilon-\nabla\cdot\pmb{F},
\ee
\be
\pmb{F}=-\lambda\nabla T
\quad {\rm with} \quad  
\lambda={16\sigma_{\rm SB}T^3/ 3\kappa\rho},
\ee
{where} $\pmb{v}$ is the velocity vector of the fluid, $p$ is the pressure,
$\rho$ is the mass density,  $T$ is the temperature, $s$ is the specific
entropy, $\pmb{F}$ is the energy flux vector, $\Phi$ is the gravitational
potential, $\epsilon$ is the  nuclear energy generation rate per gram,  $\kappa$
is the opacity, $G$ is the gravitational constant, and $\sigma_{\rm SB}$ is the
Stefan-Boltzmann constant.

We assume that the star pulsates in small amplitudes around {the hydrostatic
and radiative} equilibrium state. If pulsation amplitudes are sufficiently
small, any physical quantity $f\left(\pmb{x},t\right)$ of the star may be
represented by
\be
f(\pmb{x},t)=f^{(0)}(\pmb{x})+f^{\prime}(\pmb{x},t)+f^{(2)}(\pmb{x},t) +\cdots,
\ee
where $f^{(0)}$ denotes equilibrium { quantities}, $f^\prime$ Eulerian
perturbations of first-order,  and $f^{(2)}$ Eulerian perturbations of
second-order of the pulsation amplitude. The velocity field
$\pmb{v}\left(\pmb{x},t\right)$ may also be given by
\be
\pmb{v}(\pmb{x},t)=\pmb{v}^{(0)}(\pmb{x})+\pmb{v}^\prime(\pmb{x},t)+\pmb{v}^{(2)}(\pmb{x},t)+\cdots,
\ee
and the equilibrium state is assumed to be that of a uniformly rotating star so
that, in spherical polar coordinates $(r,\theta,\phi)$,
\be
\pmb{v}^{(0)}=r\sin\theta\Omega\pmb{e}_\phi,
\ee
where $\Omega$ is the angular velocity of rotation and assumed to be constant,
and $\pmb{e}_\phi$ is the unit vector in the azimuthal direction. In this paper,
we ignore rotational deformation of the equilibrium state so that $f^{(0)}$
depends only on the radial distance $r$ from the center of the star. We also
employ the Cowling approximation, neglecting the Euler perturbation
$\Phi^\prime$ of the gravitational potential $\Phi$.

\subsection{First-order pulsation quantities}

For a given equilibrium state of a star, we solve the linear oscillation
equation to obtain wave quantities represented by $f^\prime$. For uniformly
rotating stars, the oscillation equations, which we solve to obtain normal
modes, are found, for example, in Lee \& Saio (1987) and Lee \& Baraffe (1995).
Since separation of variables between the radial $(r)$
and angular $(\theta,\phi)$ coordinates is not possible for perturbations in
rotating stars, we use finite series expansion to represent the perturbations in
terms of  spherical harmonic functions $Y_l^m(\theta,\phi)$ (e.g., Lee \& Saio 1987).
For the Lagrangian displacement vector
$\pmb{\xi}$, we may write
\be
\pmb{\xi}(\pmb{x},t)=e^{\rmi\sigma t}\pmb{\xi}(\pmb{x})=e^{\rmi\sigma t}\left(\xi_r\pmb{e}_r+\xi_\theta\pmb{e}_\theta+\xi_\phi\pmb{e}_\phi\right),
\ee
where $\pmb{e}_r$, $\pmb{e}_\theta$, and $\pmb{e}_\phi$ are the orthonormal basis vectors in spherical polar coordinates, $\sigma$
denotes the oscillation frequency observed in an inertial frame. 
{For} a given azimuthal order $m$, the components of $\pmb{\xi}(\pmb{x})$ are given by
\be
{\xi_r}(\pmb{x})=r\sum_{j=1}^{j_{\rm max}}S_{l_j}(r)Y_{l_j}^m(\theta,\phi),
\label{eq:xiexp_r}
\ee
\be
{\xi_\theta}(\pmb{x})=r\sum_{j=1}^{j_{\rm max}}\left[H_{l_j}(r)\frac{\partial}{\partial\theta}
Y_{l_j}^m(\theta,\phi)+T_{l'_j}(r)\frac{1}{\sin\theta}\frac{\partial}{\partial\phi}Y_{l'_j}^m(\theta,\phi)
\right],
\label{eq:xiexp_theta}
\ee
\be
{\xi_\phi}(\pmb{x})=r\sum_{j=1}^{j_{\rm max}}\left[H_{l_j}(r)\frac{1}{\sin\theta}\frac{\partial}{\partial\phi}
Y_{l_j}^m(\theta,\phi) - T_{l'_j}(r)\frac{\partial}{\partial\theta}Y_{l'_j}^m(\theta,\phi)
\right],
\label{eq:xiexp_phi}
\ee
and the Eulerian pressure perturbation,
$p^\prime(\pmb{x},t)=p^\prime(\pmb{x})e^{\rmi\sigma t}$, is given by
\be
p^\prime\left(\pmb{x}\right)=\sum_{j=1}^{j_{\rm max}}p^\prime_{l_j} (r)Y_{l_j}^m\left(\theta,\phi\right),
\ee
where $l_j=2(j-1)+|m|$ and $l'_j=l_j+1$ for even modes, and $l_j=2j-1+|m|$ and
$l'_j=l_j-1$ for odd modes for $j=1~,2~,3\cdots,~j_{\rm max}$.
{The} angular dependence of $p^\prime(\pmb{x})$ is symmetric
(antisymmetric) about the equator for even (odd) modes. {In
addition, we} have
\be
\pmb{v}^\prime=\delta\pmb{v}-\pmb{\xi}\cdot\nabla\pmb{v}^{(0)}=\frac{d\pmb{\xi}}{dt}-\pmb{\xi}\cdot\nabla\pmb{v}^{(0)}=\rmi\omega\pmb{\xi},
\ee
where $\omega=\sigma+m\Omega$ is the oscillation frequency observed in the
corotating frame of the star, and $\delta\pmb{v}$ denotes the Lagrangian
perturbation of the velocity vector.

\subsection{Pulsation driven mean flows}

\subsubsection{Expansion of second-order equations}

We use a theory of wave-mean flow interaction to discuss axisymmetric flows
driven by non-axisymmetric oscillations in rotating stars. We regard the
axisymmetric flows as mean flows, which contain both
{zero$^{th}$} and second-order contributions in the oscillation
amplitude. The {zero$^{th}$} order quantities $f^{(0)}$ are
those of equilibrium state and are independent of time $t$. The second-order
quantities $f^{(2)}$ carry the time dependence of the mean flow. To derive
governing equations for the second-order quantities $f^{(2)}$, we use the zonal
averaging defined by
\be
\overline{f}=\frac{1}{2\pi}\int_0^{2\pi}fd\phi, 
\ee
and if we assume
\be
\overline {f^\prime}=0,
\ee
we have
\be
\overline{f}=\overline{f^{(0)}}+\overline{f^{(2)}}.
\ee
Here, we have ignored higher order terms $f^{(k)}$ with $k\ge 3$. Hereafter, we
simply write $f^{(0)}$ and $f^{(2)}$ for $\overline{f^{(0)}}$ and
$\overline{f^{(2)}}$. Because of the zonal averaging, $f^{(0)}$ and $f^{(2)}$
are independent of $\phi$.

When a non-axisymmetric oscillation mode is excited to attain a small but finite
amplitude, non-oscillatory fluid flows may arise as a result of second-order
effects of the oscillation. Applying the zonal averaging to the basic equations,
we obtain a set of differential equations that {govern}
second-order quantities: 
\be
\frac{{\partial \pmb{v} ^{\left( 2 \right)} }}{{\partial t}} + \pmb{v} ^{\left( 0 \right)}  \cdot \nabla \pmb{v} ^{\left( 2 \right)}  + \pmb{v} ^{\left( 2 \right)}  \cdot \nabla \pmb{v} ^{\left( 0 \right)}  + \frac{1}{{\rho ^{\left( 0 \right)} }}\nabla p^{\left( 2 \right)}  
+ g\frac{{\rho ^{\left( 2 \right)} }}{{\rho ^{\left( 0 \right)} }}
\pmb{e}_r  
=  - \overline {\pmb{v} ' \cdot \nabla \pmb{v} '}  + g\overline {\left( {\frac{{\rho '}}{{\rho ^{\left( 0 \right)} }}} \right)^2}  
\pmb{e}_r 
+ \frac{1}{{\rho ^{\left( 0 \right)} }}\overline {\frac{{\rho '}}{{\rho ^{\left( 0 \right)} }}\nabla p'} ,
\label{eq:secondmomcon}
\ee
\be
\frac{\partial\rho^{(2)}}{\partial t}+\nabla\cdot\left(\rho^{(0)}\pmb{v}^{(2)}\right)=-\overline{\nabla\cdot\left(\rho^\prime
\pmb{v}^\prime\right)},
\label{eq:secondcont}
\ee
\be
\pmb{F}^{(2)}=-\lambda^{(0)}\nabla T^{(2)}-\lambda^{(2)}\nabla T^{(0)}-\overline{\lambda^\prime\nabla T^\prime},
\label{eq:radtrans2}
\ee
\begin{eqnarray}
\rho^{(0)}T^{(0)}\left(\frac{\partial s^{(2)}}{\partial t}+\pmb{v}^{(2)}\cdot\nabla s^{(0)}\right)  
&=& - \rho^{(0)}T^{(0)}\overline{\left(\frac{T^\prime}{T^{(0)}}+\frac{\rho^\prime}{\rho^{(0)}}\right)\left( \frac{\partial s^\prime}{\partial t}
+\pmb{v}^\prime\cdot\nabla s^{(0)}+\pmb{v}^{(0)}\cdot\nabla s^\prime\right)}
-\rho^{(0)}T^{(0)}\overline{\pmb{v}^\prime\cdot\nabla s^\prime}\nonumber \\
&& + \rho^{(0)}\epsilon^{(2)}+\rho^{(2)}\epsilon^{(0)}-\overline{\nabla\cdot\pmb{F}^{(2)}}+\overline{\rho^\prime\epsilon^\prime},
\label{eq:ent2eq}
\end{eqnarray}
where  $g=-(\rho^{(0)})^{-1}\partial p^{(0)}/\partial r=GM_r/r^2$ with
$M_r=\int_0^r4\pi r'^2\rho^{(0)}dr'$. {Because} of the zonal
averaging, we have for non-axisymmetric oscillations
\be
\overline{f^\prime g^\prime}=\frac{1}{2\pi}\int_0^{2\pi}\Re\left(f^\prime(r,\theta)e^{im\phi}\right)
\Re\left(g^\prime(r,\theta)e^{im\phi}\right)d\phi=\frac{1}{2}\Re\left(f^{\prime *}g^\prime\right)=
\frac{1}{2}\Re\left(f^{\prime }g^{\prime *}\right),
\ee
where the asterisk (*) indicates complex conjugation.

\subsubsection{Solving the system of equations}

Since the second-order quantities are assumed axisymmetric, we expand the
velocity perturbation $\pmb{v}^{(2)}$ using spherical harmonic functions $Y_l^0$
as
\be
{v_r^{(2)}}(\pmb{x},t)=\sum_{k=1}^{k_{\rm max}}\hat v_{S,l_k}^{(2)}(r,t)Y_{l_k}^0(\theta,\phi),
\label{eq:expand2r}
\ee
\be
{v_\theta^{(2)}}(\pmb{x},t)
=\sum_{k=1}^{k_{\rm max}}\hat v_{H,l_k}^{(2)}(r,t)\frac{\partial}{\partial\theta}
Y_{l_k}^0(\theta,\phi),
\label{eq:expand2h}
\ee
\be
{v_\phi^{(2)}}(\pmb{x},t)
=-\sum_{k=1}^{k_{\rm max}}  \hat v_{T,l'_k}^{(2)}(r,t)\frac{\partial}{\partial\theta}Y_{l'_k}^0(\theta,\phi),
\label{eq:expand2t}
\ee
and {scalar physical quantities, such {as}} pressure
perturbation $p^{(2)}$ for example, as
\be
p^{(2)}(\pmb{x},t)=\sum_{k=1}^{k_{\rm max}}p_{l_k}^{(2)}(r,t)Y_{l_k}^0(\theta,\phi),
\label{eq:expand2pr}
\ee
where $l_k=2(k-1)$ and $l^\prime_k=l_k+1$ 
for $k=1,~2,~\cdots, ~k_{\rm max}$.
Note that for both even and odd linear modes the $r$ and $\phi$ components of the right-hand-side of equation (\ref{eq:secondmomcon}), for example,
are symmetric about the equator of the star and the $\theta$ component is antisymmetric.
Since
the expansion coefficients of $\pmb{v}^{(2)}$ contain contributions from
products $Y_{l}^{-m}Y_{n}^m$, we have to set the expansion length $k_{\rm
max}\sim 2\times j_{\rm max}$ for the second-order coefficients.
{The} time dependence is included in the expansion coefficient.

{By} substituting the expansions (\ref{eq:expand2r}) to
(\ref{eq:expand2pr}) into equations (\ref{eq:secondmomcon}) to
(\ref{eq:ent2eq}), multiplying {by a given} spherical harmonic function, 
and integrating over spherical surface, we derive a finite set of differential
equations for the expansion coefficients{, which depends on $r$ and $t$}.  To
compute the integral{s corresponding to non-linear terms,}
{such as} $\int (Y_{l_k}^0)^*(\overline{\pmb{v}' \cdot \nabla
\pmb{v}'})_r\sin\theta d\theta d\phi$, we have to evaluate angular integration
of products of three spherical harmonic functions, which can be systematically
carried out by introducing spin-weighted spherical harmonic functions
${}_sY_l^m(\theta,\phi)$, the definition and properties of which are summarized in Appendix A
(see also, e.g., Newman \& Penrose 1966; Varshalovich et al. 1988).
{
In Appendix B1, we introduce a new set of basis vectors $\pmb{e}_r$, $\pmb{e}_q$ and $\pmb{e}_{\bar q}$, 
and rewrite the velocity and displacement vectors, 
and basic equations (\ref{eq:secondmomcon}) to (\ref{eq:ent2eq}) on this
basis. Then, we derive explicit expressions of non-linear terms by using 
spin-weighted spherical harmonics (Appendix B2). This allows us to obtain
ordinary differential equations for the radial functions of the velocity field
and scalar quantities (Appendixes B3 and B4).}

Using vector notation for the dependent variables of second-order defined as
\be
\pmb{z}_1=\left(\matrix{\hat v_{S,{l_1}}^{(2)}/r\sigma_0\cr \hat v_{S,{l_2}}^{(2)}/r\sigma_0\cr \vdots \cr}\right), \quad 
\pmb{z}_2=\left(\matrix{p^{(2)}_{l_1}/gr\rho^{(0)}\cr p^{(2)}_{l_2}/gr\rho^{(0)}\cr \vdots \cr}\right), \quad
\pmb{z}_3=\left(\matrix{L^{(2)}_{r,l_1}/L^{(0)}_r\cr L^{(2)}_{r,l_2}/L^{(0)}_r\cr \vdots \cr}\right), \quad
\pmb{z}_4=\left(\matrix{T^{(2)}_{l_1}/T^{(0)}\cr T^{(2)}_{l_2}/T^{(0)}\cr \vdots \cr}\right), 
\ee
\be
\pmb{z}_h=\left(\matrix{\sqrt{\Lambda_{l_1}}\hat v_{H,{l_1}}^{(2)}/r\sigma_0\cr \sqrt{\Lambda_{l_2}}\hat v_{H,{l_2}}^{(2)}/r\sigma_0\cr \vdots \cr}\right), \quad 
\pmb{z}_t=\left(\matrix{\sqrt{\Lambda_{l'_1}}\hat v_{T,{l'_1}}^{(2)}/r\sigma_0\cr \sqrt{\Lambda_{l'_2}}\hat v_{T,{l'_2}}^{(2)}/r\sigma_0\cr \vdots \cr}\right), \quad
\pmb{\rho}^{(2)}=\left(\matrix{\rho^{(2)}_{l_1}\cr \rho^{(2)}_{l_2}\cr \vdots \cr}\right),
\ee
we rewrite equations (\ref{eq:z1}) to (\ref{eq:z4}) as
\be
r\frac{\partial\pmb{z}_1}{\partial r}=\pmbmt{\Lambda}_0^{1/2}\pmb{z}_h-\left(3+rA-\frac{V}{\Gamma_1}\right)\pmb{z}_1
-\frac{\partial}{\partial\tau}\frac{\pmb{\rho}^{(2)}}{\rho^{(0)}}+\frac{\pmb{H}^0}{\sigma_0\rho^{(0)}},
\label{eq:mean0z1}
\ee
\be
r\frac{\partial\pmb{z}_2}{\partial r}=-\sqrt{2}\bar fc_1\pmbmt{C}_A^1\pmb{z}_t-\left(rA-\frac{V}{\Gamma_1}+U-1\right)\pmb{z}_2
-\frac{\pmb{\rho}^{(2)}}{\rho^{(0)}}-c_1\frac{\partial}{\partial\tau}\pmb{z}_1+\frac{\pmb{G}_r^0}{g},
\label{eq:mean0z2}
\ee
\be
r\frac{\partial{\pmb{z}_3}}{\partial r}=-c_2\left[\frac{\partial{\pmb{z}_4}}{\partial \tau}
-\nabla_{ad}V\frac{\partial\pmb{z}_2}{\partial \tau}
+V\left(\nabla_{ad}-\nabla\right)\pmb{z}_1\right]
+c_3\left[\left(\hat\epsilon_T-\alpha_T\right)
\pmb{z}_4+\left(\hat\epsilon_p+\frac{1}{\chi_\rho}\right)\pmb{z}_2\right]
-\frac{d\ln L_r^{(0)}}{d\ln r}\pmb{z}_3-\frac{\pmbmt{\Lambda}_0}{V\nabla}\pmb{z}_4+\pmb{I}^0,
\label{eq:mean0z3}
\ee
\be
r\frac{\partial\pmb{z}_4}{\partial r}=V\nabla\left(4-\hat\kappa_T+\alpha_T\right)\pmb{z}_4
-V\nabla\left(\hat\kappa_p+\frac{1}{\chi_\rho}\right)V\pmb{z}_2-V\nabla\pmb{z}_3
-V\nabla \pmb{J}^0,
\label{eq:mean0z4}
\ee
and equations (\ref{eq:vt}) and (\ref{eq:vh}) as
\be
-\frac{\partial\pmb{z}_h}{\partial\tau}+\bar f\pmbmt{C}_B^1\pmb{z}_t=\pmbmt{\Lambda}_0^{1/2}\frac{\pmb{z}_2}{c_1}
+\frac{\left(\pmb{G}_q^0-\pmb{G}_{\bar q}^0\right)}{\sqrt{2} g c_1},
\label{eq:dzhdt}
\ee
\be
-\bar f\pmbmt{C}_B^0\pmb{z}_h-
\frac{\partial\pmb{z}_t}{\partial\tau}=-\sqrt{2}\bar f\pmbmt{C}_C^0\pmb{z}_1
+\frac{\left(\pmb{G}_q^1+\pmb{G}_{\bar q}^1\right)}{\sqrt{2} \rmi g c_1},
\label{eq:dztdt}
\ee
where the inhomogeneous terms
$\pmb{H}^0$, $\pmb{I}^0$, $\pmb{J}^0$, $\pmb{G}_X^0$, and $\pmb{G}_X^1$ with $X\equiv\left\{r,~q,~{\bar q}\right\}$ are vectors whose $k$-th components are respectively given by  $H_{l_k}^{(2)}$, 
$I^{(2)}_{l_k}$, $J_{l_k}^{(2)}$, $G_{X,l_k}^{(2)}$, and $G_{X,l^\prime_k}^{(2)}$, which are defined in
Appendix B4, and
$\pmbmt{\Lambda}_0$ and $\pmbmt{\Lambda}_0^{1/2}$ are diagonal matrices whose $k$-th
diagonal components are given by $\Lambda_{l_k}=l_k\left(l_k+1\right)$ and
$\sqrt{\Lambda_{l_k}}$, respectively.
Note that
$G_{q,l^\prime_k}^{(2)}-G_{\bar q,l^\prime_k}^{(2)}=0$ and
$G_{q,l_k}^{(2)}+G_{\bar q,l_k}^{(2)}=0$.
The symbols $\pmbmt{C}_A^1$,
$\pmbmt{C}_B^0$, $\pmbmt{C}_B^1$, and $\pmbmt{C}_C^0$ denote matrices whose $kj$ components are
respectively given by $C_A^{l_kl^\prime_j}$, $C_B^{l^\prime_kl_j}$, $C_B^{l_kl^\prime_j}$,
and $C_C^{l^\prime_kl_j}$, where 
$
C_A^{l_kl_j}\equiv C_{0(-1)1}^{l_kl_j1}+C_{01(-1)}^{l_kl_j1},
$
$
C_B^{l_kl_j}\equiv C_{1(-1)0}^{l_kl_j1}+C_{(-1)10}^{l_kl_j1},
$
$
C_C^{l_kl_j}\equiv C_{10(-1)}^{l_kl_j1}+C_{(-1)01}^{l_kl_j1},
$
and 
the definition of the coefficient $C_{abc}^{ijk}$ is given by (\ref{eq:csquare})
in Appendix B3.
The physical quantities $V$, $U$, $rA$, $\Gamma_1$, $\tau$, $\bar f$, and $\sigma_0$ in the set of equations (\ref{eq:mean0z1}) to (\ref{eq:dztdt})
are defined as
\be
V=-{d\ln p^{(0)}\over d\ln r}, \quad U=\frac{d\ln M_r}{d\ln r}, \quad rA=\frac{d\ln\rho^{(0)}}{d\ln r}-\frac{1}{\Gamma_1}\frac{d\ln p^{(0)}}{d\ln r}, \quad
\Gamma_1=\left(\frac{\partial\ln p}{\partial\ln\rho}\right)_{ad}, \quad \tau=\sigma_0 t, \quad \bar f=\sqrt{4\pi\over 3}{\Omega\over\sigma_0},
\ee
and $\sigma_0=\sqrt{GM/R^3}$, and the definition of the other physical quantities is
given in Appendix B1.

The set of differential equations derived above are partial differential
equations with $r$ and $t$ being the independent variables. Since we are
interested in mean flows driven by an unstable linear oscillation mode having a
complex eigenfrequency $\omega$, the forcing (inhomogeneous) terms in the
equations are proportional to $e^{-2\omega_{\rm I}t}$, where $\omega_{\rm I}$
denotes the imaginary part of $\omega$. To make analyses simple, we look for
solutions whose time dependence is also given by the factor $e^{-2\omega_{\rm
I}t}$, that is, we assume that the partial derivatives $\partial/\partial\tau$
{is} replaced by the growth or damping rate
$\bar\gamma=-2\bar\omega_{\rm I}$. 
This simplifying assumption may be justified when we are interested in mean flows driven by self-excited oscillation modes.
Since such self-excited oscillation modes must be continually pumped by certain destabilizing mechanisms (e.g., opacity
mechanism) even if their amplitudes are saturated by some mechanisms like nonlinear couplings between many different modes, we believe that we have to take account of the effects of this continual pumping of 
the oscillation modes by introducing the growth rate into the formulation. 
From equations (\ref{eq:dzhdt}) and
(\ref{eq:dztdt}), we obtain
\be
\pmb{z}_h
=\pmbmt{Z}_{11}\pmb{z}_1+\pmbmt{Z}_{12}\pmb{z}_2/c_1+\pmb{Z}_{13}, \label{eq:meanzh}
\ee
\be
\pmb{z}_t
=\pmbmt{Z}_{21}\pmb{z}_1+\pmbmt{Z}_{22}\pmb{z}_2/c_1+\pmb{Z}_{23}, \label{eq:meanzt}
\ee
where
\be
\pmbmt{Z}_{11}=\sqrt{2}\bar f^2\pmbmt{W}^{-1}_{10}\pmbmt{C}_B^1\pmbmt{C}_C^0, \quad 
\pmbmt{Z}_{12}=-\bar\gamma\pmbmt{W}^{-1}_{10}\pmbmt{\Lambda}_0^{1/2}, \quad
\pmb{Z}_{13}=-\bar\gamma\pmbmt{W}^{-1}_{10}\frac{\pmb{G}_q^0-\pmb{G}_{\bar q}^0}{\sqrt{2}gc_1}
-\bar f\pmbmt{W}^{-1}_{10}\pmbmt{C}_B^1\frac{\pmb{G}_q^1+\pmb{G}_{\bar q}^1}{\sqrt{2}\rmi g c_1},
\ee
\be
\pmbmt{Z}_{21}=\sqrt{2}\bar\gamma \bar f\pmbmt{W}^{-1}_{01}\pmbmt{C}_C^0, \quad
\pmbmt{Z}_{22}=\bar f\pmbmt{W}^{-1}_{01}\pmbmt{C}_B^0\pmbmt{\Lambda}_0^{1/2}, \quad
\pmb{Z}_{23}=\bar f\pmbmt{W}^{-1}_{01}\pmbmt{C}_B^0\frac{\pmb{G}_q^0-\pmb{G}_{\bar q}^0}{\sqrt{2} gc_1}
-\bar\gamma\pmbmt{W}^{-1}_{01}\frac{\pmb{G}_q^1+\pmb{G}_{\bar q}^1}{\sqrt{2}\rmi g c_1},
\ee
\be
\pmbmt{W}_{10}=\bar f^2\pmbmt{C}_B^1\pmbmt{C}_B^0+\bar\gamma^2\pmbmt{E}, \quad
\pmbmt{W}_{01}=\bar f^2\pmbmt{C}_B^0\pmbmt{C}_B^1+\bar\gamma^2\pmbmt{E},
\ee
and $\pmbmt{E}$ is the unit matrix. Substituting equations (\ref{eq:meanzh}) and
(\ref{eq:meanzt}) into equations (\ref{eq:mean0z1}) to (\ref{eq:mean0z4}),  we
finally obtain
\be
r\frac{\partial\pmb{z}_1}{\partial r}=\left[-\left(3+rA-\frac{V}{\Gamma_1}\right)\pmbmt{E}+\pmbmt{\Lambda}_0^{1/2}\pmbmt{Z}_{11}\right]\pmb{z}_1
-\left(\bar\gamma\frac{V}{\chi_\rho}\pmbmt{E}-\frac{\pmbmt{\Lambda}_0^{1/2}\pmbmt{Z}_{12}}{c_1}\right){\pmb{z}_2}
+\bar\gamma\frac{\chi_T}{\chi_\rho}\pmb{z}_4
-\bar\gamma\pmb{R}^{0}+\pmbmt{\Lambda}_0^{1/2}\pmb{Z}_{13}+\frac{\pmb{H}^0}{\sigma_0\rho},
\label{eq:meanz1}
\ee
\be
r\frac{\partial\pmb{z}_2}{\partial r} = -c_1\left(\sqrt{2}\bar f\pmbmt{C}_A^1\pmbmt{Z}_{21}+\bar\gamma\pmbmt{E}\right)\pmb{z}_1
-\left[\left(rA-\frac{V}{\Gamma_1}+U-1+\frac{V}{\chi_\rho}\right)\pmbmt{E}+\sqrt{2}\bar f\pmbmt{C}_A^1\pmbmt{Z}_{22}\right]\pmb{z}_2 
+\frac{\chi_T}{\chi_\rho}\pmb{z}_4-\sqrt{2}\bar fc_1\pmbmt{C}_A^1\pmb{Z}_{23}-\pmb{R}^{0}+\frac{\pmb{G}_r^0}{g},
\label{eq:meanz2}
\ee
\be
r\frac{\partial{\pmb{z}_3}}{\partial r}=-c_2\left[\bar\gamma\left({\pmb{z}_4}
-\nabla_{ad}V{\pmb{z}_2}\right)
+V\left(\nabla_{ad}-\nabla\right)\pmb{z}_1\right]
+c_3\left[\left(\hat\epsilon_T-\alpha_T\right)
\pmb{z}_4+\left(\hat\epsilon_p+\frac{1}{\chi_\rho}\right)\pmb{z}_2\right]
-\frac{d\ln L_r^{(0)}}{d\ln r}\pmb{z}_3-\frac{\pmbmt{\Lambda}_0}{V\nabla}\pmb{z}_4+\pmb{I}^0,
\label{eq:meanz3}
\ee
\be
r\frac{\partial\pmb{z}_4}{\partial r}=
-V\nabla\left(\hat\kappa_p+\frac{1}{\chi_\rho}\right)V\pmb{z}_2-V\nabla\pmb{z}_3
+V\nabla\left(4-\hat\kappa_T+\alpha_T\right)\pmb{z}_4
-V\nabla \pmb{J}^0,
\label{eq:meanz4}
\ee
where we have used
\be
\frac{\pmb{\rho}^{(2)}}{\rho^{(0)}}=-\frac{\chi_T}{\chi_\rho}{\pmb{z}_4}+\frac{V}{\chi_\rho}\pmb{z}_2
+\pmb{R}^{0},
\ee
and the $k$-th component of the vector $\pmb{R}^{0}$ is given by
\be
R_k=\int{}_0Y_k^0Q^{(2)}(\rho)d\Omega,
\ee
and the definition of the symbol $Q^{(2)}(\rho)$ is given by (\ref{eq:q(2)}).
The set of differential equations from (\ref{eq:meanz1}) to (\ref{eq:meanz4}) is
regarded as the mean flow equation solved in this paper.
The matrix $\hat {\pmbmt{W}}\equiv -\gamma\pmbmt{\Lambda}_0^{1/2}\pmbmt{Z}_{12}$ is a
symmetric matrix corresponding to the matrix $\pmbmt{W}$ discussed, for example, by
Lee \& Saio (1997) in the traditional approximation. In fact, replacing
$\gamma$ by $\rmi \omega$ makes $\hat {\pmbmt{W}}$ reduce to $\pmbmt{W}$. We also
note that $\sqrt{2}\bar f\pmbmt{C}_A^1\pmbmt{Z}_{22} =
\left(\pmbmt{\Lambda}_0^{1/2}\pmbmt{Z}_{11}\right)^T$.

For the boundary conditions applied at the stellar center, we require that the
functions $\pmb{z}_1$ and $\pmb{z}_2$ are regular and that $ds/dt=0$, which
leads to
\be
\bar\gamma\left(\frac{T^{(2)}}{T^{(0)}}-\nabla_{ad}V\frac{p^{(2)}}{\rho^{(0)}gr}\right)
-\frac{v_r^{(2)}}{r\sigma_0}V\left(\nabla-\nabla_{ad}\right)=-\frac{\overline{\rmi\omega\pmb{\xi}\cdot\nabla s^\prime}}{c_p}
-\bar\gamma\frac{s^{(0)}}{c_p}Q^{(2)}(s).
\ee
The outer boundary conditions applied at the stellar surface are given by the
conditions $dp/dt=0$ and $L_r=4\pi r^2\sigma_{\rm SB}T^4$, from which we derive
\be
\bar\gamma\frac{p^{(2)}}{\rho^{(0)}gr}-\frac{v_r^{(2)}}{r\sigma_0}=-\frac{\overline{\rmi \bar\omega\pmb{\xi}\cdot\nabla p^\prime}}{\rho^{(0)}gr}
\ee
and
\be
\frac{L_r^{(2)}}{L_r^{(0)}}-4\frac{T^{(2)}}{T^{(0)}}=-\frac{\overline{\pmb{\xi}\cdot\nabla L_r^{(1)}}}{L_r^{(0)}}+
4\frac{1}{T^{(0)}}\overline{\left(\pmb{\xi}\cdot\nabla\delta T^{(1)}+\frac{1}{ 2}\frac{\partial^2T^{(0)}}{\partial r^2}\xi_r\xi_r\right)}
+6\overline{\left(\frac{\delta T^{(1)}}{T^{(0)}}\right)^2}+\overline{\left(\frac{\xi_r}{r}\right)^2}+8\overline{\frac{\xi_r}{r}\frac{\delta T^{(1)}}{T^{(0)}}},
\ee
where
\be
\delta T^{(1)}=T^{(1)}+\xi_r\frac{dT^{(0)}}{dr},
\ee
and we have assumed { that} $L_r^{(0)}$ is constant in the outer envelope.

\subsubsection{Lagrangian perturbations of second-order}

Assuming { that} the Lagrangian displacement vector $\pmb{\xi}(\pmb{x})$ is
infinitesimal, we may expand the perturbed velocity field $\pmb{v}=(v_i)$ at
$\pmb{x}+\pmb{\xi}$ as
\begin{eqnarray}
v_i\left(\pmb{x}+\pmb{\xi}\right) & =& v_i\left(\pmb{x}\right)+v_{i;j}(\pmb{x})\xi_j(\pmb{x})
+\frac{1}{2}v_{i;j;k}(\pmb{x})\xi_j(\pmb{x})\xi_k(\pmb{x})+\cdots\nonumber \\
& =& v^{(0)}_i(\pmb{x})+v_i^{(1)}(\pmb{x})+v_i^{(2)}(\pmb{x})+v_{i;j}^{(0)}(\pmb{x})\xi_j(\pmb{x})+v_{i;j}^{(1)}(\pmb{x})\xi_j(\pmb{x})+\frac{1}{2}v^{(0)}_{i;j;k}(\pmb{x})\xi_j(\pmb{x})\xi_k(\pmb{x})+\cdots,
\end{eqnarray}
where the semicolon indicates the covariant derivative, and repeated
indices imply the summation from 1 to 3. {The} perturbed velocity field
$\pmb{v}(\pmb{x})$ is expanded in terms of the oscillation amplitude, which is
also assumed {to be} small, as
\be
v_i\left(\pmb{x}\right)=v_i^{(0)}\left(\pmb{x}\right)+v_i^{(1)}(\pmb{x})+v_i^{(2)}(\pmb{x})+\cdots,
\ee
{where} $v_i^{(0)}$ is the unperturbed field. Carrying out zonal averaging,
we obtain to {the} second-order of the wave amplitude
\be
\overline{v_i(\pmb{x}+\pmb{\xi})}=v_i^{(0)}+v_i^{(2)}+\overline{v_{i;j}^{(1)}(\pmb{x})\xi_j(\pmb{x})}+
\frac{1}{2}\overline{v_{i;j;k}^{(0)}(\pmb{x})\xi_j(\pmb{x})\xi_k(\pmb{x})},
\ee
where we have used $\overline{v_i^{(1)}}=0$.
Hence, the zonally averaged Lagrangian velocity perturbation of second-order is defined as
\be
\delta v_i^{(2)}\equiv\overline{v_i(\pmb{x}+\pmb{\xi})}-v_i^{(0)}(\pmb{x})=v_i^{(2)}(\pmb{x})+\overline{v_{i;j}^{(1)}(\pmb{x})\xi_j(\pmb{x})}+
\frac{1}{2}\overline{v_{i;j;k}^{(0)}(\pmb{x})\xi_j(\pmb{x})\xi_k(\pmb{x})},
\label{eq:deltav2}
\ee
where
\be
\overline{v_{i;j}^{(1)}(\pmb{x})\xi_j(\pmb{x})}=\overline{\rmi\omega \xi_{i;j}\xi_j}=\frac{1}{2}{\rm Re}\left(\rmi\omega\xi_{i;j}\xi_j^*\right).
\ee
The additional terms $\overline{v_{i;j}^{(1)}\xi_j}+
0.5\overline{v_{i;j;k}^{(0)}\xi_j\xi_k}$ in equation (\ref{eq:deltav2}) are
called Stokes corrections (or Stokes drift for velocity field).

\subsection{Angular momentum conservation in wave-mean flow interaction}

If we employ the Lagrangian mean theory of wave-meanflow interaction, in which the position vector $\hat{\pmb{x}}$
is divided into the mean $\pmb{x}=\overline{\hat{\pmb{x}}}$ and the Lagrangian displacement $\pmb{\xi}(\pmb{x},t)$, that is,
$\hat{\pmb{x}}=\pmb{x}+\pmb{\xi}(\pmb{x},t)$,
we may obtain in the Cowling approximation a mean flow equation
given by (see, e.g., Lee 2013; see also Grimshaw 1984)
\be
\tilde\rho{d\over dt}\overline{\ell\left(\pmb{x}+\pmb{\xi}\right)}=-\nabla\cdot\overline{\left(\pmb{\xi}{\partial p'\over\partial\phi}\right)},
\label{eq:meanfloweq}
\ee
where $\tilde\rho$ is the effective density, for which $\overline{\tilde\rho}=\tilde\rho$ (e.g., B\"uhler 2014), and
the total time derivative $d/dt$ is defined as
\be
{d\over dt}={\partial\over\partial t}+\overline{v_r(\pmb{x}+\pmb{\xi})}{\partial\over\partial r}
+\overline{v_\theta(\pmb{x}+\pmb{\xi})}{1\over r}{\partial\over\partial \theta}
+\overline{v_\phi(\pmb{x}+\pmb{\xi})}{1\over r\sin\theta}{\partial\over\partial \phi},
\nonumber
\ee
and, if $v_r^{(0)}=v_\theta^{(0)}=0$ as assumed for uniformly rotating stars,
\be
\overline{v_r(\pmb{x}+\pmb{\xi})}=\delta v_r^{(2)}, \quad 
\overline{v_\theta(\pmb{x}+\pmb{\xi})}=\delta v_\theta^{(2)}.
\nonumber
\ee
In the following, we set $\tilde\rho=\rho^{(0)}$ for simplicity (e.g., B\"uhler 2014).
The zonally averaged specific angular momentum in the $z$-direction at $\pmb{x}+\pmb{\xi}$ 
is given, to second order of perturbation amplitudes, by
\begin{eqnarray}
\overline{\ell(\pmb{x}+\pmb{\xi})}&=&\overline{\left[\left(\pmb{x}+\pmb{\xi}\right)\times\pmb{v}\left(\pmb{x}+\pmb{\xi}\right)\right]}\cdot\pmb{e}_z\nonumber\\
&\approx&\overline{\left(v_\phi^{(0)}+\delta v_\phi^{(1)}+\delta v_\phi^{(2)}\right)\left[\xi_\theta\cos\theta+\left(r+\xi_r\right)\sin\theta\right]}-\overline{\delta v_\theta^{(1)}\xi_\phi}\cos\theta-\overline{\delta v_r^{(1)}\xi_\phi}\sin\theta
\nonumber\\
&=&\ell^{(0)}+\ell^{(2)}
\end{eqnarray}
where $\pmb{e}_z$ is a unit vector along the $z$-axis, and
\be
\ell^{(0)}=\left(r\sin\theta\right)^2\Omega, \label{eq:l0}
\nonumber
\ee
\begin{eqnarray}
\ell^{(2)}
=r\sin\theta v_\phi^{(2)}+r\sin\theta\overline{v^{(1)}_{\phi;j}\xi_j}
+\left[\overline{\left(v_\phi^{(1)}\xi_r-\xi_\phi v_r^{(1)}\right)\sin\theta
+\left(v^{(1)}_\phi\xi_\theta-\xi_\phi v_\theta^{(1)}\right)\cos\theta}\right]
+\left[\overline{\left(\xi_r\sin\theta+\xi_\theta\cos\theta\right)^2+\xi_\phi^2}\right]\Omega,\label{eq:l2}
\end{eqnarray}
and
\be
\delta v_a^{(1)}=v_a^{(1)}+ v_{a;j}^{(0)}\xi_j\quad {\rm for}\quad a=r,~\theta,~\phi.
\ee

The mean flow equation (\ref{eq:meanfloweq}) is now given, correct to second order of perturbation amplitudes, by
\be
{\partial\over\partial t}\ell^{(2)}+\left(\delta v_r^{(2)}{\partial\over\partial r}
+{\delta v_\theta^{(2)}\over r}{\partial\over\partial\theta}\right)\ell^{(0)}=
-{1\over \rho^{(0)}}\nabla\cdot\overline{\left(\pmb{\xi}{\partial p'\over\partial\phi}\right)}.
\label{eq:meanfloweq2}
\nonumber
\ee
Making use of equations (\ref{eq:l0}) and (\ref{eq:l2}), we rewrite equation
(\ref{eq:meanfloweq2})
as
\begin{eqnarray}
&&r\sin\theta\overline{
{\partial v_{\phi;j}^{(1)}\over\partial t}\xi_j}+{1\over\rho^{(0)}}\overline{{\rho^\prime\over\rho^{(0)}}{\partial p^\prime\over\partial \phi}}
+
\left[\overline{\left({\partial v_\phi^{(1)}\over\partial t}\xi_r-\xi_\phi {\partial v_r^{(1)}\over\partial t}\right)\sin\theta
+\left({\partial v^{(1)}_\phi\over\partial t}\xi_\theta-\xi_\phi {\partial v_\theta^{(1)}\over\partial t}\right)\cos\theta}\right]
\nonumber\\
&&+2\Omega\left[\overline{\left(\xi_r\sin\theta+\xi_\theta\cos\theta\right)
\left(v_r^{(1)}\sin\theta+v_\theta^{(1)}\cos\theta\right)
+\xi_\phi v_\phi^{(1)}}\right]
+
2r\Omega \left(\overline{v_{r:j}^{(1)}\xi_j}\sin^2\theta+\overline{v_{\theta;j}^{(1)}\xi_j}\sin\theta\cos\theta\right)
\nonumber\\
&=&-{1\over \rho^{(0)}}\nabla\cdot\overline{\left(\pmb{\xi}{\partial p'\over\partial\phi}\right)},
\label{eq:lhsmfe2}
\end{eqnarray}
where we have used, to eliminate the second order perturbations $v_r^{(2)}$, $v_\theta^{(2)}$, and $v_\phi^{(2)}$,
the $\phi$-component of the equation of motion:
\be
\sin\theta{\partial\over\partial t}v_\phi^{(2)}+2\Omega\sin\theta\left(\cos\theta v_\theta^{(2)}
+\sin\theta v_r^{(2)}\right)=-\sin\theta\overline{v_{\phi;j}^{(1)}v_j^{(1)}}
+{1\over\rho^{(0)}r}\overline{{\rho^\prime\over\rho^{(0)}}{\partial p^\prime\over\partial \phi}}.
\label{eq:phicomp2}
\ee
The left-hand-side of (\ref{eq:lhsmfe2}) is given by
the sum of products of first order perturbations associated with the oscillation mode.
Using the perturbed continuity equation $\rho^\prime=-\nabla\cdot(\rho^{(0)}\pmb{\xi})$ and 
the $\phi$-component of the equation of motion for
first order perturbations,
we rewrite the term $(\rho^{(0)})^{-2}\overline{\rho'\partial p'/\partial\phi}$ in equation (\ref{eq:lhsmfe2}) as
\begin{eqnarray}
{1\over\rho^{(0)}}\overline{{\rho^\prime\over\rho^{(0)}}{\partial p^\prime\over\partial\phi}}&=&-{1\over\rho^{(0)}}\nabla\cdot\overline{\left(\pmb{\xi}{\partial p^\prime\over\partial\phi}\right)}+\overline{\pmb{\xi}\cdot\nabla\left({1\over\rho^{(0)}}{\partial p^\prime\over\partial\phi}\right)}\nonumber\\
&=&-{1\over\rho^{(0)}}\nabla\cdot\overline{\left(\pmb{\xi}{\partial p^\prime\over\partial\phi}\right)}
-\overline{\pmb{\xi}\cdot\nabla\left[r\sin\theta{\partial v_\phi^{(1)}\over\partial t}+2r\sin\theta\Omega\left(\sin\theta v_r^{(1)}
+\cos\theta v_\theta^{(1)}\right)\right]},
\label{eq:rhoprimepprime}
\end{eqnarray}
which we substitute into equation (\ref{eq:lhsmfe2}) to
prove the identity.

We may use the meanflow equation (\ref{eq:lhsmfe2}) or (\ref{eq:meanfloweq2}) along with equation (\ref{eq:phicomp2}) 
to check numerical consistency.
To simplify the computation, we integrate equation (\ref{eq:meanfloweq2}) over spherical surface to obtain
\be
{\partial\over\partial t}\left<\ell^{(2)}\right>+2r\Omega\left<\left(\delta v_r^{(2)}\sin^2\theta
+{\delta v_\theta^{(2)}}\sin\theta\cos\theta\right)\right>= m{1\over 2\pi r^2\rho^{(0)}}{\partial\over\partial r}W(r),
\label{eq:l2eq}
\nonumber
\ee
where
\be
W(r)=\pi r^2{\rm Im}\left(\left<\xi_r^* p^\prime\right>\right),
\ee
and
\be
\left<f\right>=\int_0^\pi\int_0^{2\pi}f\sin\theta d\theta d\phi.
\nonumber
\ee
The function $W(r)$ may be regarded as a work function (e.g., Unno et al 1989), and 
$dW/dr>0$ and $dW/dr<0$ respectively indicate the excitation and damping regions for the 
oscillation mode.
We use equation (\ref{eq:l2eq}), in stead of (\ref{eq:meanfloweq2}), to see the numerical consistency.

\begin{figure}
\begin{center}
\resizebox{0.5\columnwidth}{!}{
\includegraphics{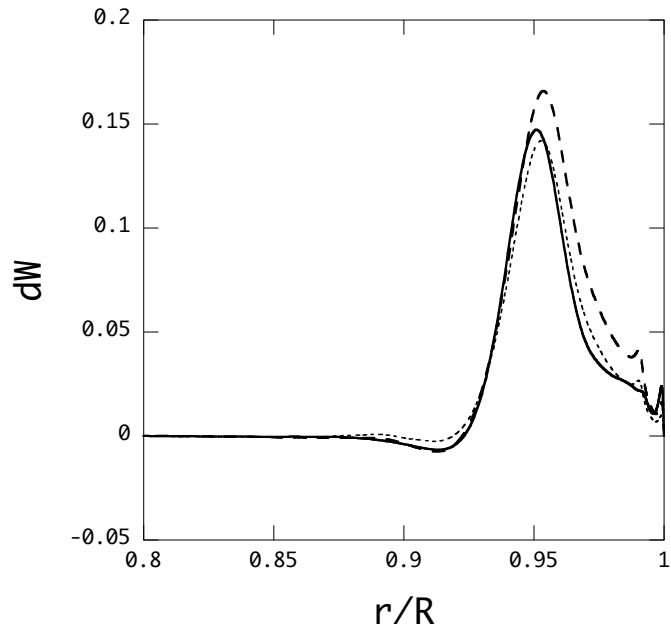}}
\end{center}
\caption{$d{\cal W}$ for the prograde $l=|m|=2$ $p$-mode of the $\beta$ Cephei star model (solid line) and for
the retrograde $l'=m=2$ $r_{36}$-mode (dashed line) and the prograde $l=|m|=2$ $g_{30}$-mode (dotted line) of the SPB star model,
where we assumed $\bar\Omega=0.1$ for the $p$- and $g$-modes and $\bar\Omega=0.4$ for the $r$-mode.
We plot $5\times d{\cal W}$ for the $p$-mode.
} 
\end{figure}

For later use, we rewrite (\ref{eq:l2eq}) into a non-dimensional form
$
{\cal L}={\cal R},
$
where
\be
{\cal L}=c_1\bar\Omega\bar\gamma\left<\ell_0^{(2)}\right>+2c_1\bar\Omega\left<\left({\delta v_r^{(2)}\over r\sigma_0}\sin\theta
+{\delta v_\theta^{(2)}\over r\sigma_0}\cos\theta\right)\sin\theta\right>,
\label{eq:ll}
\ee
\be
{\cal R}={m\over 2}\left<r{\partial\over\partial r}{\rm Im}\left({\xi^*_r\over r}{p'\over\rho g r}\right)
+\left({d\ln\rho\over d\ln r}+U+2\right){\rm Im}\left({\xi^*_r\over r}{p'\over\rho g r}\right)\right>
\equiv md{\cal W},
\label{eq:rdw}
\ee
and
\be
\left<\ell_0^{(2)}\right>=\left<\ell^{(2)}\right>/r^2\Omega.
\nonumber
\ee
It may be useful to plot the function $d{\cal W}$, as defined by equation (\ref{eq:rdw}), to indicate the locations of mode excitation and damping regions
in the outer envelope of the stars for the oscillation modes discussed in this paper.
Figure 1 shows $d{\cal W}$ for the prograde $l=|m|=2$ $g_{30}$-mode and retrograde $l^\prime=m=2$ 
$r_{36}$-mode of
a $6M_\odot$ SPB star model and for a prograde $l=|m|=2$ $p$-mode of a $15M_\odot$ $\beta$ Cephei model, where
we assume $\bar\Omega=0.1$ for the $g$- and $p$-modes and $\bar\Omega=0.4$ for the $r$-mode,
and the physical parameters of the SPB star and $\beta$ Cephei star models are given in \S 3.1 and \S 3.2,
respectively.
As shown by the figure, the strong excitation occurs in the region at $r/R\sim 0.95$ and a damping takes place at $r/R\sim 0.92$ below the 
excitation region.

\section{Applications to pulsating massive stars}

To show an application of the above formalism to rotating
massive stars, we apply it to rotating, slowly pulsating B (SPB) stars and $\beta$ Cephei stars.
As {background equilibrium} models for oscillation calculation, we use
stellar models computed with a standard stellar evolution code, where no effects
of rotational deformation are considered in evolution calculation. The opacity
tables used for evolution and oscillation calculations are those computed by Iglesias \& Rogers (1996).
In this paper, we are interested in
axisymmetric mean flows driven by non-axisymmetric oscillation modes of rotating
stars, where the oscillation modes are excited by the $\kappa$-mechanism
associated with the iron opacity bump. Non-adiabatic oscillations of uniformly
rotating stars are computed with the method employed in Lee \& Saio (1987).
The quantity $f$ in that latter paper is set to 1
so that the effects of the centrifugal acceleration and of
the corresponding rotational deformation of the equilibrium structure are not
taken into account in our computations of the oscillations (see e.g., Lee \& Baraffe 1995).

\begin{figure}
\begin{center}
\resizebox{0.33\columnwidth}{!}{
\includegraphics{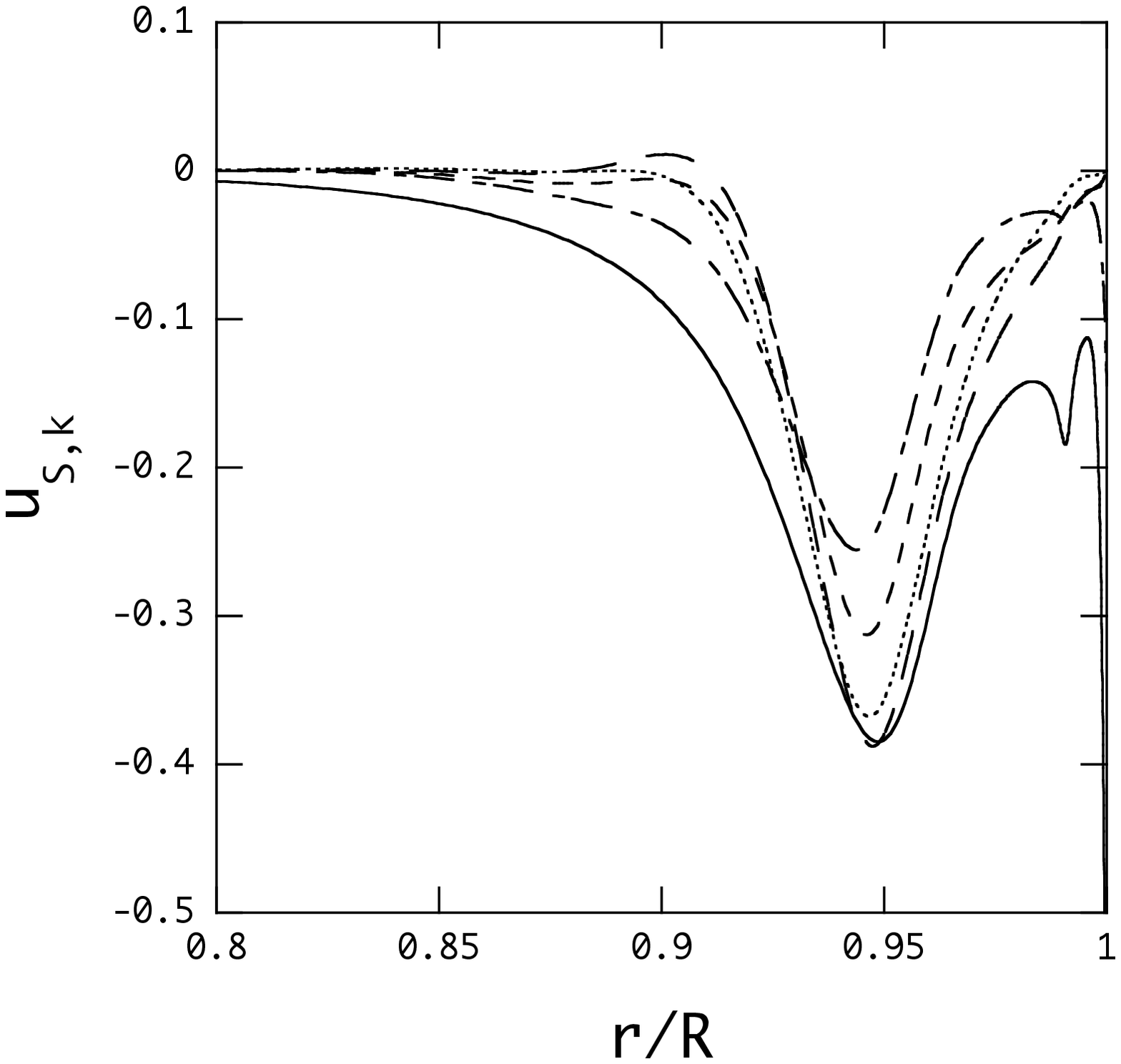}}
\resizebox{0.33\columnwidth}{!}{
\includegraphics{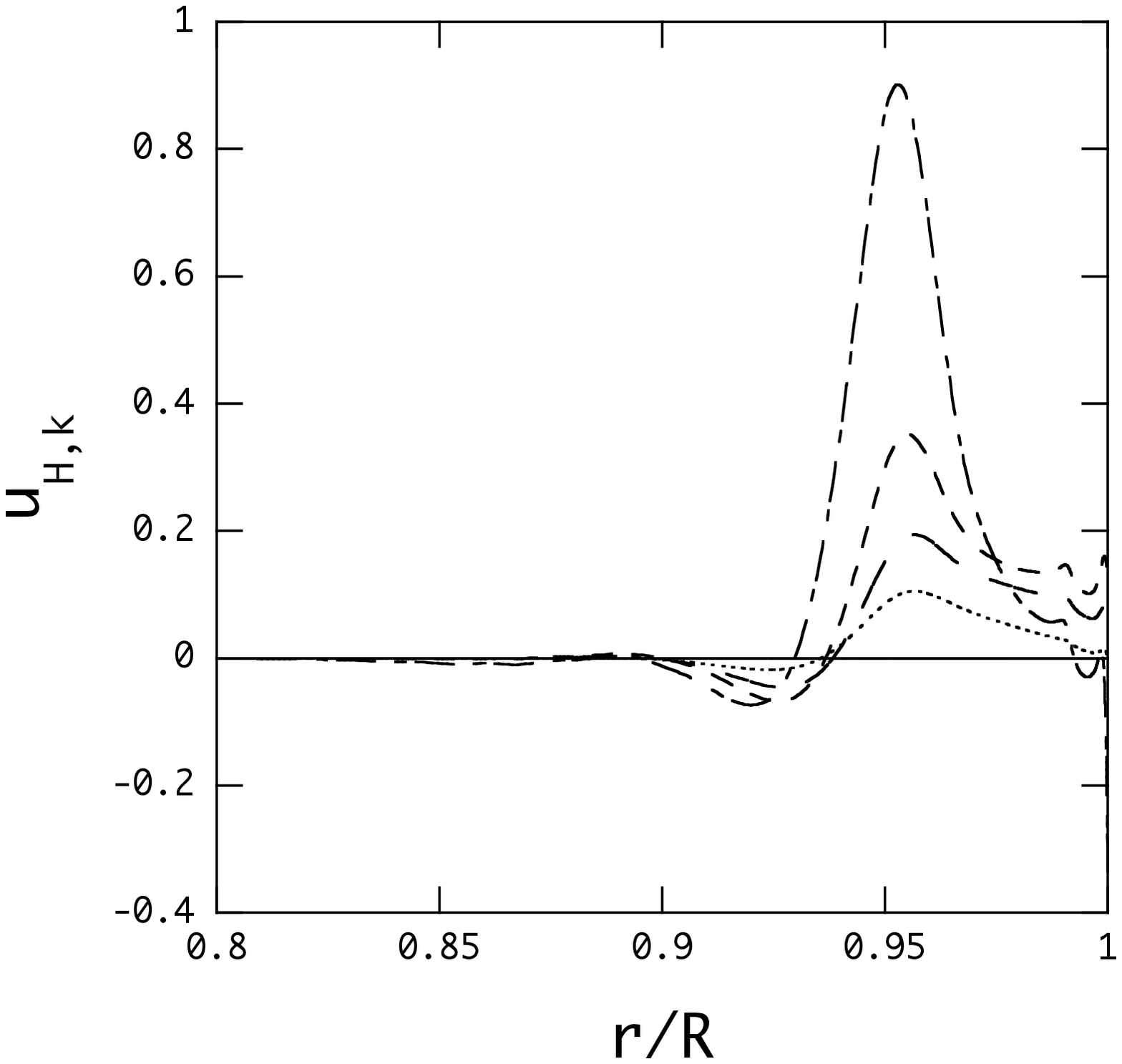}}
\resizebox{0.33\columnwidth}{!}{
\includegraphics{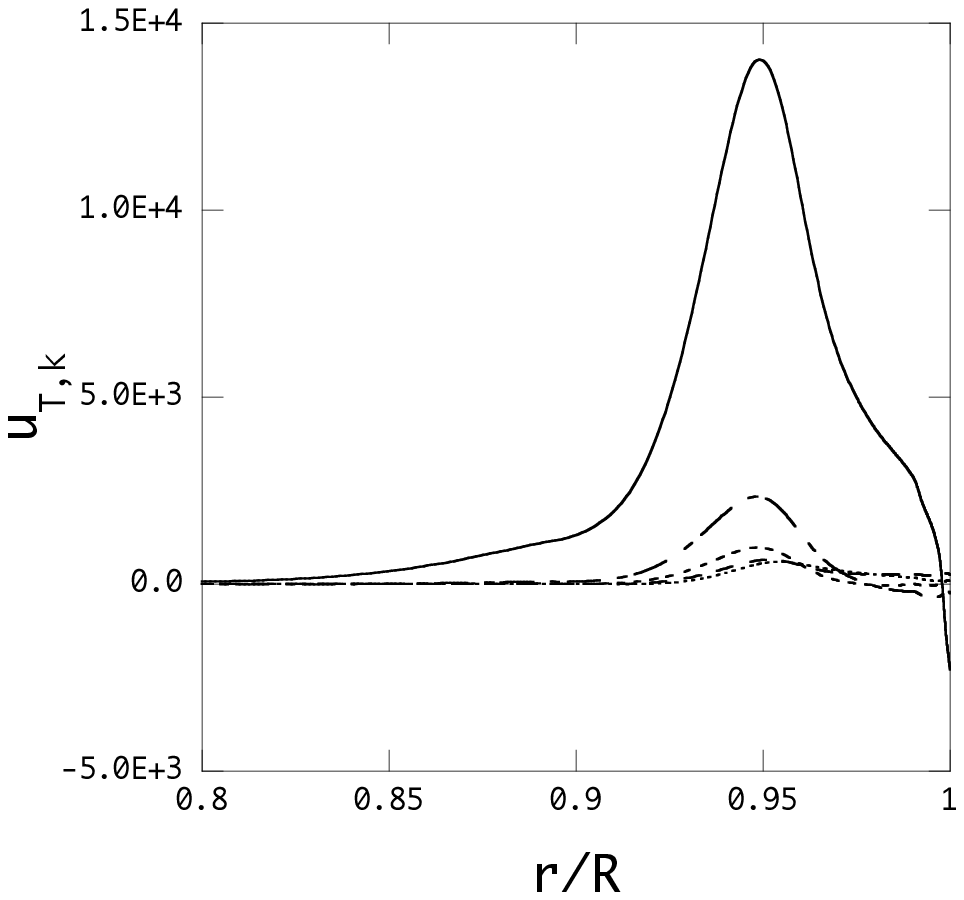}}
\end{center}
\begin{center}
\resizebox{0.33\columnwidth}{!}{
\includegraphics{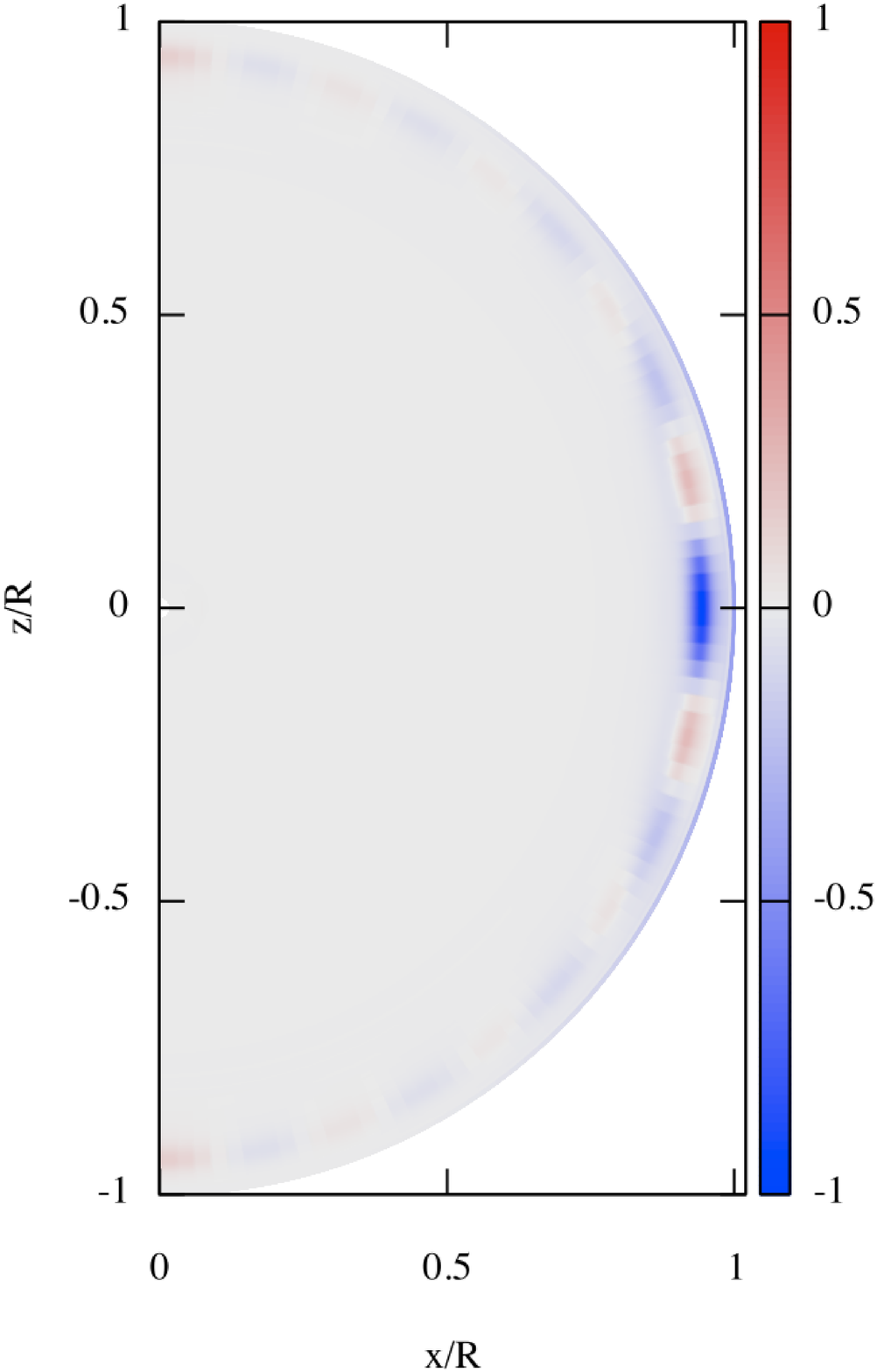}}
\resizebox{0.33\columnwidth}{!}{
\includegraphics{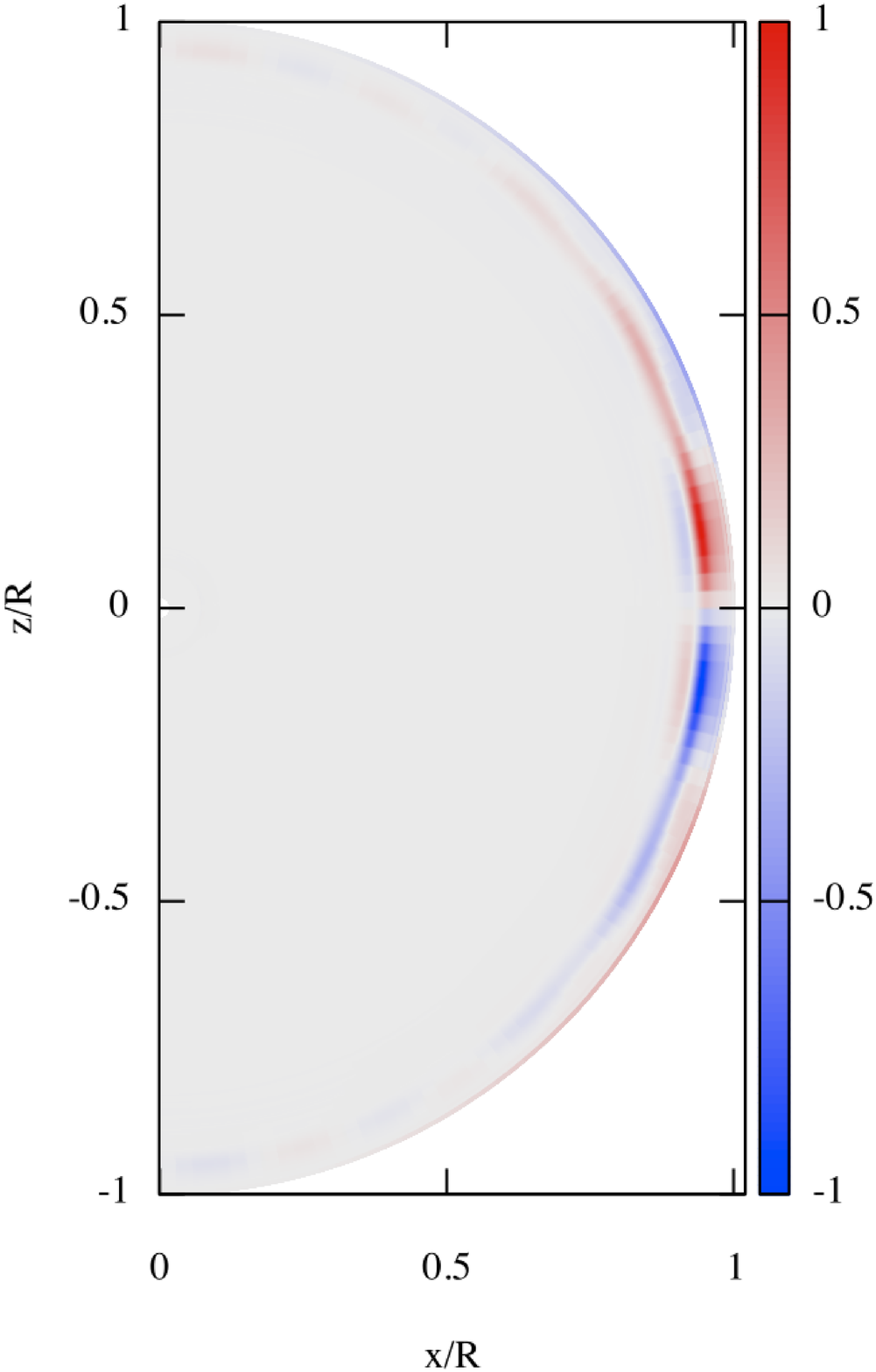}}
\resizebox{0.33\columnwidth}{!}{
\includegraphics{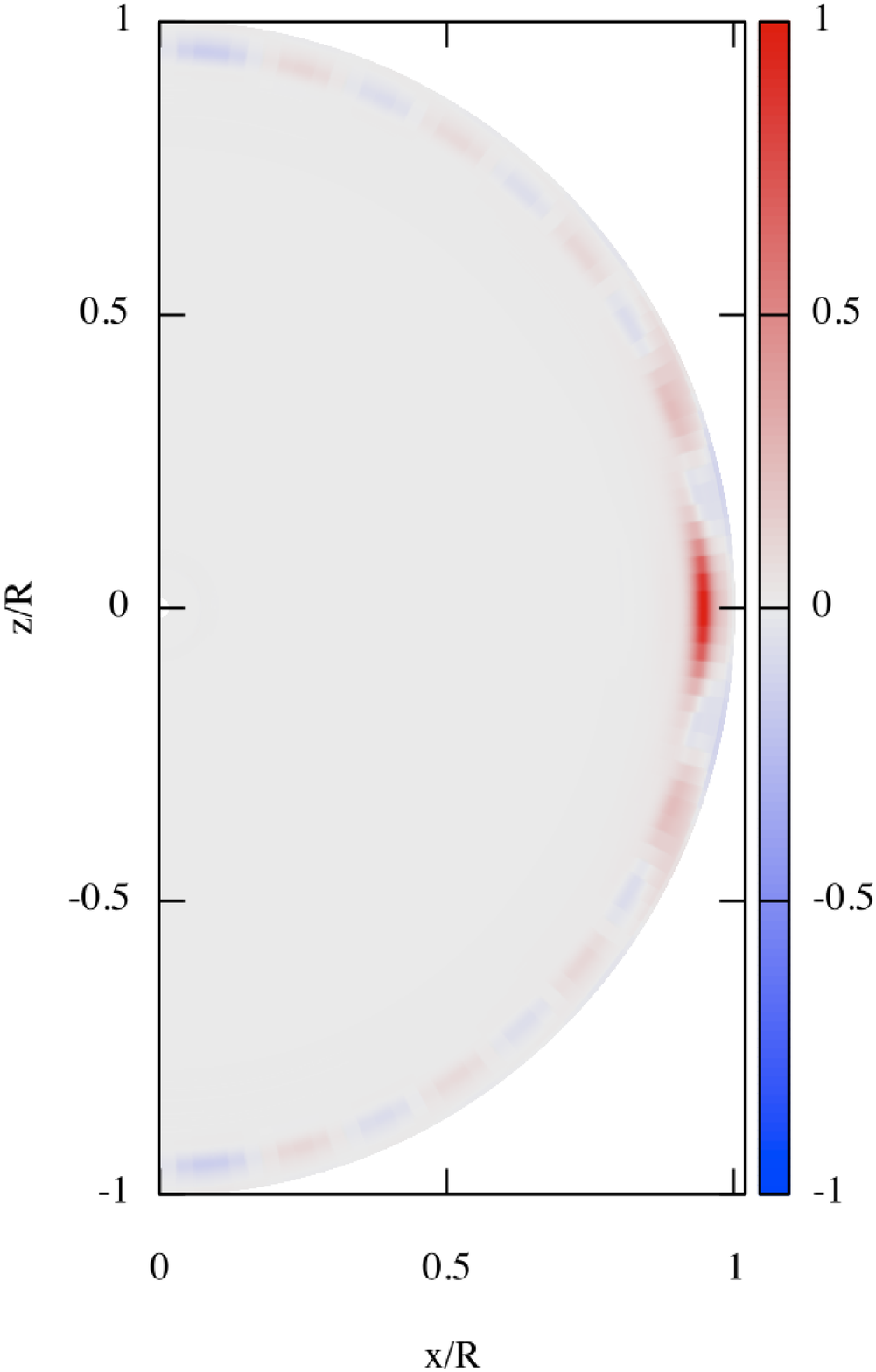}}
\end{center}
\caption{{\bf Top:} second-order perturbations $u_{S,k}\equiv\hat v_{S,l_k}^{(2)}/R\sigma_0$,
$u_{H,k}\equiv\hat v_{H,l_k}^{(2)}/R\sigma_0$, and
$u_{T,k}\equiv\hat v_{T,l_k}^{(2)}/R\sigma_0$
as a function of $r/R$ for the even retrograde $g_{64}$-mode of $l=|m|=2$ of a $6M_\odot$ main sequence star model at $\bar\Omega=0.1$,
where the solid lines, dash-dotted lines, dashed lines, long dashed lines, and dotted lines
represent respectively the components corresponding to $k=1,~3,~5,~7$, and $9$. {\bf Bottom:} density maps of the functions $v_r^{(2)}(r,\theta)/R\sigma_0$,
$v_\theta^{(2)}(r,\theta)/R\sigma_0$, and $v_\phi^{(2)}(r,\theta)/R\sigma_0$ for the mode. The functions are normalized by their maximum amplitudes. Red and blue colors correspond to positive and negative values, respectively.}
\end{figure}

\subsection{Slowly Pulsating B-type Stars}

In {SPB} stars, low frequency $g$- and $r$-modes are
excited by the opacity bump mechanism 
(Dziembowski et al. 1993; Gautschy \& Saio
1993; Lee 2006; Aprilia et al. 2011).
Here, we
calculate mean flows driven by unstable low frequency $g$- and $r$-modes of a
$6M_\odot$ {main-sequence} star, whose physical parameters are
{typical of a SPB star:} $\log(L/L_\odot)=3.2328$, $\log T_{\rm
eff}=4.1982$, $R/R_\odot=5.55$, and $X_c=0.1237$, $X=0.7$, and $Z=0.02$.
Because good numerical consistency in the sense discussed in \S 2.4 is attained only for $\bar\Omega\ltsim 0.1$ for the $g$-modes,
we restrict our discussion of $g$-modes to the case of $\bar\Omega=0.1$.
Since most of the unstable $g$-modes of the SPB model have frequencies $|\bar\omega|\gtsim~\bar\Omega$ for $\bar\Omega\sim 0.1$, 
we can use the expansion length $j_{\rm max}\sim 5$ to obtain sufficiently accurate eigenfunctions, and
we set $k_{\rm max}=10$ for second order perturbation calculations.

\begin{figure}
\begin{center}
\resizebox{0.33\columnwidth}{!}{
\includegraphics{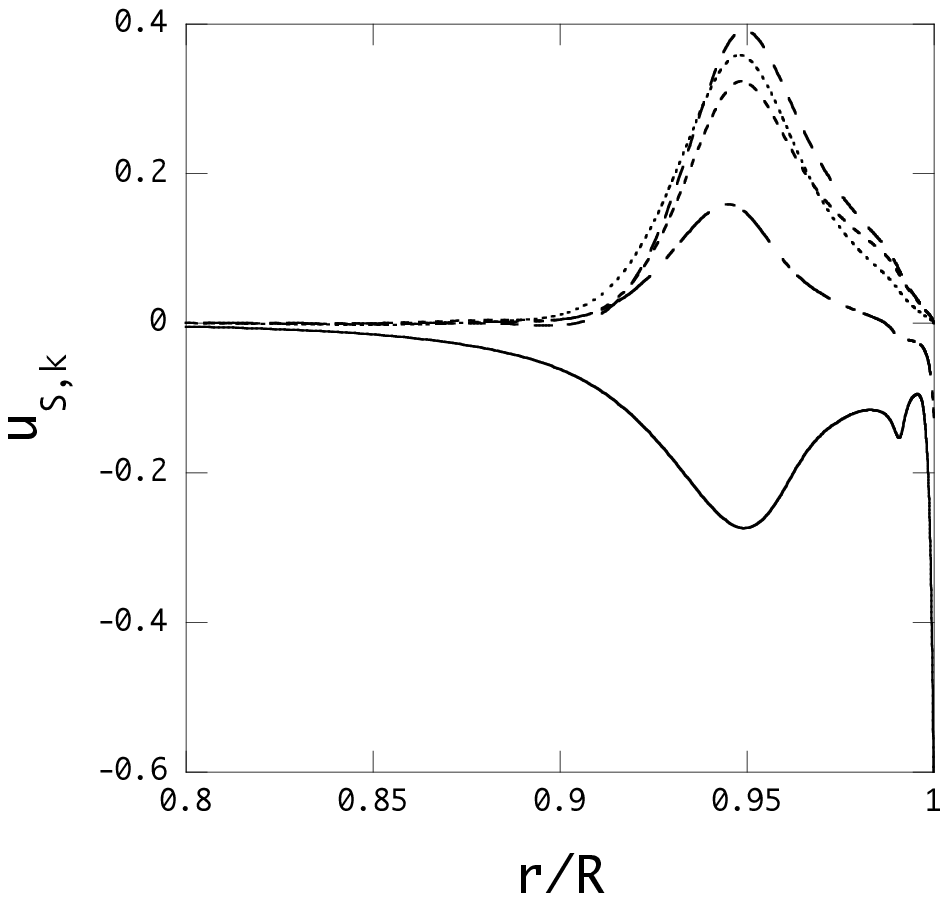}}
\resizebox{0.33\columnwidth}{!}{
\includegraphics{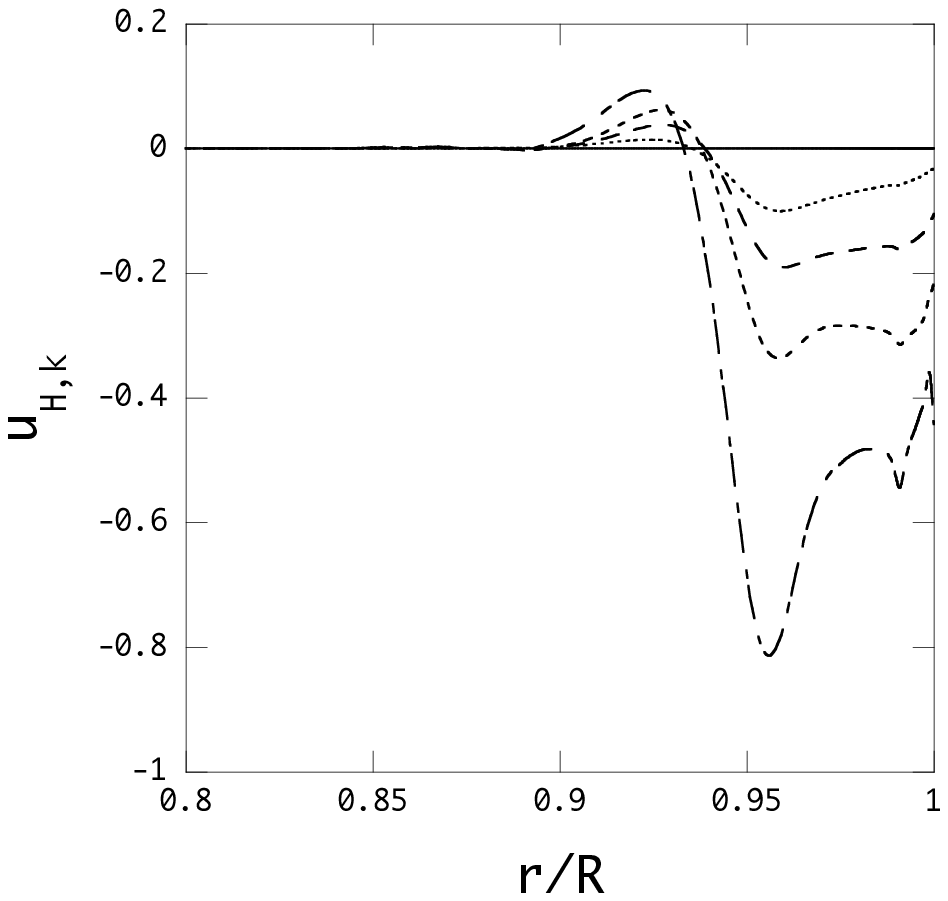}}
\resizebox{0.33\columnwidth}{!}{
\includegraphics{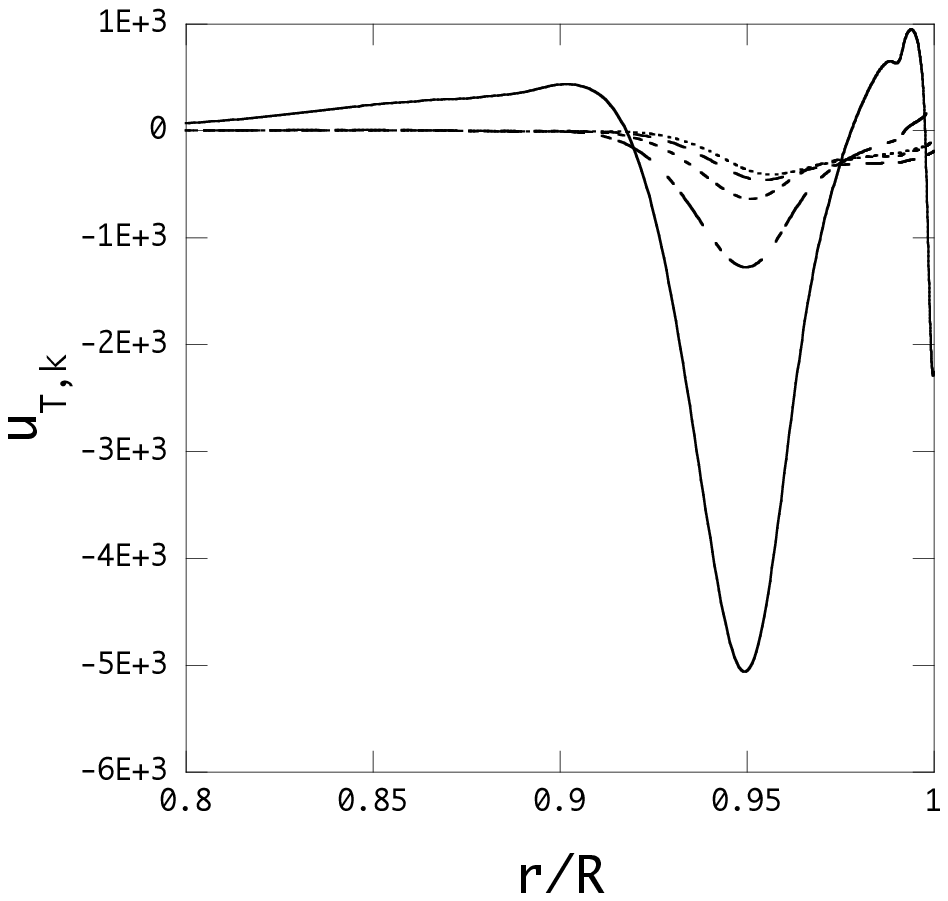}}
\end{center}
\begin{center}
\resizebox{0.33\columnwidth}{!}{
\includegraphics{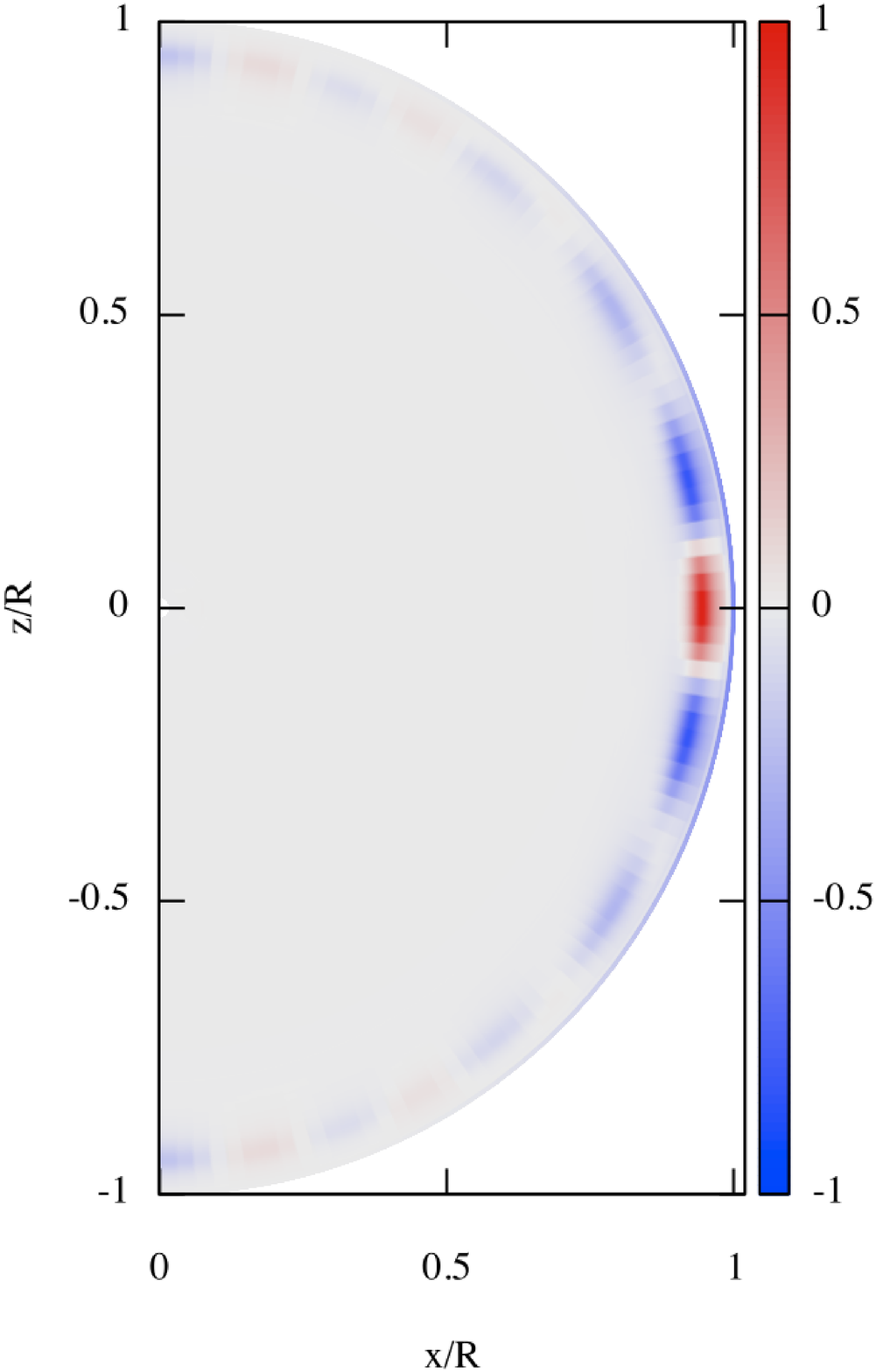}}
\resizebox{0.33\columnwidth}{!}{
\includegraphics{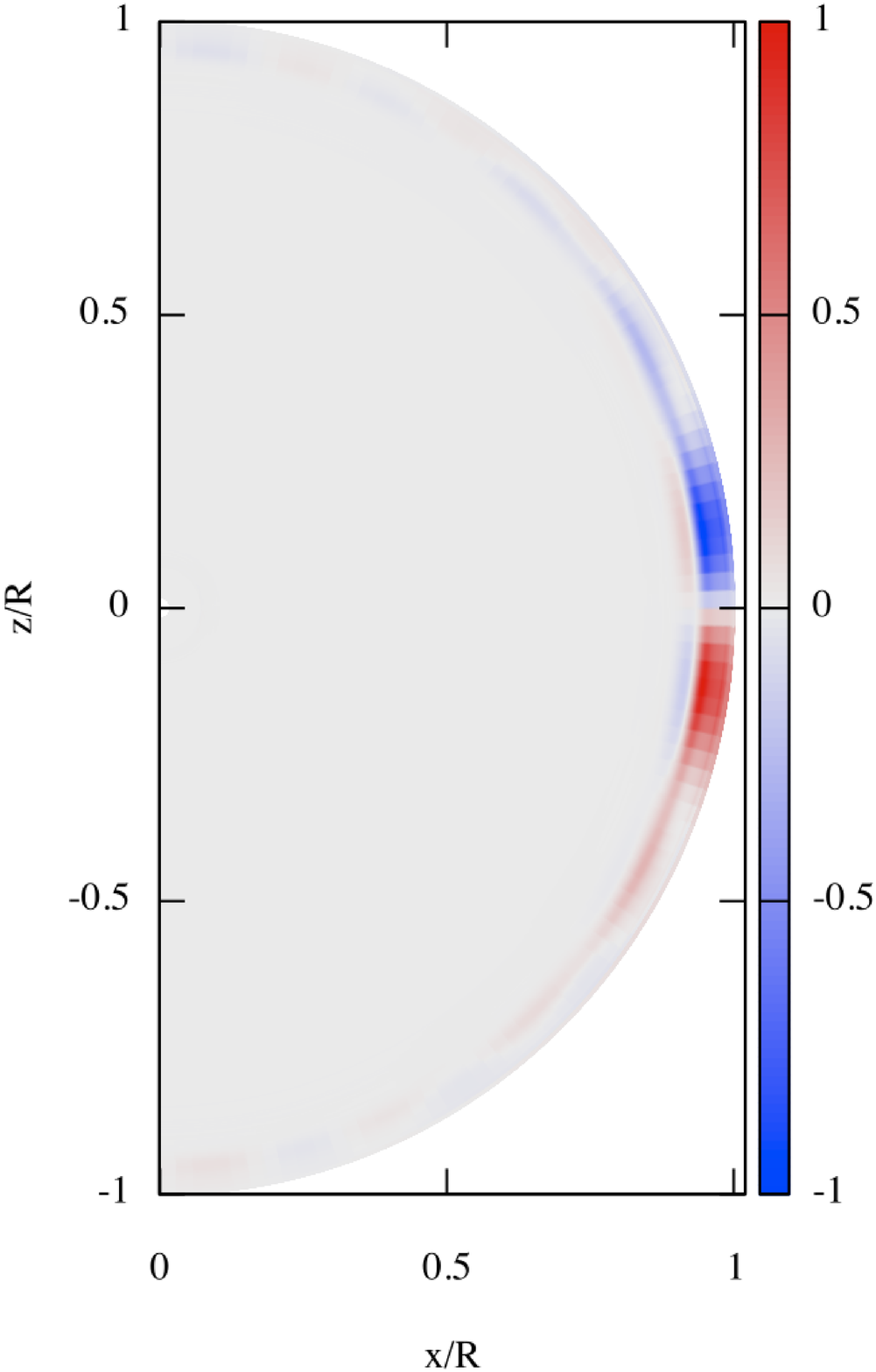}}
\resizebox{0.33\columnwidth}{!}{
\includegraphics{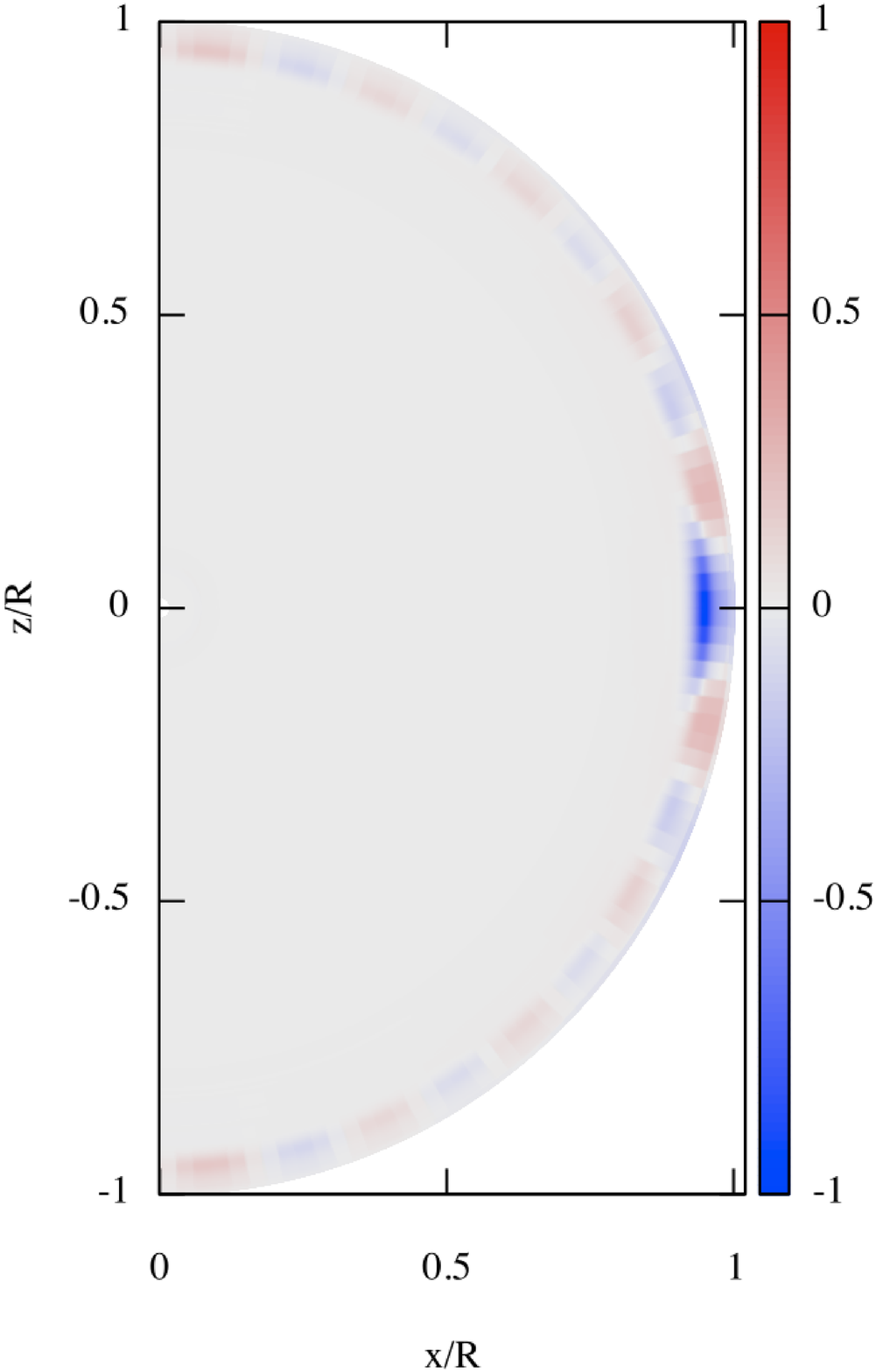}}
\end{center}
\caption{Same as Figure 2 but for the even prograde $g_{30}$-mode of
$l=|m|=2$.
} 
\end{figure}

\subsubsection{$g$-modes}

Let us first discuss mean flows driven by low frequency $g$-modes. As
numerically shown by Aprilia et al. (2011), both prograde and retrograde low
frequency $g$-modes are excited by the opacity bump mechanism {in} rapidly
rotating SPB stars. {Most} of $l=|m|$ retrograde $g$-modes of
the star, that {would be} unstable if the star were
non-rotating, are stabilized by rapid rotation as a result of their coupling
with stable $l>|m|$ $g$-modes. {As a consequence,} only a small
number of them survive as unstable modes in rapidly rotating SPB stars. This is
not the case {of} prograde $l=|m|$ $g$-modes and most of them remain
unstable even at rapid rotation.

In Figure 2, we plot, in the top panels, the expansion coefficients $u_{S,k}\equiv \hat
v_{S,l_k}^{(2)}/R\sigma_0$, $u_{H,k}\equiv \hat v_{H,l_k}^{(2)}/R\sigma_0$, and
$u_{T,k}\equiv \hat v_{T,l_k}^{(2)}/R\sigma_0$  as a function of $r/R$ for the
even retrograde $g_{30}$-mode of $l=|m|=2$ 
for $\bar\Omega=0.1$, where the normalized
eigenfrequency $\bar\omega\equiv\omega/\sigma_0$ is
$\bar\omega=(0.4672,-5.7\times10^{-6})$. Here, the amplitude normalization of
the linear mode is given by $S_{l_1}=1$ at the surface. Note that $l=|m|$ implies the
mode is an even mode. As shown by Figure 1, the functions $u_{S,k}$, $u_{H,k}$,
and $u_{T,k}$ have large amplitudes only in the outer envelope layers. Note also that
the peaks of the functions at $r/R\sim 0.95$ correspond to the place at which the mode excitation strongly occurs.
The reason for this behavior is that, in addition to the fact that mode amplitudes can be large
near the surface because of low density, thermal diffusion grows
below the surface of upper main-sequence stars. Therefore,
deposition/extraction of momentum may take place in these layers (e.g., Lee et al. 2014).
These wave-mean flow interactions thus
drive both zonal and meridional flows there (see e.g. Mathis et al. 2013 in the
case of low-mass stars). The amplitude of the zonal component $u_{T,k}$ is much
larger than those of the meridional ones, $u_{S,k}$ and $u_{H,k}$, by several
orders of magnitude. Therefore, the kinetic energy of the induced azimuthal
rotational motion is larger than that of generated meridional circulation.

The reason that the azimuthal flow is much larger than the radial/latitudinal flows can be simply understood from 
the propagation of the mode pattern. A purely adiabatic non-axisymmetric mode is a standing wave in both the 
radial and latitudinal directions. However, the mode propagates around the equator of the star and can be thought 
of as a traveling wave in the azimuthal direction. Consequently, the mean flows created by the wave are much 
larger in the azimuthal direction, because the traveling wave propagates in one direction (either prograde or 
retrograde). In the radial/latitudinal direction, the standing wave, which is composed of traveling waves 
propagating in opposite directions, contains components that cancel each other out such that there is no net flow 
in these directions. In other words, since the modes contain angular momentum only in the z-direction, the flows 
they induce are only in the azimuthal direction.
For a weakly non-adiabatic mode such as those discussed here, the modes also produce non-zero mean flows in the 
radial/latitudinal directions. However, the azimuthal flows will still be much larger.

Figure 2 also shows for the $g_{30}$-mode the density maps of the components
$v_r^{(2)}(r,\theta)/R\sigma_0$,  $v_\theta^{(2)}(r,\theta)/R\sigma_0$, and
$v_\phi^{(2)}(r,\theta)/R\sigma_0$, in the bottom panels, where these
functions, which depend on both radial distance and latitude, are normalized by their
maximum amplitudes. {The} vertical axis is the rotation axis
($x=y=0$), and the equatorial plane is given by $z=0$. The radial and the
azimuthal velocity components ($v_r^{(2)}(r,\theta)/R\sigma_0$ and
$v_\phi^{(2)}(r,\theta)/R\sigma_0$) are symmetric about the equator, while 
{the latitudinal one} ($v_\theta^{(2)}(r,\theta)/R\sigma_0$) is antisymmetric.
As shown by the maps for the retrograde $g_{30}$ mode, the pulsation driven mean flows near the surface have the velocity components $v_\theta^{(2)}$ toward the equatorial plane and the component $v_r^{(2)}$ heading inwards.
The $\phi$ component $v_\phi^{(2)}$ is positive, indicating acceleration of the rotation near the equatorial
surface.
As suggested by the density maps, the amplitudes of pulsation driven mean flows tend to be confined in the equatorial region.
This reflects the properties of low frequency modes of rotating stars, that is, their amplitudes are also
confined in the equatorial region, particularly for rapidly rotating case (see, e.g., Berthomieu et al 1978;
Bildsten et al 1996, Lee \& Saio 1997; Townsend 2003ab).

We see that pulsations induce a differential rotation
$\delta\Omega\left(r,\theta\right)=v_\phi^{(2)}(r,\theta)/(r\sin\theta)$, which
is a function of both radial distance and latitude.
This is the signature of pulsation-driven transport of
angular momentum both in the vertical and in the horizontal directions (e.g., Mathis 2009).
It is the first time
that such wave-driven differential rotation is computed in 2-D. Indeed, previous
computations treated reduced cases of the radial shellular rotation
(e.g., Talon \& Charbonnel 2005; Mathis et al. 2013)
or focused on the
dynamics in the equatorial plane
(e.g., Rogers et al. 2013).
Like in previous computations performed for low-mass stars by 
Talon \& Charbonnel (2005) and Mathis et al. (2013),
we see in massive stars that the
meridional circulation can be multicellular both in radius and in latitude.
These patterns differ from properties of meridional circulations driven by
large-scale differential rotation in models where waves are not taken into
account. Such a computation is of interest for development of 2-D models of
rotating stars (Rieutord 2006),
in which waves
must be taken into account in a near future. In the case of the $g_{30}$-mode,
the amplitudes of the functions are confined in the equatorial regions. In
the context of active massive stars, such as Be stars, it is interesting to
note that for the retrograde mode, the velocity perturbations
$v_\phi^{(2)}$ at the
surface becomes positive in the narrow equatorial region, while $v_r^{(2)}$
is negative. The positive
zonal component can then help the surface layers to reach the
critical angular velocity needed to eject matter in the
circumstellar environment.

In Figure 3, we show, in the top panels, the expansion coefficients $u_{S,k}$, $u_{H,k}$,
and $u_{T,k}$ for the even prograde $g_{30}$-mode of $l=|m|=2$ at
$\bar\Omega=0.1$, for which $\bar\omega=(0.4025,-7.9\times10^{-6})$. 
The corresponding 2-D density maps of the velocity components
$v_r^{(2)}(r,\theta)/R\sigma_0$,  $v_\theta^{(2)}(r,\theta)/R\sigma_0$, and
$v_\phi^{(2)}(r,\theta)/R\sigma_0$ are given in the bottom panels in the same figure. 
As in the
previous case, the functions $u_{S,k}$, $u_{H,k}$, and $u_{T,k}$ have
large amplitudes only in the outer surface layers. 
As in the previously studied case of the retrograde
$g_{30}$-mode, the amplitude of $u_{T,k}$ is much larger than those of
$u_{S,k}$ and $u_{H,k}$. From these density maps, 
the flow patterns driven by the prograde $g_{30}$-mode in the surface equatorial region
is opposite to those by the retrograde $g_{30}$ mode, that is,
the radial velocity $v_r^{(2)}$ is towards 
the surface and the components $v_\theta^{(2)}$ show flows out of the equatorial plane.
This different
behavior between retrograde and prograde $g$-modes shows that it is necessary to sum over all the modes
that are excited in a given star to be able to understand the
total wave-driven zonal and meridional mean flows and their net
effect. This is the reason why $r$- and $p$- modes have also
to be considered.

\subsubsection{$r$-modes}

\begin{figure}
\begin{center}
\resizebox{0.33\columnwidth}{!}{
\includegraphics{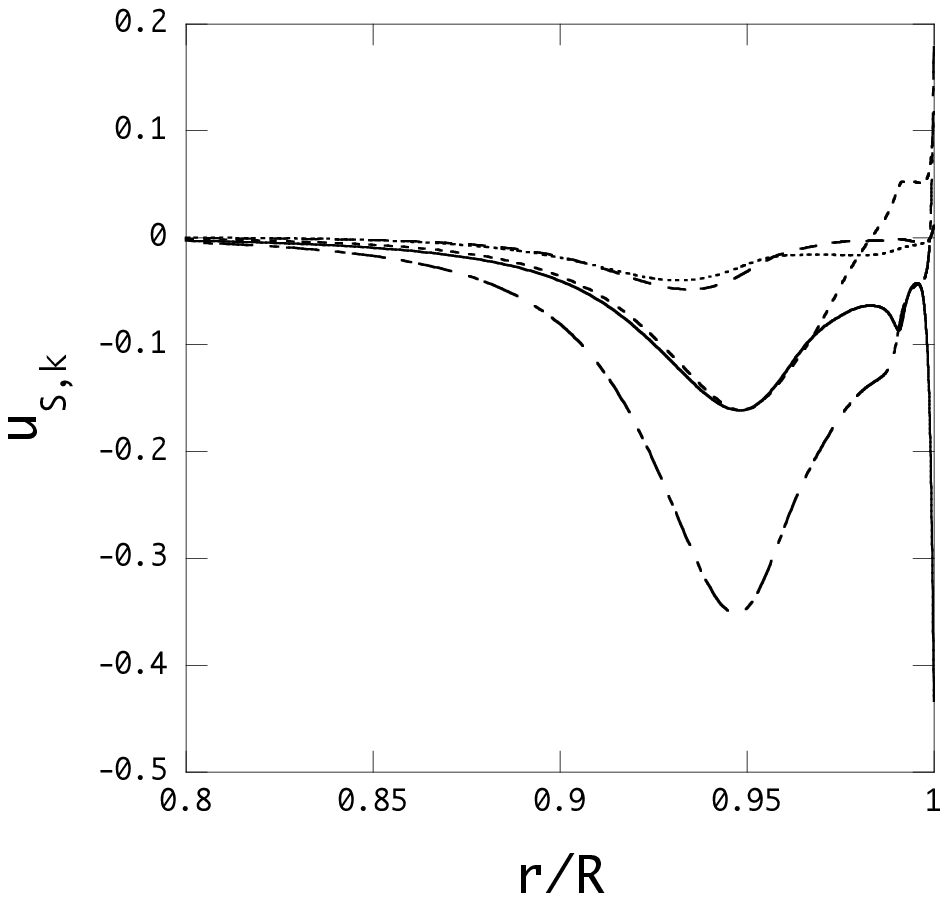}}
\resizebox{0.33\columnwidth}{!}{
\includegraphics{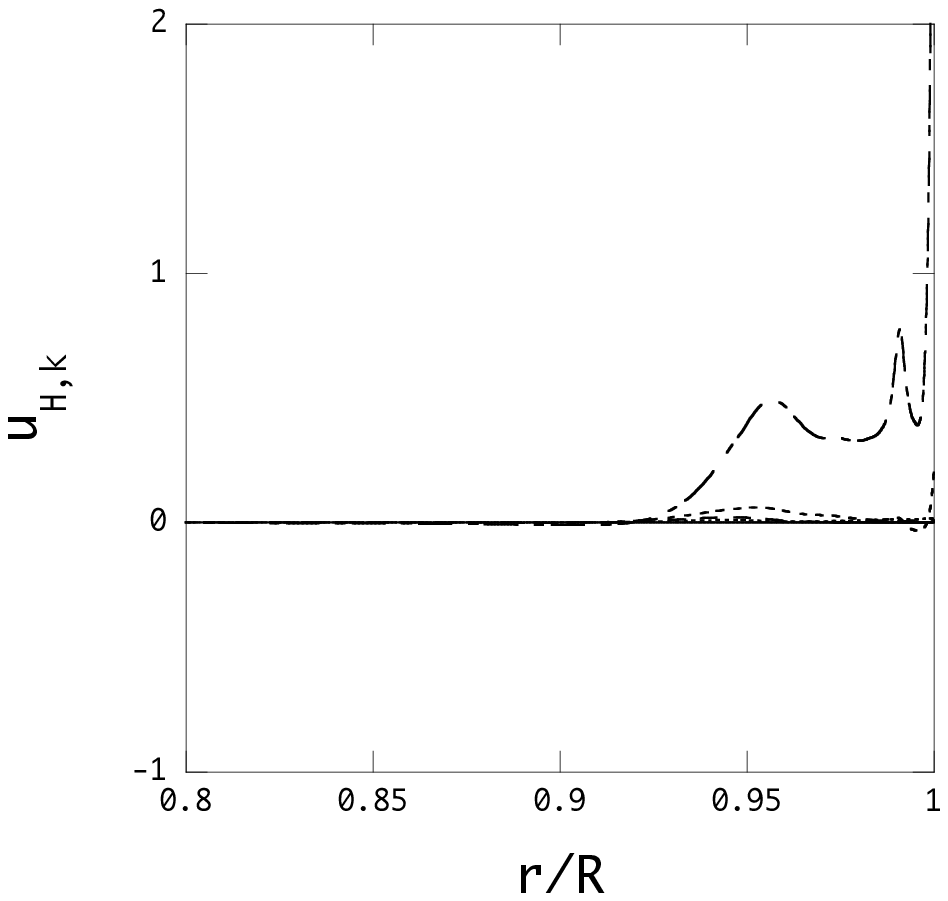}}
\resizebox{0.33\columnwidth}{!}{
\includegraphics{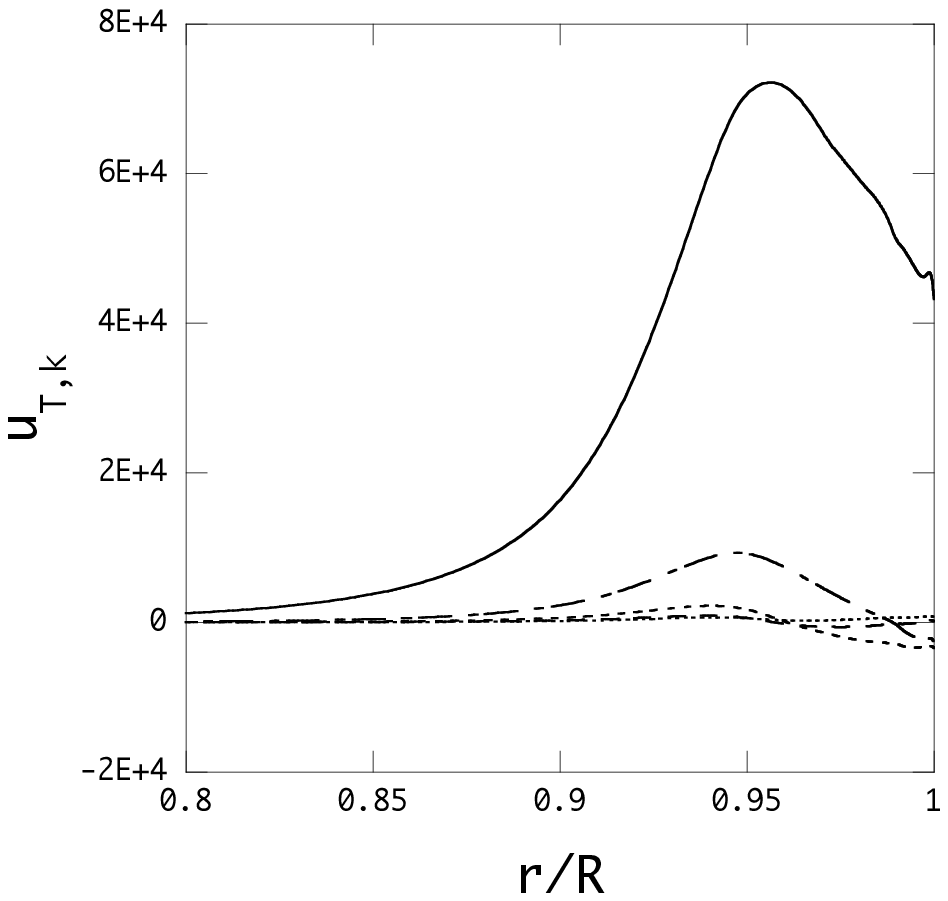}}
\end{center}
\begin{center}
\resizebox{0.33\columnwidth}{!}{
\includegraphics{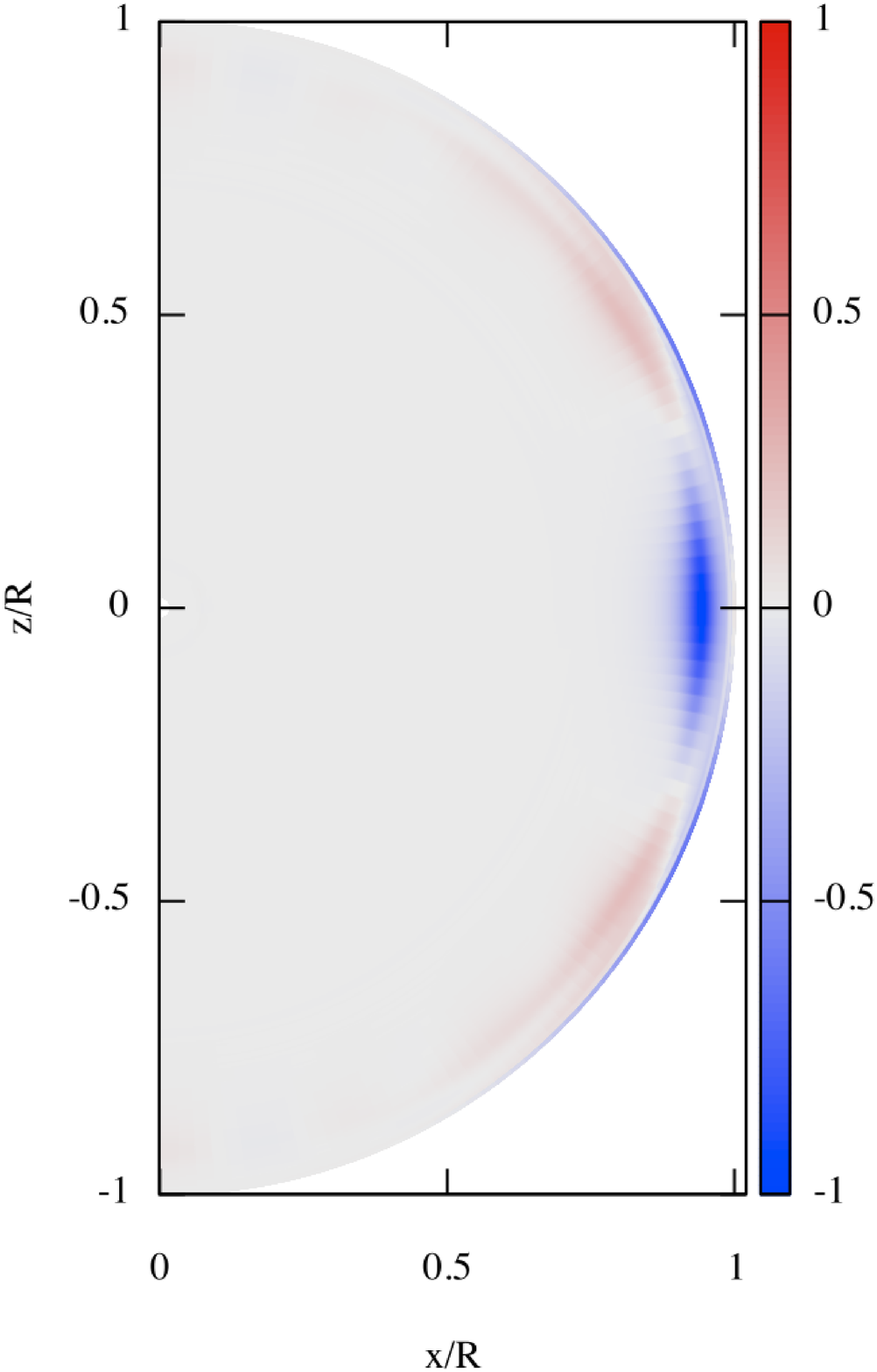}}
\resizebox{0.33\columnwidth}{!}{
\includegraphics{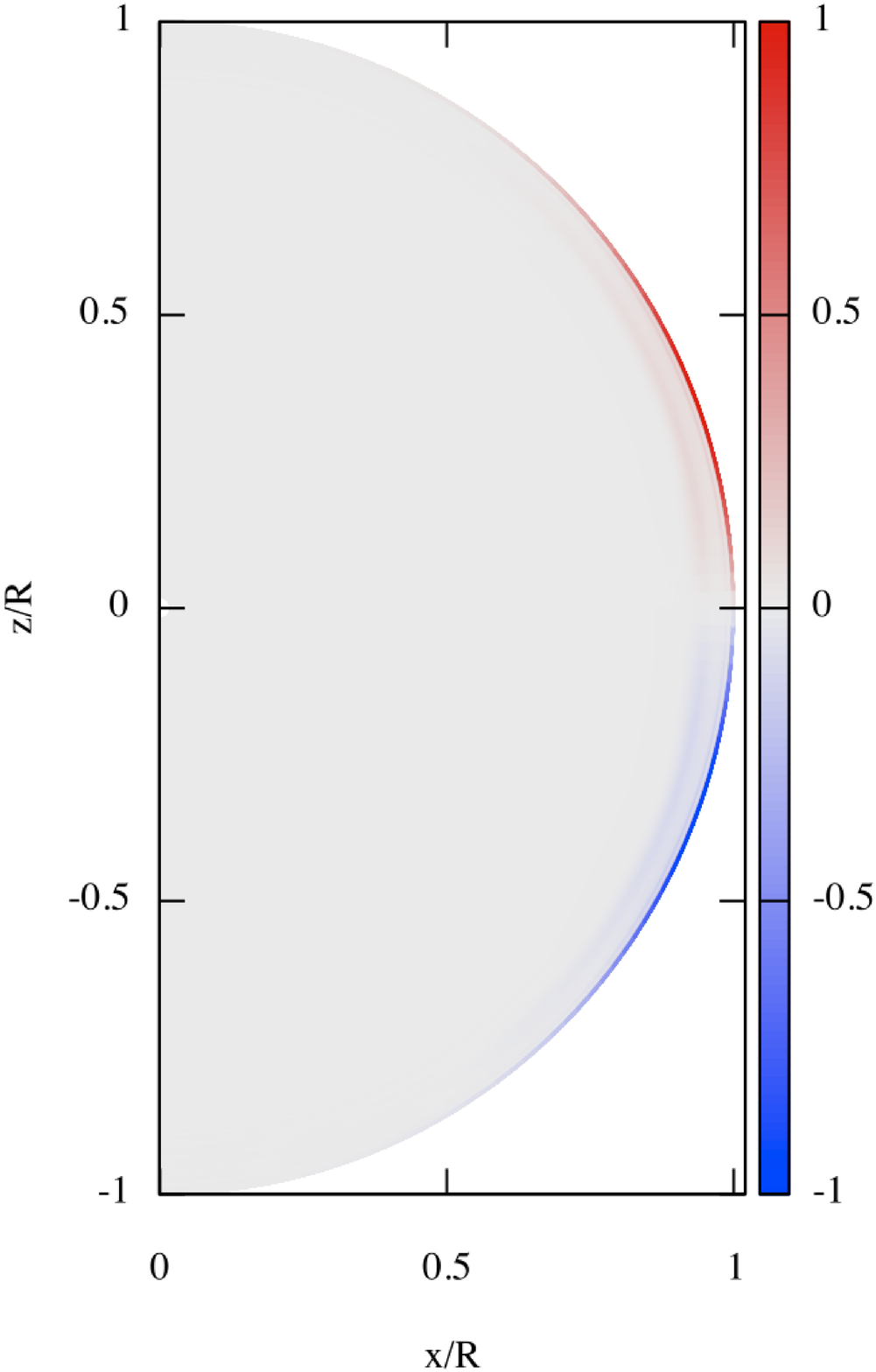}}
\resizebox{0.33\columnwidth}{!}{
\includegraphics{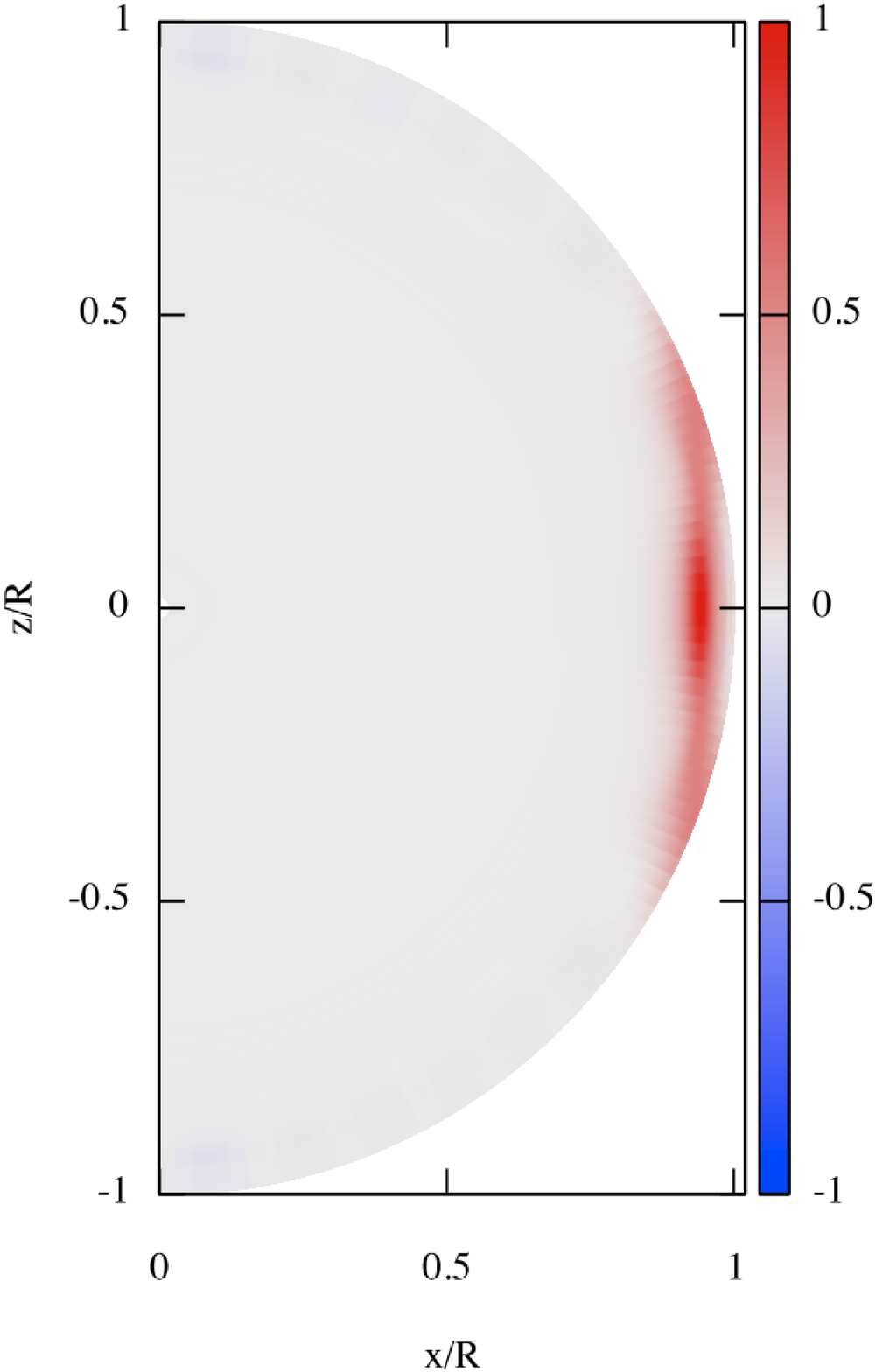}}
\end{center}
\caption{Same as Figure 2 but for the $l^\prime=|m|=2$ odd $r_{36}$-mode at $\bar\Omega=0.4$.
}
\end{figure}

$r$-modes are rotationally induced retrograde modes and constitute a
subclass of inertial modes, for which the Coriolis acceleration is the
restoring force. In the slow rotation limit (i.e., $\Omega\rightarrow 0$),
$r$-modes associated with the degree $l^\prime$ and the azimuthal index $m$ have
an asymptotic co-rotating frame frequency given by
$\omega=2m\Omega/\left[l^\prime(l^\prime+1)\right]$, and the displacement vector
$\pmb{\xi}$ is dominated by {its} toroidal component $\left(\rmi
T_{l^\prime}\right)$. Since numerous $l^\prime=|m|$ $r$-modes of odd parity are
{also} excited by the opacity bump mechanism in SPB stars (e.g., Lee 2006),
it is necessary to examine how $r$-modes drive mean flows in rotating SPB
stars.

The top panels of Figure 4 show the expansion coefficients $u_{S,k}$, $u_{H,k}$, and $u_{T,k}$ as
a function of $r/R$, while the bottom panels of the figure give the corresponding 2-D density maps of
the three components $v_r^{(2)}(r,\theta)/R\sigma_0$,
$v_\theta^{(2)}(r,\theta)/R\sigma_0$, and $v_\phi^{(2)}(r,\theta)/R\sigma_0$ for
the odd $r_{36}$-mode of $l^\prime=m=2$ at $\bar\Omega=0.4$, 
for which $\bar\omega=(0.2199,-2.1\times10^{-6})$. As in the case of g-modes, the
amplitudes of the expansion coefficients become large in the outer envelope.
{Moreover, the zonal flow ($u_{T,k}$) is still much larger than the
meridional one (given by $u_{S,k}$ and $u_{H,k}$). 
As shown by the 2-D density maps, gross flow patterns driven by the $r$-mode are similar to
those by the retrograde $g_{30}$-mode, although the zonal acceleration near the surface
is much more prominent than that for the retrograde $g$-mode.

\subsubsection{Lagrange velocity perturbations}

The vertical Lagrangian velocity perturbation $\delta v_r^{(2)}$ can be an
important quantity when we consider radial transport of gaseous
matter in the interior of stars. In Figure 5, we show the 2-D density maps
of $\delta v_r^{(2)}/R\sigma_0$ for the even retrograde and prograde $g_{30}$-modes of $l=|m|=2$ 
at $\bar\Omega=0.1$ and for the odd $r_{36}$-mode of $l^\prime=|m|=2$ at
$\bar\Omega=0.4$. Comparing the bottom left panels of Figures 2, 3,
and 4 with the corresponding ones in Figure 5, we see that 
there appear no essential differences between $v_r^{(2)}$ and $\delta v_r^{(2)}$ for the 
low frequency modes.
When zonal flows in the surface equatorial layers are accelerated by the retrograde modes, 
the radial flows are driven inwards.

\begin{figure}
\begin{center}
\resizebox{0.33\columnwidth}{!}{
\includegraphics{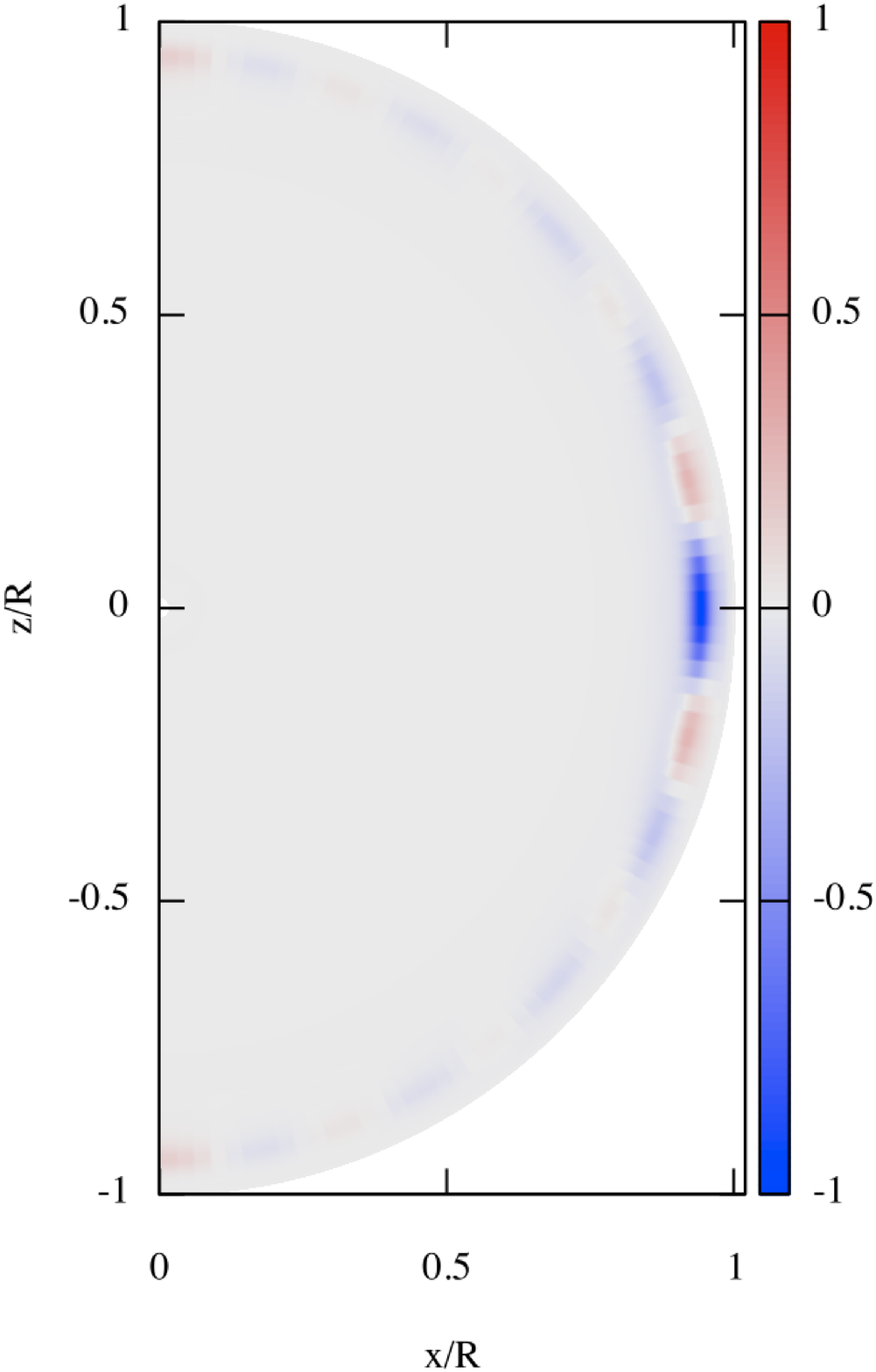}}
\resizebox{0.33\columnwidth}{!}{
\includegraphics{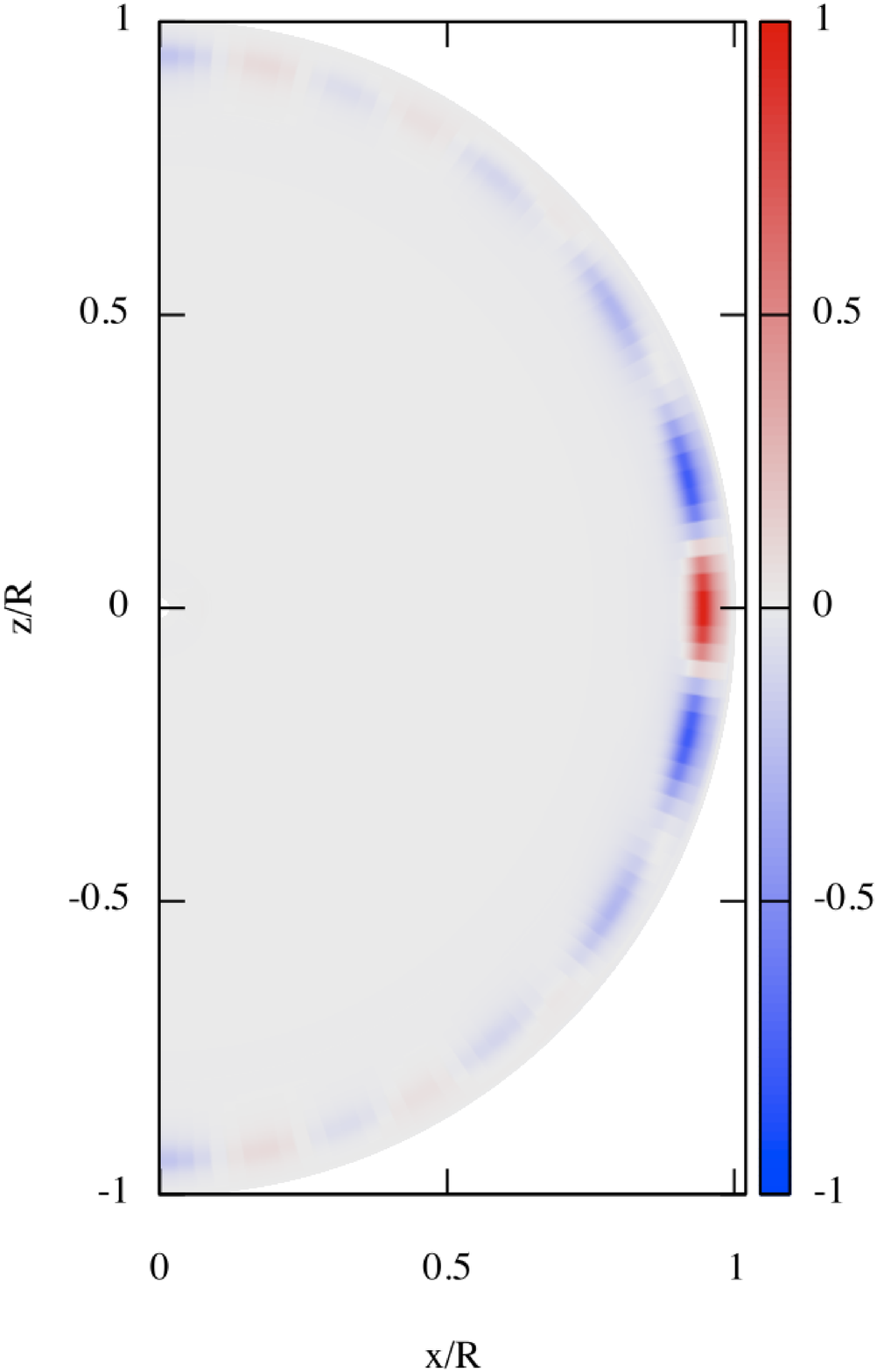}}
\resizebox{0.33\columnwidth}{!}{
\includegraphics{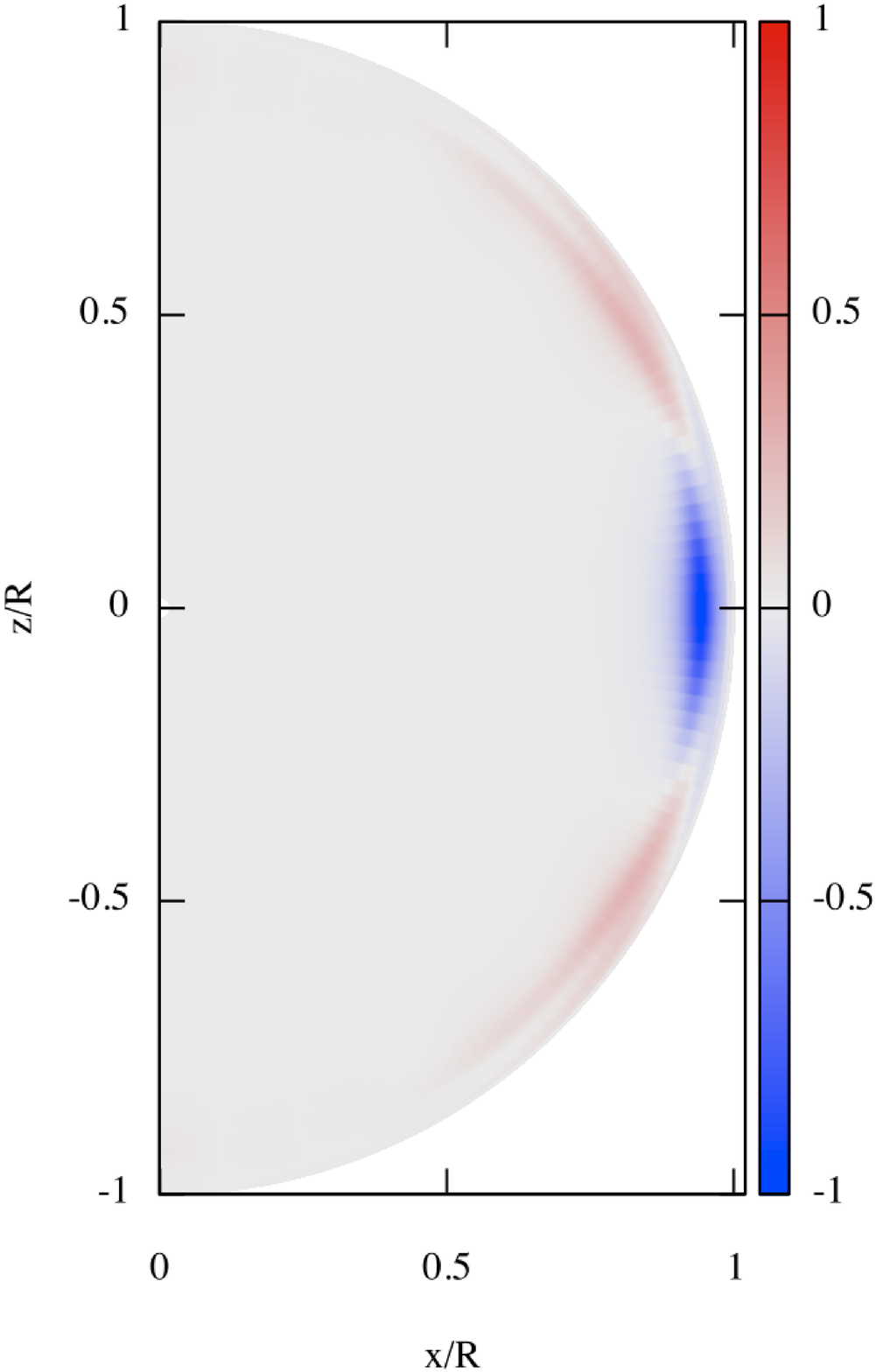}}
\end{center}
\caption{Density maps of the functions $\delta v_r^{(2)}(r,\theta)/r\sigma_0$, from left to right panels, for the retrograde $g_{30}$-mode and
the prograde $g_{30}$-mode of $l=|m|=2$ at $\bar\Omega=0.1$,
and for the $l^\prime=|m|=2$ $r_{36}$-mode at $\bar\Omega=0.4$ for the $6M_\odot$ main sequence model. The functions are normalized by their maximum amplitudes. Red and blue colors correspond to positive and negative values respectively.}
\end{figure}

\subsection{$\beta$ Cephei stars}

In addition to exciting $g$- and $r$-modes in cool B stars (SPB
stars), the iron opacity bump mechanism also excites
$p$-modes in hotter B stars ($\beta$ Cephei stars). We
calculate mean flows driven by unstable low radial order $p$-modes of a
$15M_\odot$ main-sequence model, whose physical parameters
are {typical of a $\beta$ Cep star:} $\log(L/L_\odot)=4.5928$,
$\log T_{\rm eff}=4.394$, $R/R_\odot=10.8$, and $X_c=0.130$. For this model,
$l=|m|$ low radial oder modes with normalized frequency
$\bar\omega$ between 2 and 5 become pulsationally unstable. Since the method of
calculation used in this paper for $p$-modes of rotating stars is not
necessarily appropriate for very rapid rotation
$\bar\Omega\sim 1$, as suggested by full 2-D computations of $p$-modes in
rapidly rotating stars (Ligni\`eres et al. 2006;
Reese et al. 2006),
we {restrict the present} discussion to the case of slow
rotation rates $\bar\Omega\sim 0.1$.

Figure 6 and 7 are respectively
for the retrograde $l=m=2$ $p$-mode of $\bar\omega=(2.584,-7.9\times10^{-7})$ and
for the prograde $l=|m|=2$ $p$-mode of $\bar\omega=(2.634,-4.5\times10^{-7})$. Note
that, if we count the number of radial nodes of the eigenfunction $S_{l_1}$ in
the way described in Unno et al (1989),
we obtain $k\equiv n_p-n_g=-3$,
suggesting that the mode should be classified as a $g_3$-mode, where $n_g$ and
$n_p$ are the numbers of $g$-type and $p$-type nodes of the eigenfunction
$S_{l_1}$. However, we simply call these modes a $p$-mode because it behaves as a
$p$-mode in the envelope.

As shown by Figures 6 and 7, and as in the case of SPB stars
studied above, the zonal component ($u_{T,k}$) is
dominant over meridional ones ($u_{S,k}$ and $u_{H,k}$).
The flow patterns $v_r^{(2)}$ and $v_\theta^{(2)}$ in the surface layers are quite similar between
the retrograde and prograde $p$-modes, which is contrary to the case of low frequency $g$-modes.
This is because the Coriolis terms proportional to $m\Omega/\omega$ are not important to determine
these velocity components.
Zonal acceleration in the surface equatorial region occurs for the retrograde mode, although
deceleration takes place for the prograde mode.
The Lagrangian velocity perturbations $\delta v_r^{(2)}$ of the $p$-modes are depicted
in Figure 8, which shows that the additional terms $\overline{v_{i;j}^{(1)}\xi_j}$ can be
significant enough, that is, 
even if $v_r^{(2)}$ is positive in the surface equatorial region, $\delta v_r^{(2)}$ becomes
negative.

\begin{figure}
\begin{center}
\resizebox{0.33\columnwidth}{!}{
\includegraphics{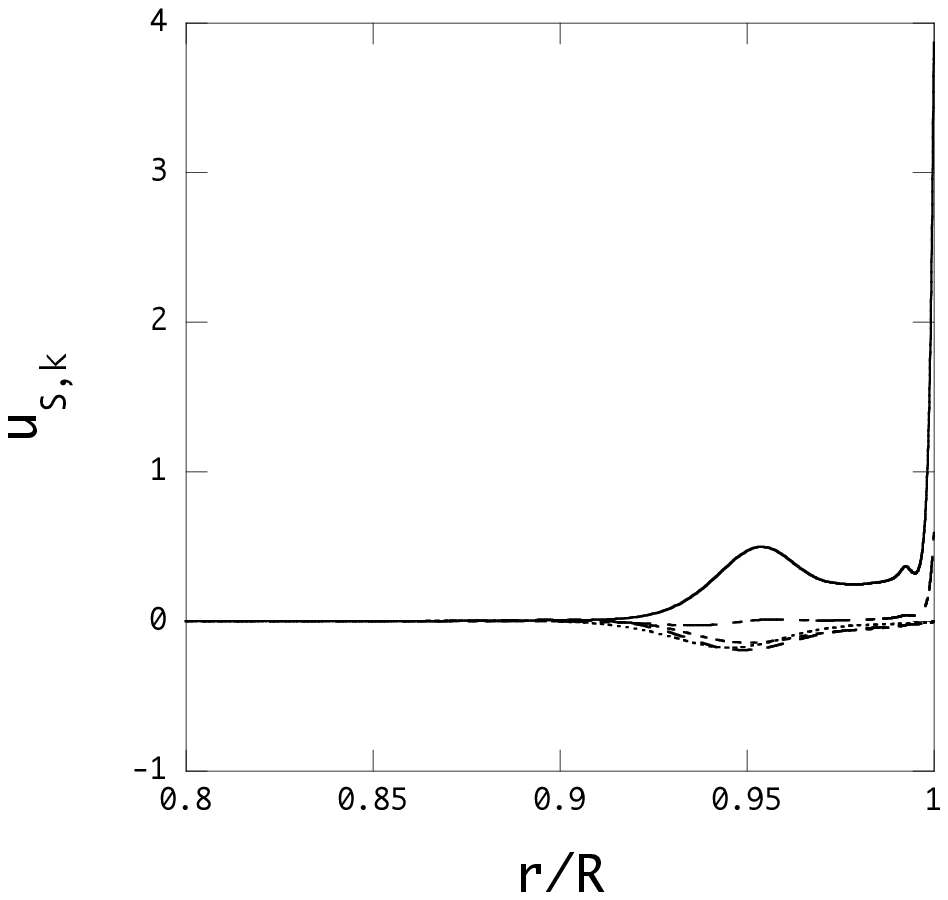}}
\resizebox{0.33\columnwidth}{!}{
\includegraphics{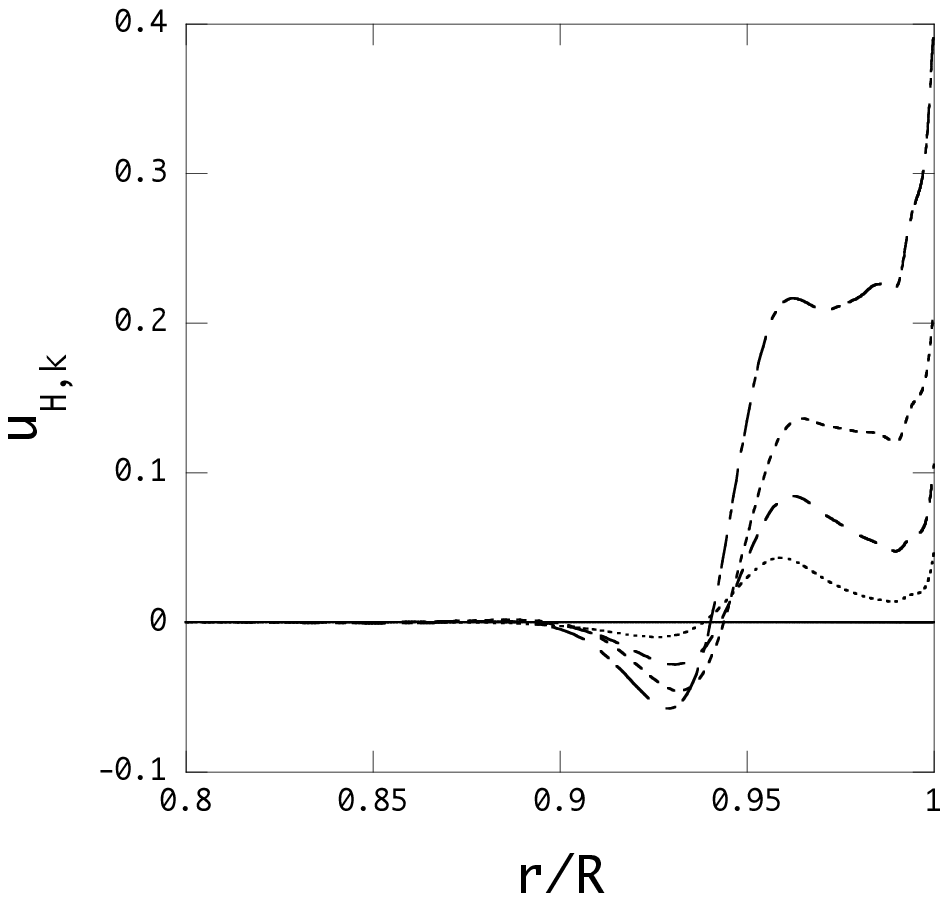}}
\resizebox{0.33\columnwidth}{!}{
\includegraphics{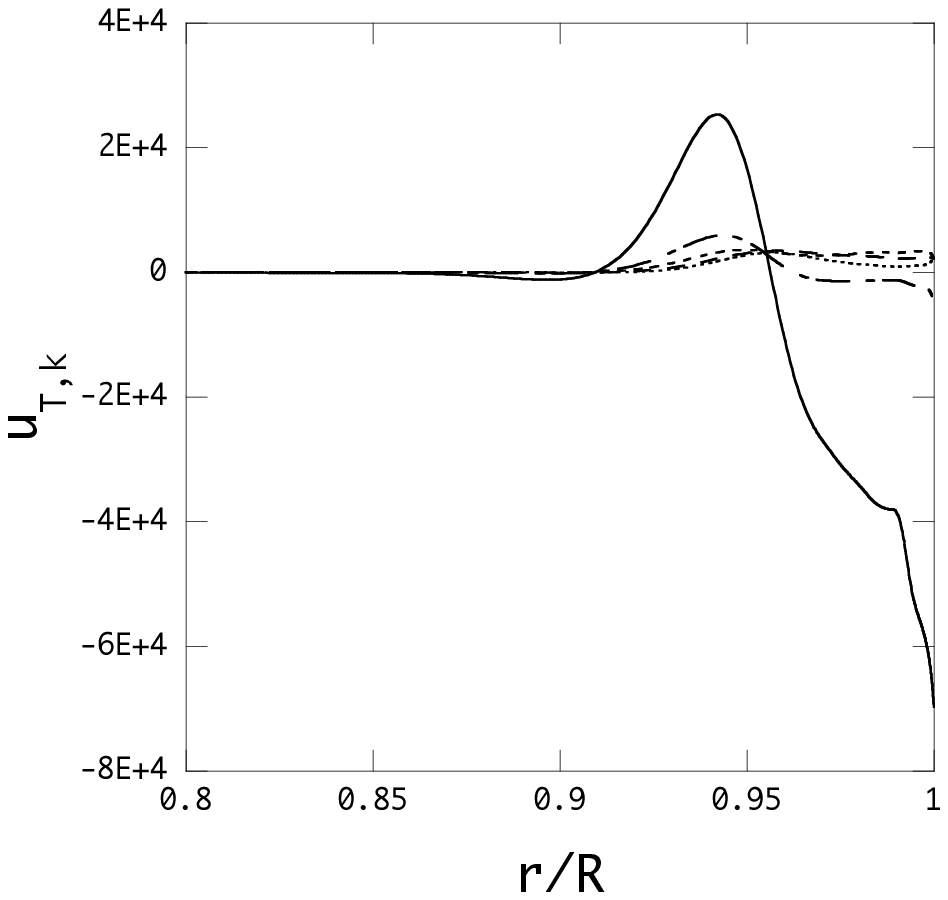}}
\end{center}
\begin{center}
\resizebox{0.33\columnwidth}{!}{
\includegraphics{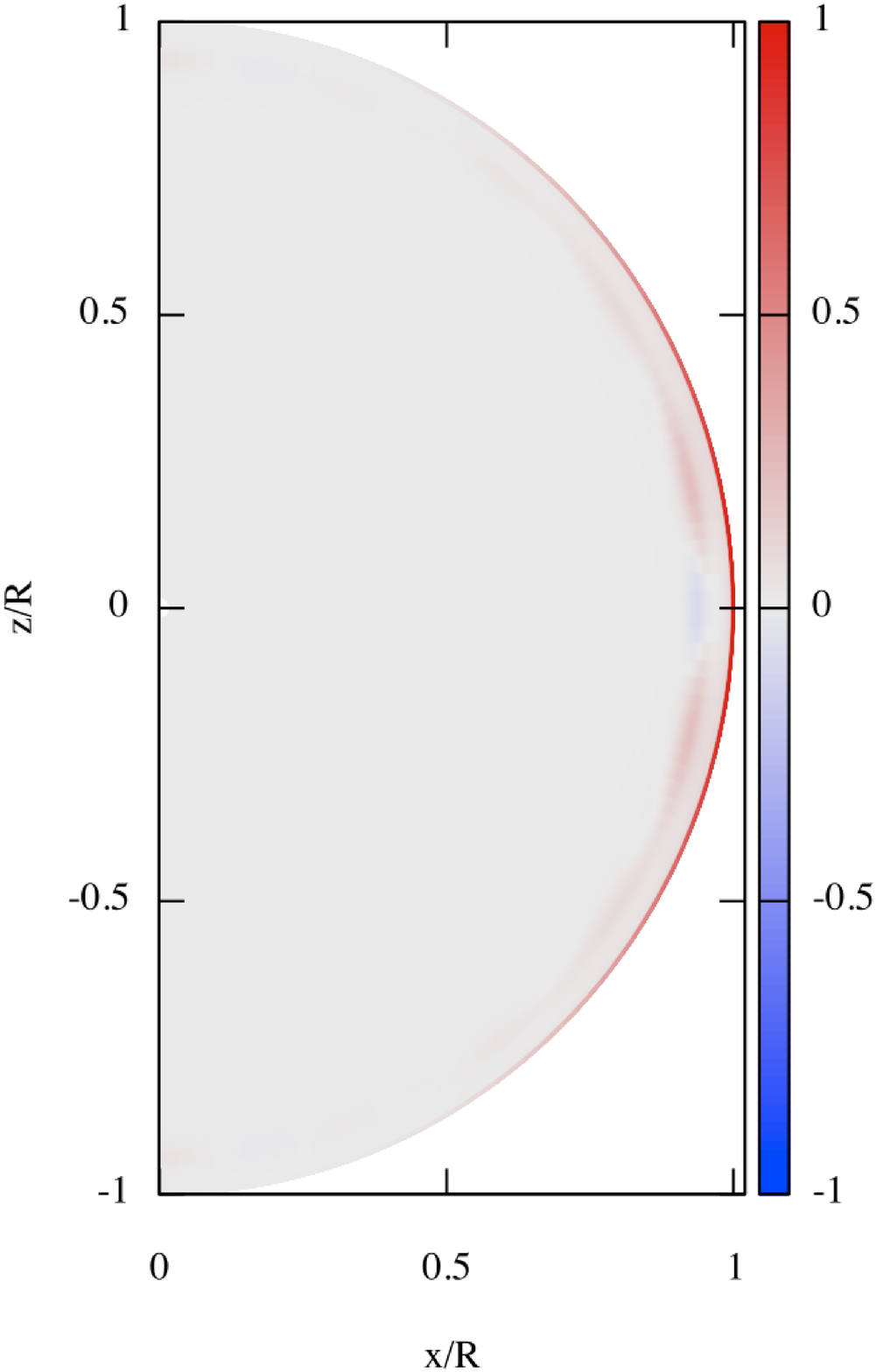}}
\resizebox{0.33\columnwidth}{!}{
\includegraphics{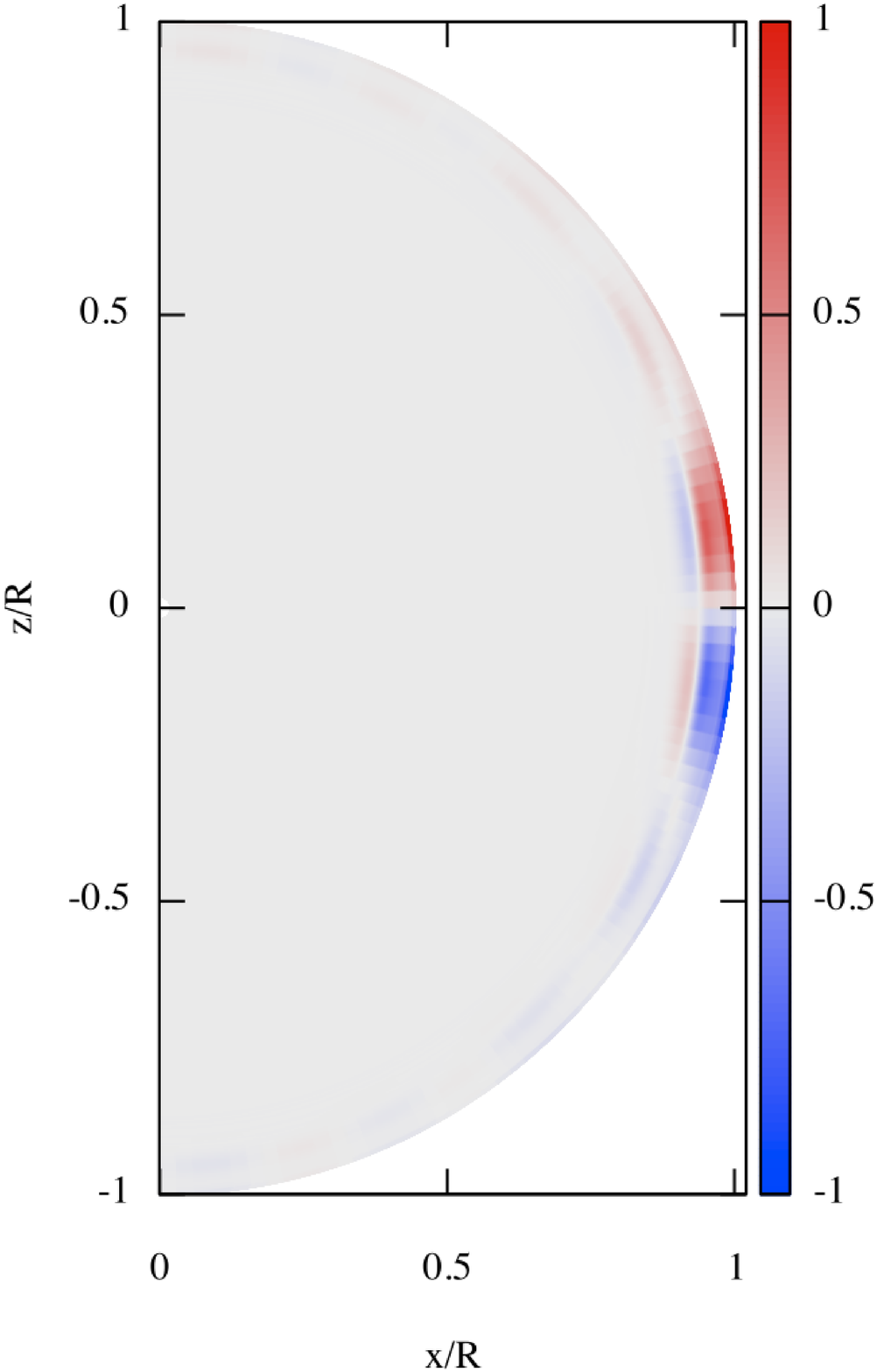}}
\resizebox{0.33\columnwidth}{!}{
\includegraphics{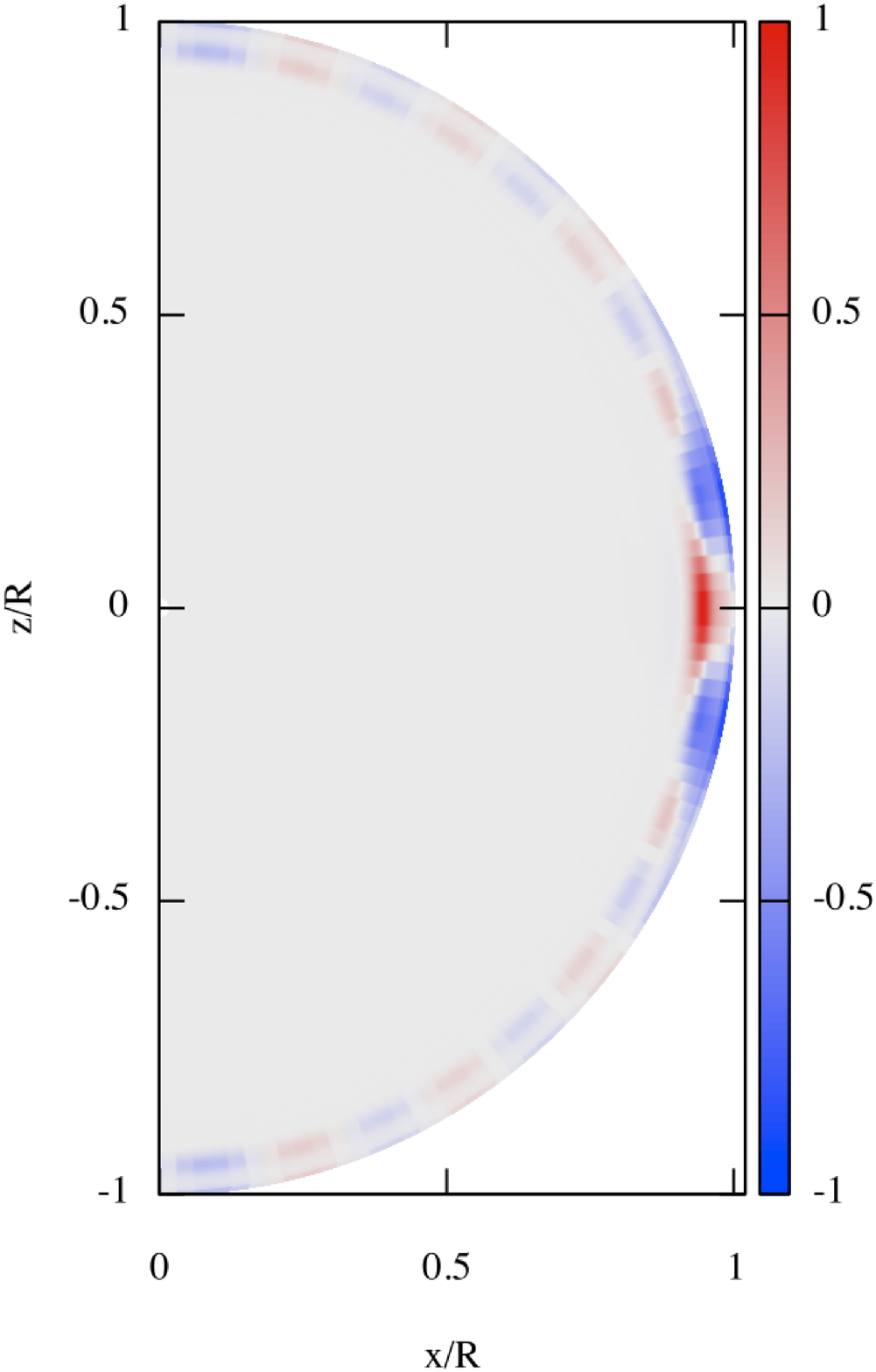}}
\end{center}
\caption{Same as Figure 2 but for the $l=|m|=2$ even retrograde $p$-mode
of the $15M_\odot$ main sequence star model where $\bar\Omega=0.1$.
}
\end{figure}

\begin{figure}
\begin{center}
\resizebox{0.33\columnwidth}{!}{
\includegraphics{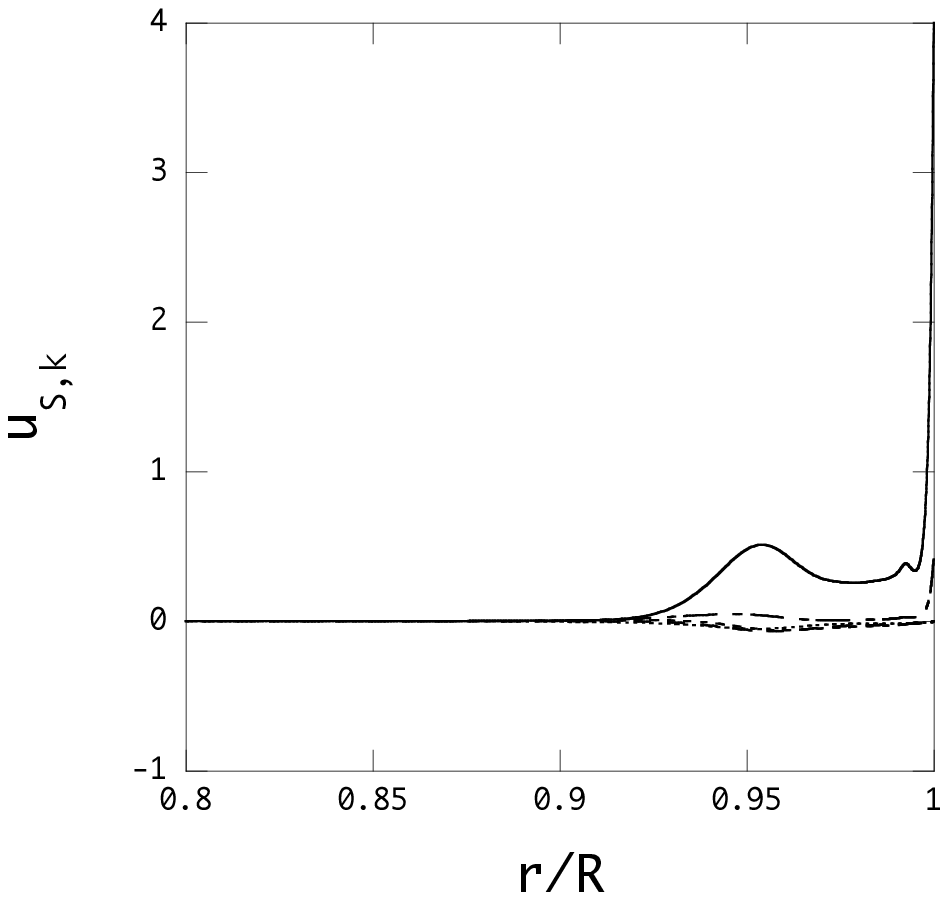}}
\resizebox{0.33\columnwidth}{!}{
\includegraphics{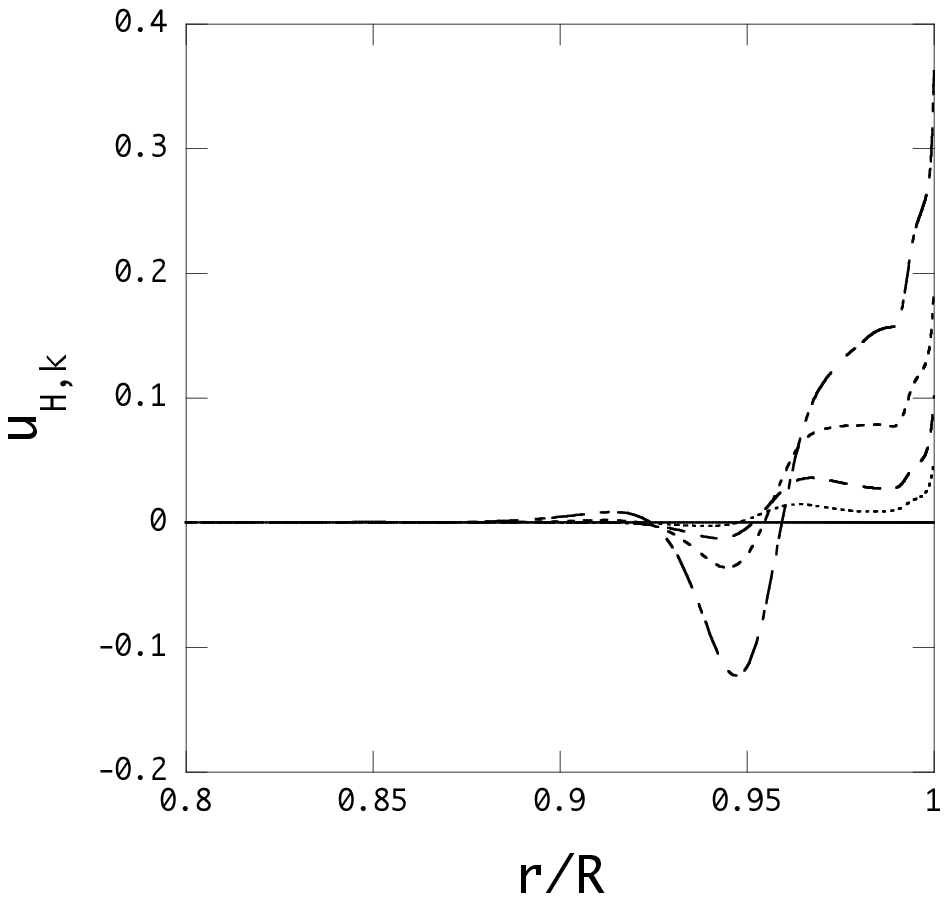}}
\resizebox{0.33\columnwidth}{!}{
\includegraphics{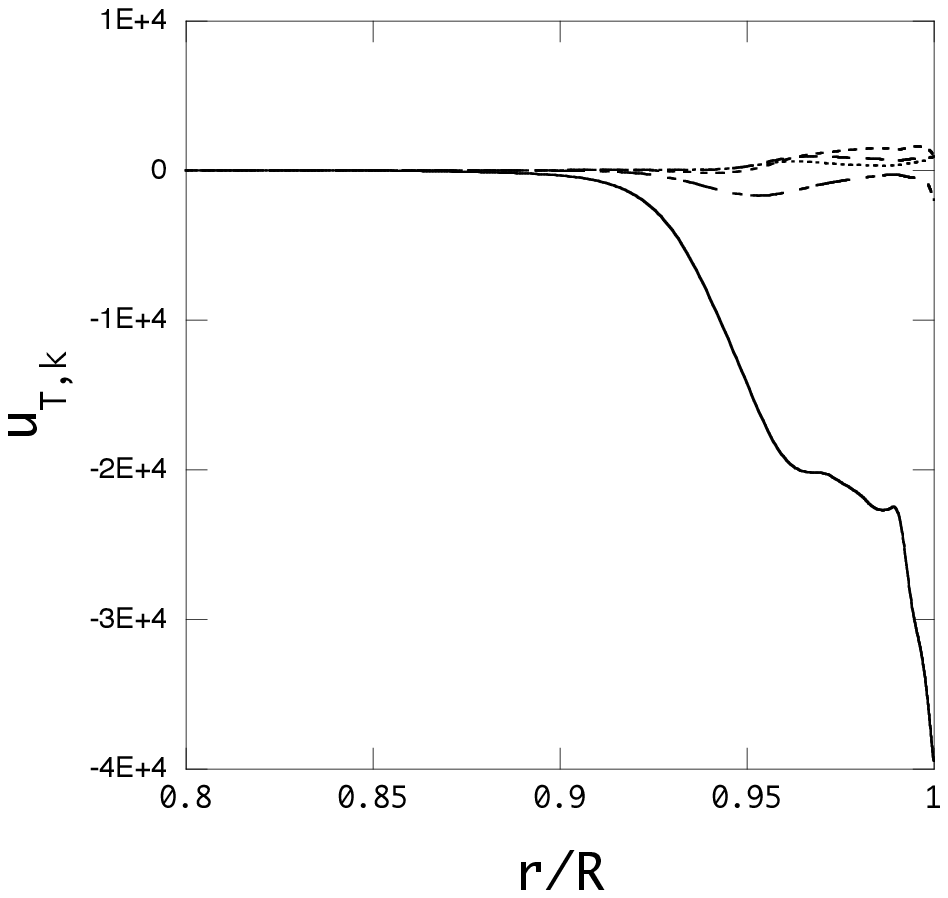}}
\end{center}
\begin{center}
\resizebox{0.33\columnwidth}{!}{
\includegraphics{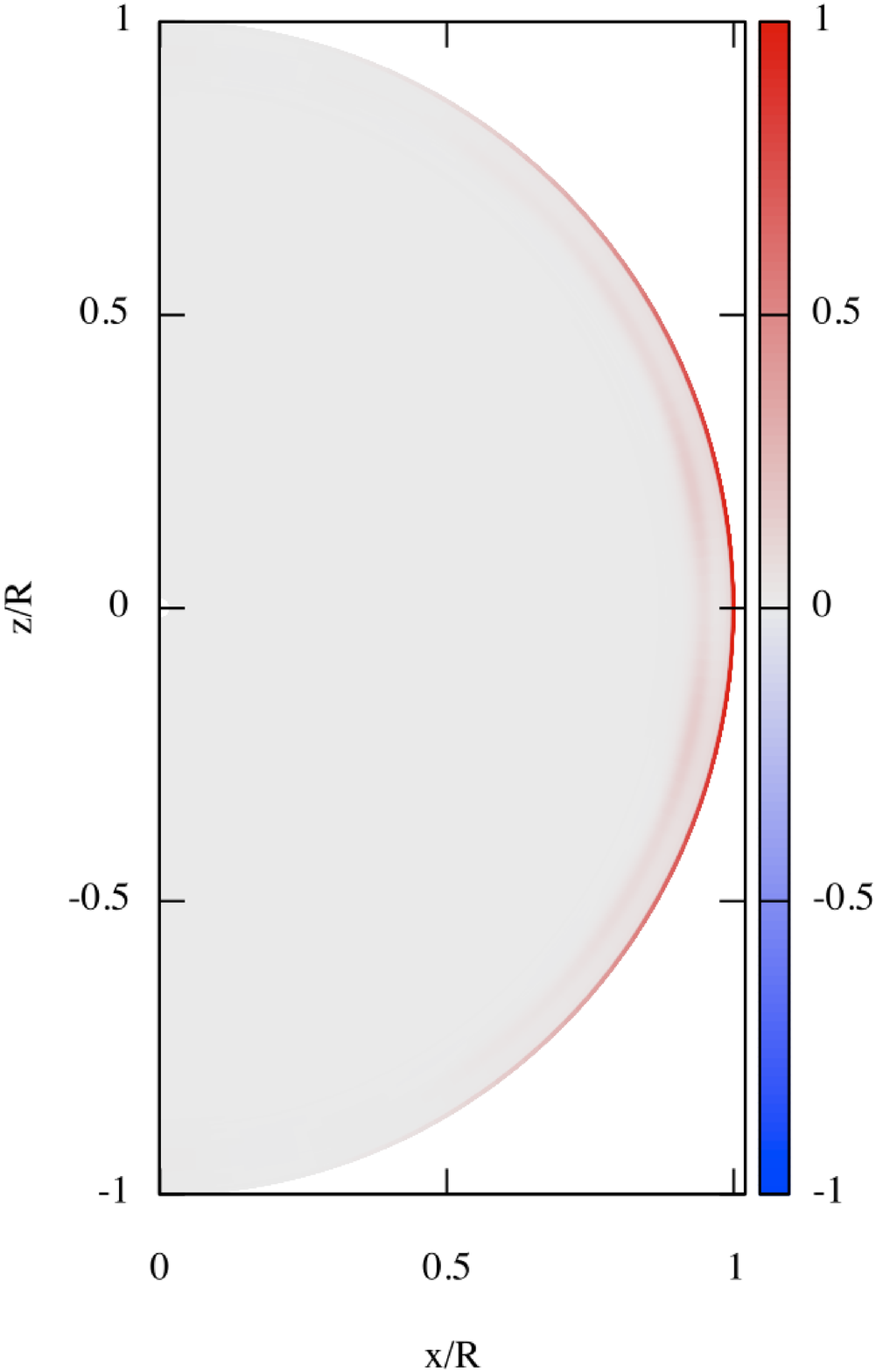}}
\resizebox{0.33\columnwidth}{!}{
\includegraphics{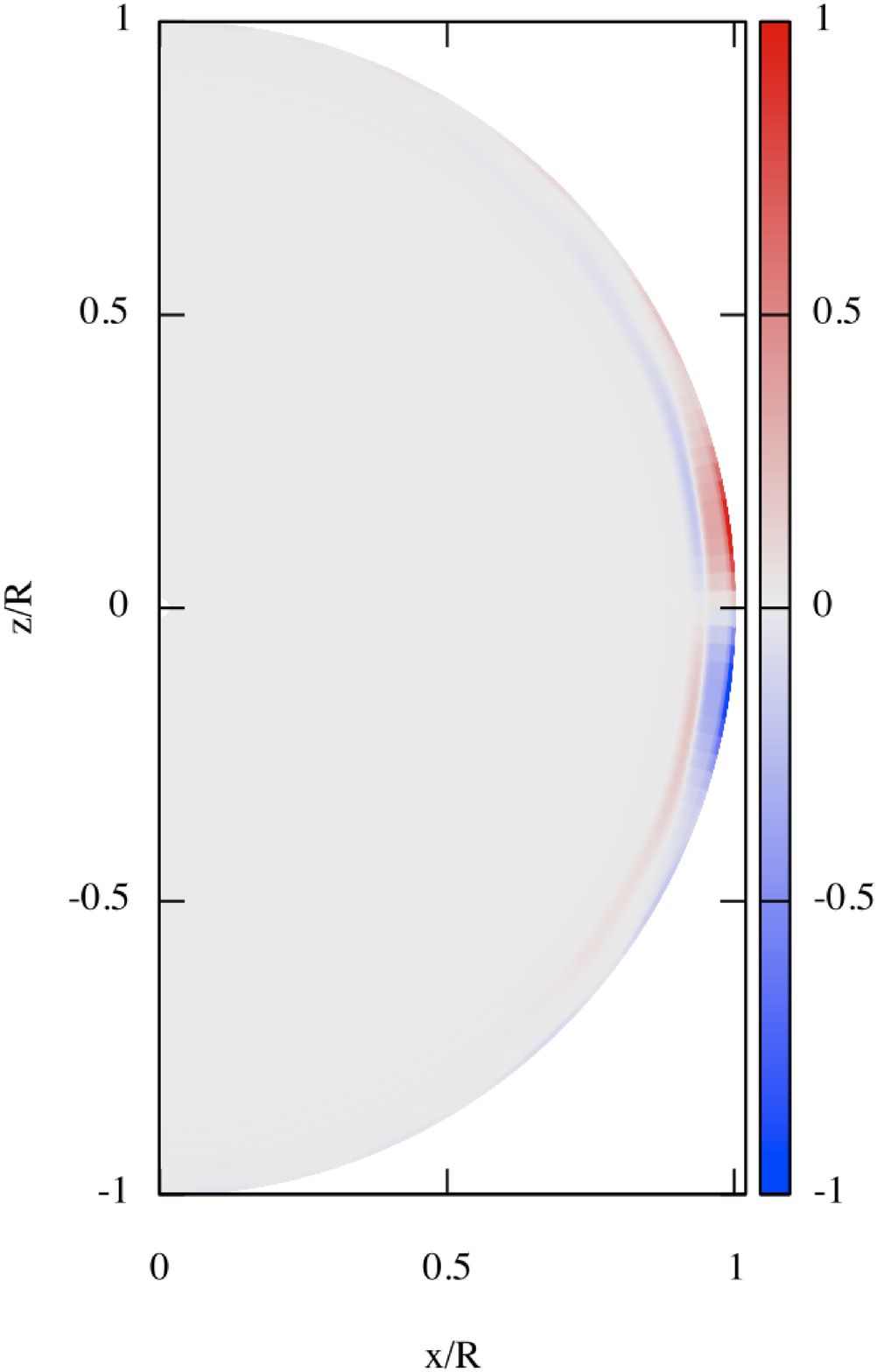}}
\resizebox{0.33\columnwidth}{!}{
\includegraphics{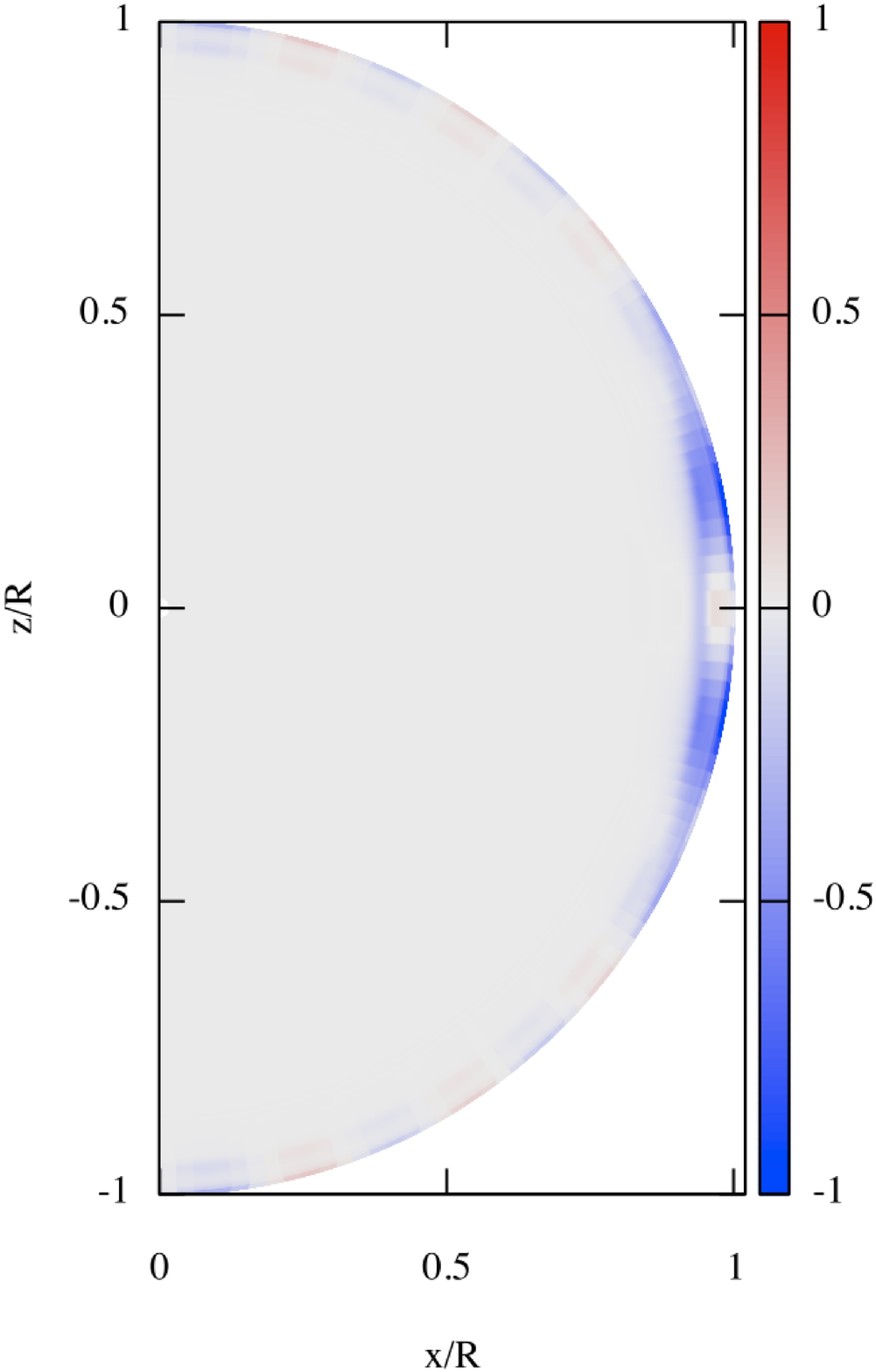}}
\end{center}
\caption{Same as Figure 6 but for the $l=|m|=2$ even prograde $p$-mode.
}
\end{figure}

\begin{figure}
\begin{center}
\resizebox{0.33\columnwidth}{!}{
\includegraphics{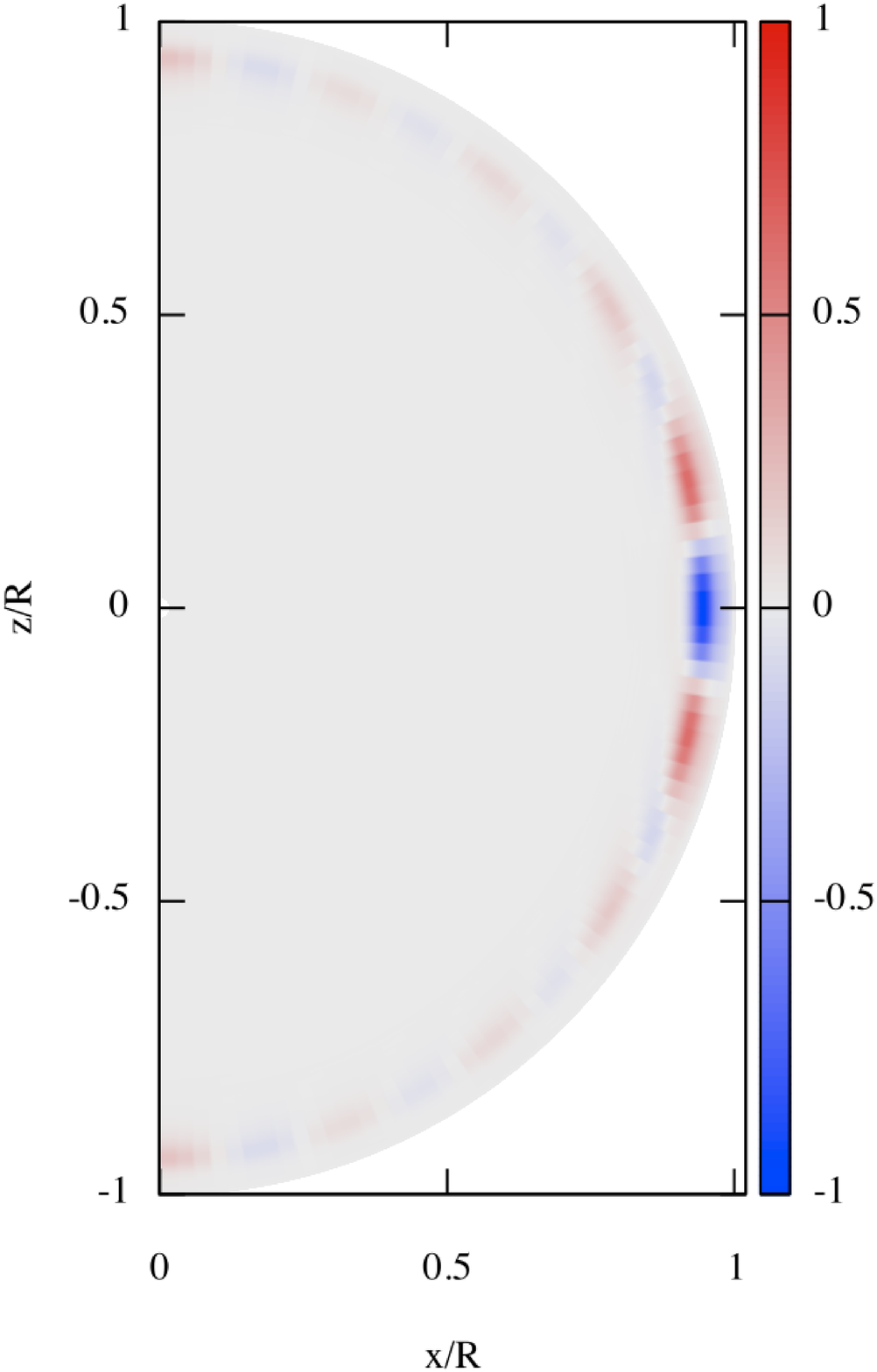}}
\resizebox{0.33\columnwidth}{!}{
\includegraphics{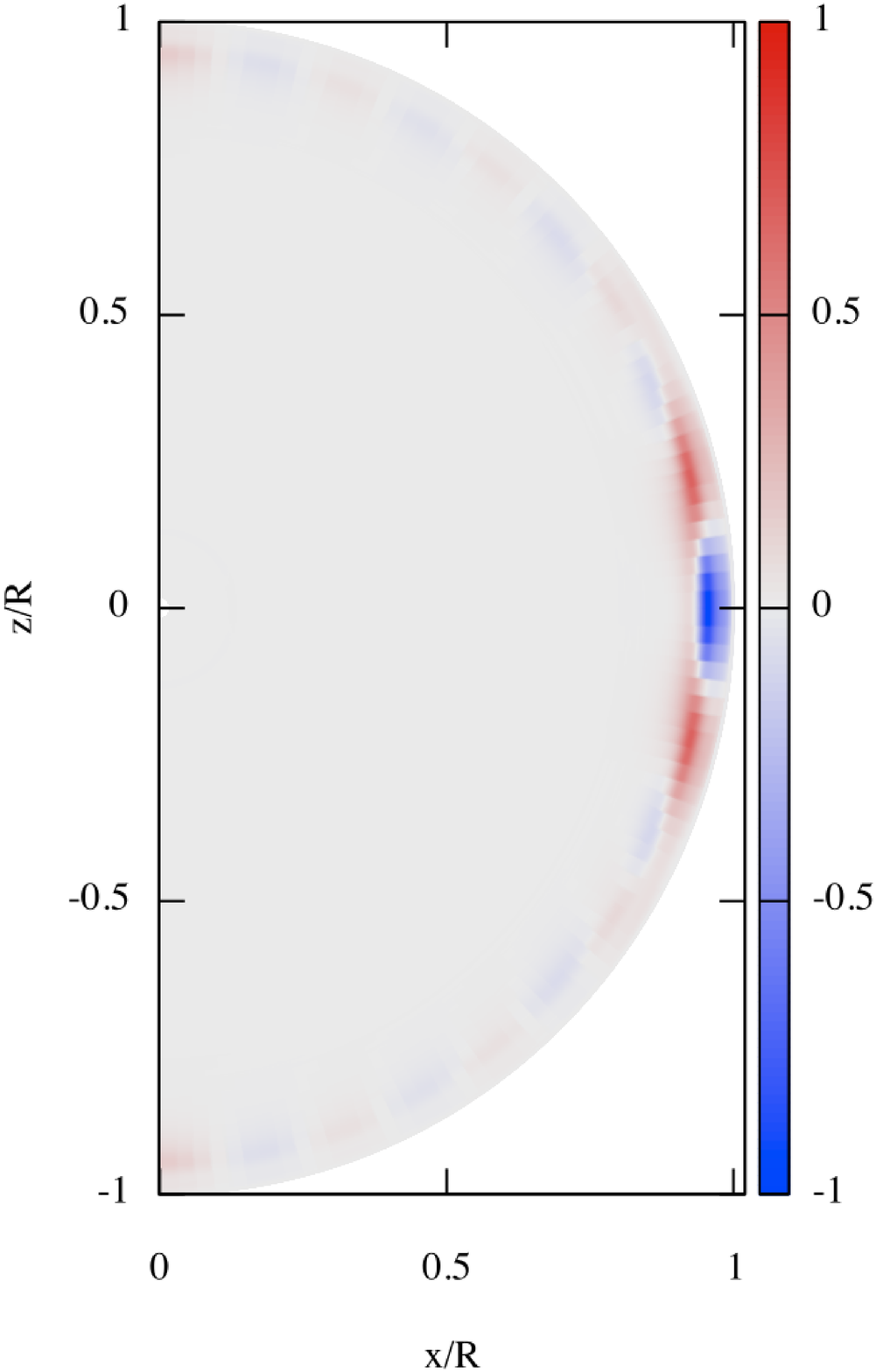}}
\end{center}
\caption{Same as Figure 5 but for even retrograde and prograde $p$-modes of
$l=|m|=2$ of the $15M_\odot$ main sequence star model where $\bar\Omega=0.1$.
}
\end{figure}

\section{Discussion and Conclusion}

\begin{figure}
\begin{center}
\resizebox{0.33\columnwidth}{!}{
\includegraphics{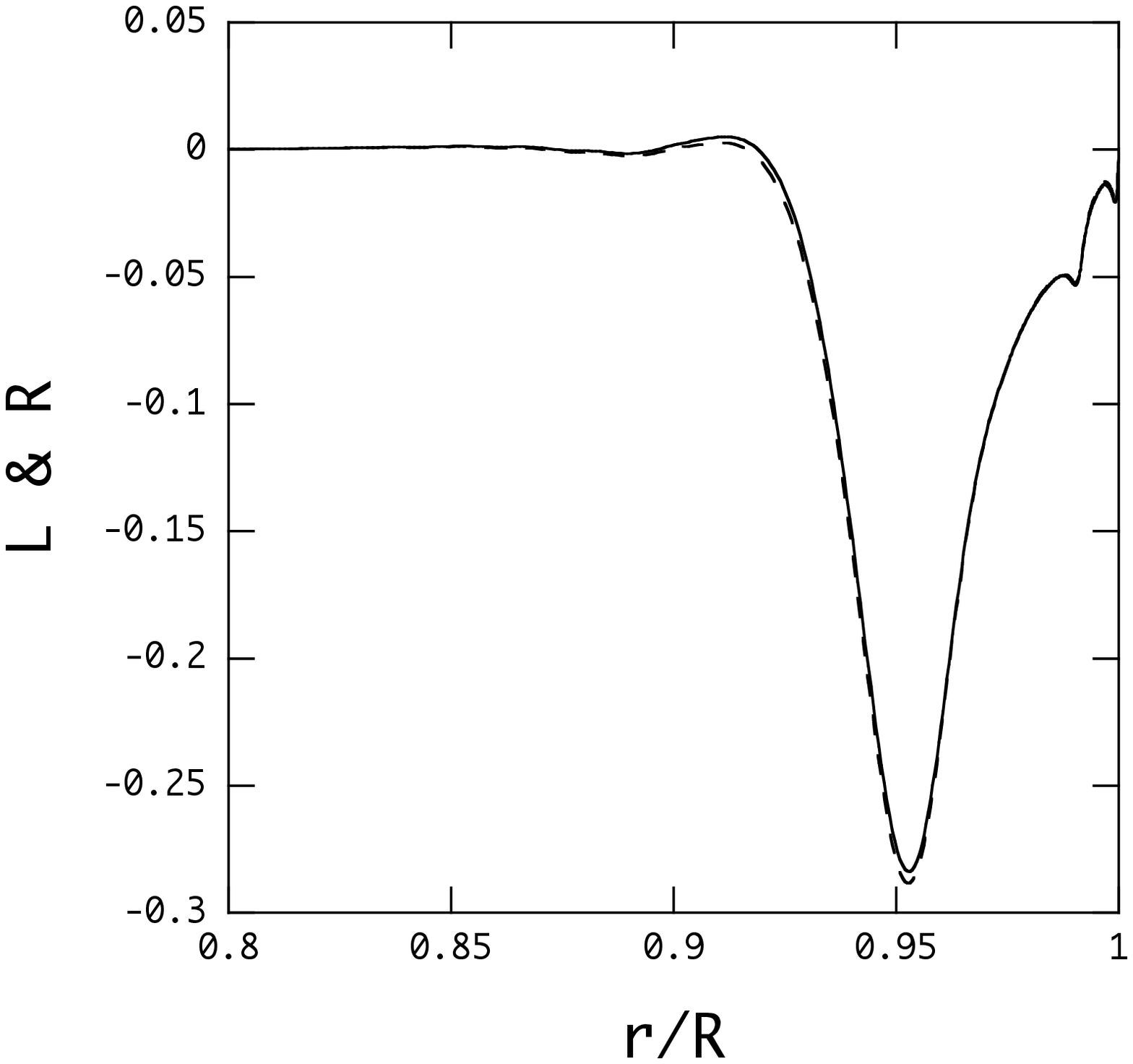}}
\resizebox{0.33\columnwidth}{!}{
\includegraphics{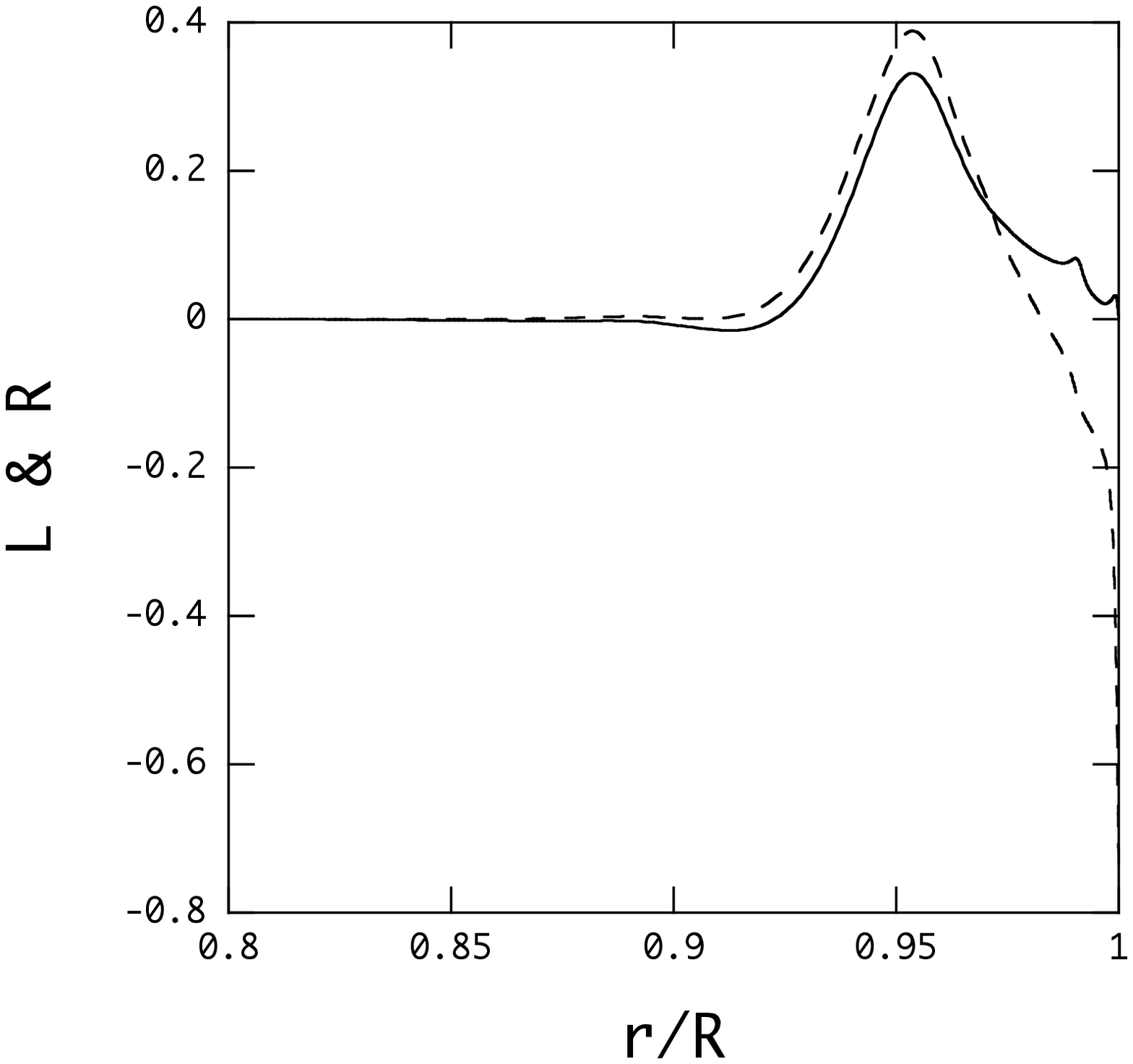}}
\resizebox{0.33\columnwidth}{!}{
\includegraphics{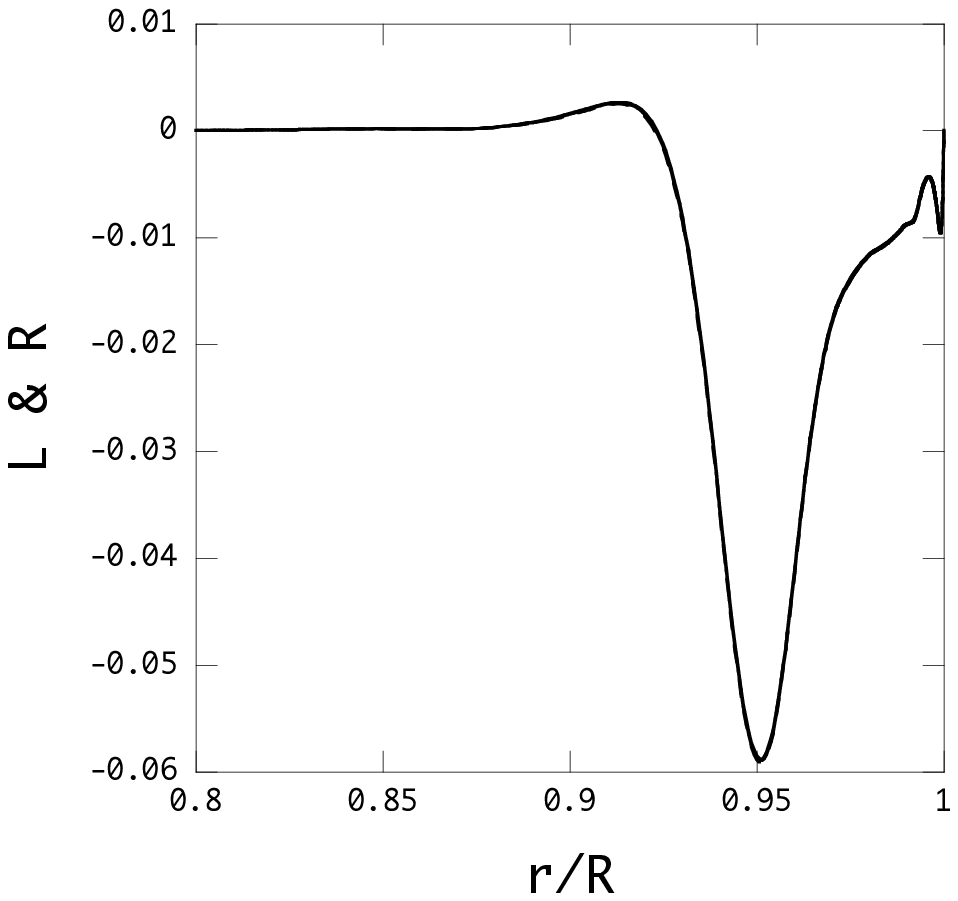}}
\end{center}
\caption{$\cal L$ (dashed line) and $\cal R$ (solid line) 
for
the $g_{30}$-mode and $r_{36}$-mode of the SPB star model and for the $p$-mode of the $\beta$ Cephei star model,
from left to right panels, where we use $\bar\Omega=0.1$ for the $g$- and $p$-modes and $\bar\Omega=0.4$
for the $r$-mode.
}
\end{figure}

As a consistency cheque of our mean flow computations, in Figure 9 we plot the functions $\cal L$ and $\cal R$
for the prograde $p$- and $g$-modes, and the retrograde $r$-modes depicted in Figure 1.
Note that equation (\ref{eq:phicomp2}) is well satisfied for these modes.
The two functions agree very well for the $g$- and $p$-modes, but they show
disagreement in the outer most layers for the $r_{36}$-mode.
Although the reason of the disagreement for the $r$-mode is not well understood,
possible reasons might be that the number of radial nodes of the eigenfunctions is large, which makes it
difficult to correctly compute the eigenfunctions, and that
the toroidal components of the displacement vector are significantly dominant over the the other components
although equation (\ref{eq:phicomp2}) is determined by the phase differences between the less dominant eigenfunctions.

Assuming an initial uniform rotation, we have derived the governing
equations for zonal and meridional axisymmetric mean flows driven by
unstable non-axisymmetric oscillation modes in rapidly
rotating massive main-sequence stars, where the
magnitude of mean flows are assumed to be of second-order of
the oscillation amplitudes. The governing equations are a set
of coupled linear ordinary differential equations for the second-order
quantities with inhomogeneous terms, coming from products of the eigenfunctions
of linear oscillation modes.
To demonstrate the applicability of the formalism and its 
importance for astrophysical studies, we have computed zonal and
meridional axisymmetric mean flows driven by non-axisymmetric $g$- and
$r$-modes in SPB stars and $p$-modes in $\beta$ Cephei stars, where oscillation
modes are assumed to be excited by the iron opacity bump mechanism.

For most of the oscillation modes considered in this paper, these mean
flows have large amplitudes only in the surface regions of the stars, particularly
in the regions where the oscillation modes are excited. 
The first interesting point to note is that for low frequency retrograde $g$-modes and
$r$-modes excited by the opacity bump mechanism, 
$v_\phi^{(2)}$ can be positive in the surface equatorial regions, indicating
that surface fluid elements could be accelerated  
in the same direction as the surface stellar rotation.
The velocity fields generated in the surface layers are related to
the Reynolds stresses of the waves and transported fluxes
that drive exchanges between waves (oscillations) and zonal and meridional mean
flows 
(e.g., Bretherton 1969; Andrews \& McIntyre 1978a; Holton 1982; Mathis et al. 2013; Belkacem et al. 2015a).
For the zonal component,
we provide for the first time the 2-D geometry of the
wave-driven differential rotation given by
$v_\phi^{(2)}(r,\theta)/\left(r\sin\theta\sigma_0\right)$. This provides us
with information on the transport of
angular momentum by waves both in the vertical and in the latitudinal directions
(e.g., Andrews \& McIntyre 1978a; Mathis 2009).

Moreover, we computed for the first time 2-D wave-driven meridional
circulation. It can have multi-cellular pattern in the radial direction, while
its latitude-dependence depends on the studied modes. If we use Eq. (1) to
roughly estimate the magnitude of {the rotationally-driven} meridional
circulation, we have $v_{r:\rm MC}/R\sigma_0\sim 10^{-7}\times \bar\Omega^2$ for
a $10M_\odot$ ZAMS star. This number suggests that pulsation-driven mean
meridional flows have comparable magnitude
to rotationally-driven meridional circulation if the
oscillation modes have amplitudes $S_{l_1}\sim 10^{-4}$ to $10^{-3}$ at the
surface, depending on the mode. To estimate the magnitudes of velocity fields of
pulsation-driven mean flows, we need to theoretically estimate the amplitudes of
pulsation, which has always been a difficult problem. One possible way to
estimate oscillation amplitudes for the $\kappa-$mechanism is to employ a weak
non-linear theory of oscillations (see e.g., Lee 2012),
although it is another very difficult task to apply {this} theory to
oscillation modes in rapidly rotating stars. As suggested by Lee (2012),
the amplitudes of this magnitude $S_{l_1}\sim
10^{-4}$ to $10^{-3}$ is probably attainable in SPB stars. 
This is the
reason why wave-driven mean zonal and meridional mean flows must be taken into
account {when studying} the evolution of rotating massive
stars, as {it is done for} low-mass stars
(e.g., Talon \& Charbonnel 2005; Mathis et al. 2013).

In this paper, we have
applied several simplifying assumptions to make the formulation and calculation
of mean flows tractable. {First,} we used the radiative transfer equation to
derive the governing equation for the second-order variables, even for
convective regions in the interior. This crude treatment may be justified for
massive main-sequence stars having only weak surface convection layers, but may
not be justified for low-mass stars with thick surface convection zones.
Note that $\kappa$-driven pulsations do not have any
contribution to mean flow in the convective core of massive stars, unless the
effect of the viscous force on inertial modes is taken into account. However,
the induced transport of momentum is negligible compared to transport by
convective eddies (Browning et al. 2004).
In addition, we
ignored the effects of centrifugal force on the equilibrium structure,
oscillation calculation, and mean flow computation for rapidly rotating stars.
This neglect of the centrifugal force effects may not be a serious drawback for
the mean flows driven by low frequency $g$- and $r$-modes. However,
{these effects have to be taken into account for mean flows
driven by $p$-modes} in stars rotating as rapidly as $\bar\Omega\sim1$
{(not treated in this paper)}. Next, we assumed { an
initial} uniform rotation, and neglected the possible existence of critical
layers and breaking regions, at which waves suffer strong dissipation and
efficient exchange of momentum between mean flows and waves may take place 
(Rogers et al. 2013; Alvan et al. 2013).
These processes will be studied in the near future.
Finally, low-frequency modes could also be
excited stochastically by turbulent convection in the core of massive
stars 
(e.g., Rogers et al. 2013; Lee et al. 2014; Mathis et al. 2014).
In particular, stochastically excited
waves appear to be very important for the ejections of matter by Be stars
(Neiner et al. 2012; Lee et al. 2014).
The theory developed in
this work will be applied to these stochastically-excited
modes in the future.

This work shows that pulsation-driven flows can be as
significant as rotation-driven flows in pulsating massive stars. Therefore, it
is important to take them into account when studying the rotational and chemical
evolution of massive stars. In addition, pulsation-driven flows can transport
angular momentum to the surface layers, resulting in acceleration or deceleration of rotation
velocity of the fluid of stars.
This angular momentum transport mechanism might help to form a circumstellar
gaseous disc around Be stars, 
although the magnitudes and directions of the mean flows in the radial direction suggested in this paper
are not necessarily favorable for the mechanism to be viable.
In this paper, we have treated the oscillations as standing waves in the radial direction.
If we assume low frequency modes
become progressive in the surface layers because of their very low frequencies as suggested by Ishimatsu \& Shibahashi (2013),
the way of transport of angular momentum by the waves could be different from that by standing waves.
This possibility will be pursued in a future paper concerning angular momentum transport by
stochastically excited oscillations of massive stars.

\section*{Acknowledgements}
We thank the anonymous referee for his/her detailed, critical and constructive comments
on the original manuscript. The third paragraph in \S 3.1.1 is due to the referee.



\begin{thebibliography}{}



\bibitem[\protect\citeauthoryear{}{}]{} Alvan L, Mathis S., Decressin T., 2013, A\&A, 553, 86
\bibitem[\protect\citeauthoryear{}{}]{} Ando H., 1983, PASJ, 35, 343
\bibitem[\protect\citeauthoryear{}{}]{} Ando H., 1986, A\&A, 163, 97
\bibitem[\protect\citeauthoryear{}{}]{} Andrews D.G., McIntyre M.F., 1976, J. Atoms. Sci., 33, 2031
\bibitem[\protect\citeauthoryear{}{}]{} Andrews D.G., McIntyre M.F., 1978a, J. Atoms. Sci., 35, 175
\bibitem[\protect\citeauthoryear{}{}]{} Andrews D.G., McIntyre M.F., 1978b, J. Fluid Mech., 89, 609
\bibitem[\protect\citeauthoryear{}{}]{} Aprilia, Lee U., Saio H., 2011, MNRAS, 412, 2265
\bibitem[\protect\citeauthoryear{}{}]{} Beck, P. G., et al., 2012, Nature, 481, Issue 7379, 55
\bibitem[\protect\citeauthoryear{}{}]{} Belkacem K., et al. 2015, eprint arXiv: 1505.05447
\bibitem[\protect\citeauthoryear{}{}]{} Belkacem K., et al. 2015, eprint arXiv:1505.05452 
\bibitem[\protect\citeauthoryear{}{}]{} Berthomieu G., Gonczi G., Graff Ph., Provost J., Rpcca A., 1978, A\&A, 70, 597
\bibitem[\protect\citeauthoryear{}{}]{} Bildsten L., Ushomirsky G., Cutler C., 1996, ApJ, 460, 827
\bibitem[\protect\citeauthoryear{}{}]{} Bretherton F.P., 1969, Journal of Fluid Mechanics, 36, 785
\bibitem[\protect\citeauthoryear{}{}]{} Browning M.K., Brun A.S., Toomre J., 2004, ApJ, 601, 512
\bibitem[\protect\citeauthoryear{}{}]{} B\"uhler O., 2014, Waves and Mean Flows (Cambridge University Press, Cambridge)
\bibitem[\protect\citeauthoryear{}{}]{} Craik A.D.D., 1985, Wave interactions and fluid flows (Cambridge University Press, Cambridge)
\bibitem[\protect\citeauthoryear{}{}]{} Decressin T., et al., 2009, A\&A, 495, 271
\bibitem[\protect\citeauthoryear{}{}]{} Deheuvels S., et al., 2012, ApJ, 756, 19 
\bibitem[\protect\citeauthoryear{}{}]{} Deheuvels S., et al. 2014, A\&A, 564, 27
\bibitem[\protect\citeauthoryear{}{}]{} Dunkerton T., 1980, Rev. Geophys. Sp. Phys., 18, 387
\bibitem[\protect\citeauthoryear{}{}]{} Dziembowski W.A., Moskalik P., Pamyatnykh A.A., 1993, MNRAS. 265, 588
\bibitem[\protect\citeauthoryear{}{}]{} Edmonds A.R., 1968, Angular Momentum in Quantum Mechanics (Princeton University Press,
Princeton, NJ)
\bibitem[\protect\citeauthoryear{}{}]{} Ekstr\"om S., et al., 2008, A\&A, 478, 467
\bibitem[\protect\citeauthoryear{}{}]{} Fuller J., Cantiello M., Brown B., 2014, ApJ, 796, 17
\bibitem[\protect\citeauthoryear{}{}]{} Garc\'ia, R. A., et al., 2007, Science, 316, Issue 5831, 1591
\bibitem[\protect\citeauthoryear{}{}]{} Gautschy A., Saio H., 1993, MNRAS, 267, 1071
\bibitem[\protect\citeauthoryear{}{}]{} Goldreich P., Nicholson P.D., 1989, ApJ, 342, 1075
\bibitem[\protect\citeauthoryear{}{}]{} Granada A., et al., 2013, A\&A, 553, 25
\bibitem[\protect\citeauthoryear{}{}]{} Grimshaw R., 1984, Ann. Rev. Fluid. Mech., 16, 11
\bibitem[\protect\citeauthoryear{}{}]{} Hypolite D., Rieutord M., 2014, A\&A, 572, 15
\bibitem[\protect\citeauthoryear{}{}]{} Iglesias C.A., Rogers F.J., 1996, ApJ, 464, 943
\bibitem[\protect\citeauthoryear{}{}]{} Ishimatsu H., Shibahashi H., 2013, in Progress in Physics of the Sun and Stars: A New Are in
Helio- and Asteroseismology ed. H. Shibahashi \& A.E. Lynas-Gray (APS Conference Proceedings Vol 479, San Francisco)
\bibitem[\protect\citeauthoryear{}{}]{} Kippenhahn, Weigert, \& Weiss, 2012, Stellar Structure and Evolution (Springer-Verlag, Berlin)
\bibitem[\protect\citeauthoryear{}{}]{} Kumar P., Talon S., Zahn J.P., 1999, ApJ, 520, 859
\bibitem[\protect\citeauthoryear{}{}]{} Lee U., 2006, MNRAS, 365, 677
\bibitem[\protect\citeauthoryear{}{}]{} Lee U., 2012, MNRAS, 420, 2387
\bibitem[\protect\citeauthoryear{}{}]{} Lee U., 2013, PASJ, 65, 122
\bibitem[\protect\citeauthoryear{}{}]{} Lee U., Baraffe I., 1995, A\&A, 301, 419
\bibitem[\protect\citeauthoryear{}{}]{} Lee U., Neiner C., Mathis S., 2014, MNRAS, 443, 1515
\bibitem[\protect\citeauthoryear{}{}]{} Lee U., Saio H., 1987, MNRAS, 225, 643
\bibitem[\protect\citeauthoryear{}{}]{} Lee U., Saio H., 1993, MNRAS, 261, 415
\bibitem[\protect\citeauthoryear{}{}]{} Lee U., Saio H., 1997, ApJ, 491, 839
\bibitem[\protect\citeauthoryear{}{}]{} Lee U., Saio H., Osaki Y., 1991, MNRAS, 250, 432
\bibitem[\protect\citeauthoryear{}{}]{} Ligni\`eres F., Rieutord M., Reese D., 2006, A\&A, 455, 607
\bibitem[\protect\citeauthoryear{}{}]{} Lindzen R.S., 1981, J. Geophys. Res., 86, 9707
\bibitem[\protect\citeauthoryear{}{}]{} Maeder A., \& Zahn J.P., 1998, A\&A, 334, 1000
\bibitem[\protect\citeauthoryear{}{}]{} Mathis S., 2009, A\&A, 506, 811
\bibitem[\protect\citeauthoryear{}{}]{} Mathis S., \& de Brye N., 2012, A\&A, 540, 37
\bibitem[\protect\citeauthoryear{}{}]{} Mathis S., \& Zahn J.P., 2004, A\&A, 425, 229
\bibitem[\protect\citeauthoryear{}{}]{} Mathis S., Talon S., Pantillon F.P., Zahn J.P., 2008, Solar Phys., 251, 101
\bibitem[\protect\citeauthoryear{}{}]{} Mathis S., et al., 2013, A\&A, 558, 11
\bibitem[\protect\citeauthoryear{}{}]{} Mathis S., Neiner C., TranMinh N., 2014, A\&A, 565, 47
\bibitem[\protect\citeauthoryear{}{}]{} Meynet G., \& Maeder A., 1997, A\&A, 321, 465
\bibitem[\protect\citeauthoryear{}{}]{} Meynet G., \& Maeder A., 2000, A\&A, 361, 101
\bibitem[\protect\citeauthoryear{}{}]{} Mosser B., et al. 2012, A\&A, 548, 10
\bibitem[\protect\citeauthoryear{}{}]{} Neiner C., et al., 2012, A\&A, 546, 47
\bibitem[\protect\citeauthoryear{}{}]{} Newman E.T., Penrose R., 1966, J. Math. Phys., 7, 863
\bibitem[\protect\citeauthoryear{}{}]{} Pantillon F.P., Talon S., Charbonnel C., 2007, A\&A, 474, 155
\bibitem[\protect\citeauthoryear{}{}]{} Pedlosky J., 1982, Geophysical Fluid Dynamics (Springer-Verlag, Berilin)
\bibitem[\protect\citeauthoryear{}{}]{} Press W.H., 1981, ApJ, 245, 286
\bibitem[\protect\citeauthoryear{}{}]{} Reese D., Ligni\`eres F., Rieutord M., 2006, A\&A, 455, 621
\bibitem[\protect\citeauthoryear{}{}]{} Rieutord M., 2006, A\&A, 451, 1025
\bibitem[\protect\citeauthoryear{}{}]{} Rivinius T., Carciofi A.C., Martayan C., 2013, AAR, 21, 69
\bibitem[\protect\citeauthoryear{}{}]{} Rogers T. M., MacGregor K. B., 2010, MNRAS, 401, 191
\bibitem[\protect\citeauthoryear{}{}]{} Rogers T. M., MacGregor K. B., Glarzmaier G. A., 2008, MNRAS, 387, 616
\bibitem[\protect\citeauthoryear{}{}]{} Rogers T. M., et al., 2013, ApJ, 772, 21
\bibitem[\protect\citeauthoryear{}{}]{} Rose M.E., 1957, Elementary Theory of Angular Momentum (Wiley \& Sons)
\bibitem[\protect\citeauthoryear{}{}]{} Schenk A.K., Arras P., Flanagan \'E.\'E., Teukolsky S.A., Wasserman I., 2002, Phys. Rev. D, 65, 024001
\bibitem[\protect\citeauthoryear{}{}]{} Schatzman E., 1993, A\&A, 279, 431
\bibitem[\protect\citeauthoryear{}{}]{} Schatzman E., 1996, JFM, 322, 355
\bibitem[\protect\citeauthoryear{}{}]{} Schou J., et al., 1998, ApJ, 505, 390
\bibitem[\protect\citeauthoryear{}{}]{} Talon S., \& Charbonnel C., 2005, A\&A, 440, 981
\bibitem[\protect\citeauthoryear{}{}]{} Talon S., \& Charbonnel C., 2008, A\&A, 482, 597
\bibitem[\protect\citeauthoryear{}{}]{} Talon S., Kumar P., Zahn J.P., 2002, ApJ, 574, L175
\bibitem[\protect\citeauthoryear{}{}]{} Talon S., Zahn J.-P., Maeder A., Meynet G., 1997, A\&A, 322, 209
\bibitem[\protect\citeauthoryear{}{}]{} Townsend R.H.D., MNRAS, 340, 1020
\bibitem[\protect\citeauthoryear{}{}]{} Townsend R.H.D., MNRAS, 343, 125
\bibitem[\protect\citeauthoryear{}{}]{} Unno W., Osaki Y., Ando Y., Saio H., Shibahashi H., 1989, Nonradial oscillations of Stars, 2nd edn. (University of Tokyo Press)
\bibitem[\protect\citeauthoryear{}{}]{} Varshalovich D., Moskalev A., Khersonskii V.K., 1988, Quantum Theory of Angular Momentum
(World Scientific Publishing)
\bibitem[\protect\citeauthoryear{}{}]{} Zahn J.P., 1975, A\&A, 41, 329
\bibitem[\protect\citeauthoryear{}{}]{} Zahn J.P., 1977, A\&A, 57, 383
\bibitem[\protect\citeauthoryear{}{}]{} Zahn J.P., 1992, A\&A, 265, 115
\bibitem[\protect\citeauthoryear{}{}]{} Zhan J.P., Talon S., Mathias J., 1997, A\&A, 322, 320


\end{thebibliography}


\begin{appendix}

\section{Spin-Weighted Spherical Harmonic Functions}

Spin-weighted spherical harmonic functions ${}_sY_l^m(\theta,\phi)$ may be defined
as eigenfunctions of the differential equation
\be
\eth\tilde\eth{}_sY_l^m=-(l+s)(l-s+1){}_sY_l^m,
\ee
or
\be
\tilde\eth\eth{}_sY_l^m=-(l-s)(l+s+1){}_sY_l^m,
\ee
where the differential operators $\eth$ and $\tilde\eth$ are defined for ${}_sY_l^m$ with $-l\le s\le l$ as
 (Newman \& Penrose 1966)
\be
\eth{}_sY_l^m\equiv-\left(\frac{\partial}{\partial\theta}+\frac{\rmi}{\sin\theta}\frac{\partial}{\partial\phi}-s\cot\theta\right){}_sY_l^m
=\sqrt{(l-s)(l+s+1)}{}_{s+1}Y_l^m,
\ee
\be
\tilde\eth{}_sY_l^m\equiv-\left(\frac{\partial}{\partial\theta}-\frac{\rmi}{\sin\theta}\frac{\partial}{\partial\phi}+s\cot\theta\right){}_sY_l^m
=-\sqrt{(l+s)(l-s+1)}{}_{s-1}Y_l^m.
\ee
For a given value of $s$, the function ${}_sY_l^m$ is normalized as
\be
\int \left({}_sY_{l_1}^{m_1}\right)^*{}_sY_{l_2}^{m_2}\sin\theta d\theta d\phi=\delta_{l_1l_2}\delta_{m_1m_2},
\ee
where
\be
\left({}_sY_{l}^{m}\right)^*=(-1)^{s+m}{}_{-s}Y_l^{-m}.
\ee
We note that ${}_0Y_l^m=Y_l^m$, and it is convenient to use
\be
\tilde\eth{}_0Y_l^m=-\sqrt{\Lambda_l}{}_{-1}Y_l^m, \quad \eth{}_{-1}Y_l^m=\sqrt{\Lambda_l}{}_0Y_l^m, \quad \eth\tilde\eth{}_0Y_l^m=-\Lambda_l{}_0Y_l^m,
\ee
\be
\eth{}_0Y_l^m=\sqrt{\Lambda_l}{}_1Y_l^m, \quad \tilde\eth\eth{}_0Y_l^m=-\Lambda_l{}_0Y_l^m,
\ee
\be
\eth{}_{-1}Y_1^0=\sqrt{2}{}_0Y_1^0, \quad \eth{}_1Y_1^0=0, \quad \tilde\eth{}_{-1}Y_1^0=0, \quad
\tilde\eth{}_1Y_1^0=-\sqrt{2}{}_0Y_1^0,
\ee
where $\Lambda_l=l(l+1)$.

Angular integration of a product of three spin-weighted spherical harmonic functions can be 
evaluated by using the formula given by
\begin{eqnarray}
\left[\matrix{l_1 & l_2 & l_3\cr m_1 & m_2 & m_3 \cr s_1& s_2 & s_3 \cr}\right]&\equiv&\int{}_{s_1}Y_{l_1}^{m_1}{}_{s_2}Y_{l_2}^{m_2}{}_{s_3}Y_{l_3}^{m_3} d\Omega\nonumber\\
&=&4\pi\sqrt{{2l_1+1\over4\pi}}
\sqrt{{2l_2+1\over4\pi}}\sqrt{{2l_3+1\over4\pi}}\left(\matrix{l_1 & l_2 & l_3\cr -s_1 & -s_2 & -s_2\cr}\right)\left(\matrix{l_1 & l_2 & l_3\cr m_1 & m_2 & m_2\cr}\right),
\end{eqnarray}
where $d\Omega=\sin\theta d\theta d\phi$, and 
$\left(\matrix{l_1 & l_2 & l_3\cr m_1 & m_2 & m_2\cr}\right)$
is Wigner 3-$j$ symbol (Rose 1957; Edmonds 1968).
This integration has none zero values only when $m_1+m_2+m_3=0$, $s_1+s_2+s_3=0$, and $l_1+l_2\ge l_3\ge |l_2-l_1|$.

\section{Second-order equations for mean flows and Spin-Weighted Spherical Harmonics}

\subsection{Introducing the basis associated to Spin-Weighted Spherical Harmonics}

We introduce a new set of basis vectors $\pmb{e}_q$ and $\pmb{e}_{\bar q}$ defined by
\be
\pmb{e}_q=\frac{\pmb{e}_\theta+\rmi\pmb{e}_\phi}{\sqrt{2}}, \quad \pmb{e}_{\bar q}=\frac{\pmb{e}_\theta-\rmi\pmb{e}_\phi}{\sqrt{2}},
\ee
for which
\be
\pmb{e}_r\cdot\pmb{e}_q=\pmb{e}_r\cdot\pmb{e}_{\bar q}=0,\quad \pmb{e}_q\cdot\pmb{e}_q=\pmb{e}_{\bar q}\cdot\pmb{e}_{\bar q}=0, \quad \pmb{e}_q\cdot\pmb{e}_{\bar q}=1.
\ee
Using $\pmb{e}_r$, $\pmb{e}_q$ and $\pmb{e}_{\bar q}$, we may rewrite the displacement vector $\pmb{\xi}(\pmb{x})$ as
\be
\pmb{\xi}(\pmb{x})=\xi_r\pmb{e}_r+\xi_q\pmb{e}_q+\xi_{\bar q}\pmb{e}_{\bar q},
\ee
where
\be
\xi_q=\frac{\xi_\theta-\rmi\xi_\phi}{\sqrt{2}}, \quad \xi_{\bar q}=\frac{\xi_\theta+\rmi\xi_\phi}{\sqrt{2}}.
\ee
The differential operator $\nabla$ may be rewritten as
\be
\nabla
=\pmb{e}_r\frac{\partial}{\partial r}+\pmb{e}_q\frac{\partial}{\partial \bar q}+\pmb{e}_{\bar q}\frac{\partial}{\partial q},
\ee
where
\be
\frac{\partial}{\partial q}\equiv\pmb{e}_{ q}\cdot\nabla=\frac{1}{\sqrt{2}r}\left(\frac{\partial}{\partial\theta}
+\rmi\frac{1}{\sin\theta}\frac{\partial}{\partial\phi}\right),
\ee
\be
\frac{\partial}{\partial\bar q}\equiv\pmb{e}_{\bar q}\cdot\nabla=\frac{1}{ \sqrt{2}r}\left(\frac{\partial}{\partial\theta}
-\rmi\frac{1}{\sin\theta}\frac{\partial}{\partial\phi}\right).
\ee
We find that
\be
\frac{\partial}{\partial q}{}_0Y_l^m=-\frac{1}{\sqrt{2}r}\eth{}_0Y_l^m=-\frac{1}{\sqrt{2}r}\sqrt{\Lambda_l}{}_1Y_l^m, 
\ee
\be
\frac{\partial}{\partial \bar q}{}_0Y_l^m=-\frac{1}{\sqrt{2}r}\tilde\eth{}_0Y_l^m=
\frac{1}{\sqrt{2}r}\sqrt{\Lambda_l}{}_{-1}Y_l^m.
\ee

Using the basis vectors $\pmb{e}_q$ and $\pmb{e}_{\bar q}$ and
spin-weighted spherical harmonics ${}_sY_l^m$, we rearrange the series expansions (\ref{eq:xiexp_r}) to (\ref{eq:xiexp_phi}) as
\be
\xi_r=r\sum_{j=1}^{j_{\rm max}}S_{l_j}Y_{l_j}^m,
\ee
\be
\xi_q=-\frac{1}{\sqrt{2}}r\sum_{j=1}^{j_{\rm max}}\left(H_{l_j}\tilde\eth{}_0Y_{l_j}^m+
\rmi T_{l'_j}\tilde\eth{}_0Y_{l'_j}^m\right),
\ee
\be
\xi_{\bar q}=-\frac{1}{\sqrt{2}}r\sum_{j=1}^{j_{\rm max}}\left(H_{l_j}\eth{}_0Y_{l_j}^m-
\rmi T_{l'_j}\eth{}_0Y_{l'_j}^m\right),
\ee
and similarly the expansions (\ref{eq:expand2r}) to (\ref{eq:expand2t}) as
\be
v_r^{(2)}=\sum_{k=1}^{k_{\rm max}}\hat v_{S,l_k}^{(2)}{}_0Y_{l_k}^0,
\ee
\be
v_q^{(2)}=-\frac{1}{\sqrt{2}}\sum_{k=1}^{k_{\rm max}}\left(\hat v_{H,l_k}^{(2)}\tilde\eth{}_0Y_{l_k}^0+
\rmi\hat v_{T,l'_k}^{(2)}\tilde\eth{}_0Y_{l'_k}^0\right),
\ee
\be
v_{\bar q}^{(2)}=-\frac{1}{\sqrt{2}}\sum_{k=1}^{k_{\rm max}}\left(\hat v_{H,l_k}^{(2)}\eth{}_0Y_{l_k}^0
-\rmi\hat v_{T,l'_k}^{(2)}\eth{}_0Y_{l'_k}^0\right).
\ee

If we write $\pmb{v}^{(0)}=\rmi rf{}_{-1}Y_1^0\pmb{e}_q+ \rmi rf{}_1Y_1^0\pmb{e}_{\bar q}$ 
using $f=\sqrt{4\pi/3}~\Omega$ and $\sin\theta=\sqrt{8\pi/ 3}{}_1Y_1^0=-\sqrt{8\pi/ 3}{}_{-1}Y_1^0$,
the Coriolis term in equation (\ref{eq:secondmomcon}) reduces to 
\be
\pmb{v}^{(2)}\cdot\nabla\pmb{v}^{(0)}+\pmb{v}^{(0)}\cdot\nabla\pmb{v}^{(2)}=2\rmi f
\left[-\left(v_q^{(2)}{}_1Y_1^0+v_{\bar q}^{(2)}{}_{-1}Y_1^0\right)\pmb{e}_r+\left(v_r^{(2)}{}_{-1}Y_1^0-v_q^{(2)}{}_0Y_1^0\right)\pmb{e}_q
+\left(v_r^{(2)}{}_1Y_1^0+v_{\bar q}^{(2)}{}_0Y_1^0\right)\pmb{e}_{\bar q}\right],
\ee
and hence the $r$, $q$, and $\bar q$-components of the momentum equation (\ref{eq:secondmomcon}) 
are written as
\be
\frac{\partial v_r^{(2)}}{\partial t}-2\rmi f
\left(v_q^{(2)}{}_1Y_1^0+v_{\bar q }^{(2)}{}_{-1}Y_1^0\right)
+\frac{1}{\rho^{(0)}}\frac{\partial p^{(2)}}{\partial r}+g\frac{\rho^{(2)}}{\rho^{(0)}}=G_r^{(2)},
\label{eq:rcompmom}
\ee
\be
\frac{\partial v_q^{(2)}}{\partial t}+2\rmi f
\left(v_r^{(2)}{}_{-1}Y_1^0-v_q^{ (2)}{}_{0}Y_1^0\right)
+\frac{1}{\rho^{(0)}}\frac{\partial p^{(2)}}{\partial\bar q}=G_q^{(2)},
\label{eq:qcompmom}
\ee
\be
\frac{\partial v_{\bar q}^{(2)}}{\partial t}+2\rmi f
\left(v_r^{(2)}{}_{1}Y_1^0+v_{\bar q}^{ (2)}{}_{0}Y_1^0\right)
+\frac{1}{\rho^{(0)}}\frac{\partial p^{(2)}}{\partial q}=G_{\bar q}^{(2)},
\label{eq:barqcompmom}
\ee
where
\be
\pmb{G}^{(2)}
=- \overline {\pmb{v} ' \cdot \nabla \pmb{v} '}  + g\overline {\left( {\frac{{\rho '}}{{\rho ^{\left( 0 \right)} }}} \right)^2}  
\pmb{e}_r + \frac{1}{{\rho ^{\left( 0 \right)} }}\overline {\frac{{\rho '}}{{\rho ^{\left( 0 \right)} }}\nabla p'}
\equiv G_r^{(2)}\pmb{e}_r+G_q^{(2)}\pmb{e}_q+G_{\bar q}^{(2)}\pmb{e}_{\bar q}.
\label{eq:vecg2}
\ee

We rewrite equations (\ref{eq:secondcont}), (\ref{eq:rcompmom}),  (\ref{eq:ent2eq}), and
the radial component of equation (\ref{eq:radtrans2}) into a non-dimensional form:
\be
r\frac{\partial}{\partial r}\frac{v_r^{(2)}}{ r\sigma_0}=-\frac{\partial}{\partial\tau}\frac{\rho^{(2)}}{\rho^{(0)}}
-\nabla_H\cdot\frac{\pmb{v}_H^{(2)}}{r\sigma_0}-\left(3+\frac{d\ln\rho^{(0)}}{ d\ln r}\right)\frac{v_r^{(2)}}{r\sigma_0}+\frac{H^{(2)}}{\rho^{(0)}\sigma_0},
\label{eq:dvr2dr}
\ee
\be
r\frac{\partial}{\partial r}\frac{p^{(2)}}{\rho^{(0)} gr}=-
c_1\frac{\partial}{\partial \tau}\frac{v_r^{(2)}}{r\sigma_0}
+2\rmi c_1\bar f
\left(\frac{v_q^{(2)}}{r\sigma_0}{}_1Y_1^0
+\frac{v_{\bar q}^{(2)}}{r\sigma_0}{}_{-1}Y_1^0\right)-\frac{d\ln\rho^{(0)} gr}{ d\ln r}\frac{p^{(2)}}{\rho^{(0)} gr}-\frac{\rho^{(2)}}{\rho^{(0)}}+\frac{G_r^{(2)}}{g},
\label{eq:dp2dr}
\ee
\begin{eqnarray}
r\frac{\partial}{\partial r}\frac{L_r^{(2)}}{L_r^{(0)}} &= &-c_2\left[\frac{\partial}{\partial \tau}\frac{T^{(2)}}{T^{(0)}}
-\nabla_{ad}\frac{\partial}{\partial \tau}\frac{p^{(2)}}{p^{(0)}}+\frac{v_r^{(2)}}{ r\sigma_0}V\left(\nabla_{ad}-\nabla\right)\right]\nonumber \\
&& +c_3\left[\left(\hat\epsilon_p+\frac{1}{\chi_\rho}\right)\frac{p^{(2)}}{ p^{(0)}}+
\left(\hat\epsilon_T-\alpha_T\right)\frac{T^{(2)}}{T^{(0)}}\right]
-\frac{d\ln L_r^{(0)}}{d\ln r}\frac{L_r^{(2)}}{L_r^{(0)}}+\frac{1}{ V\nabla}\nabla_H^2\frac{T^{(2)}}{T^{(0)}}+I^{(2)},
\label{eq:dlr2dr}
\end{eqnarray}
\be
r\frac{\partial}{\partial r}\frac{T^{(2)}}{T^{(0)}}=V\nabla\left(4-\hat\kappa_T+\alpha_T\right)\frac{T^{(2)}}{T^{(0)}}
-V\nabla\left(\hat\kappa_p+\frac{1}{\chi_\rho}\right)\frac{p^{(2)}}{p^{(0)}}-V\nabla\frac{L_r^{(2)}}{L_r^{(0)}}-V\nabla J^{(2)},
\label{eq:dt2dr}
\ee
where  
\be
\sigma_0=\sqrt{GM/ R^3}, \quad  \tau=\sigma_0t, \quad \bar\Omega={\Omega/\sigma_0}, 
\quad \bar f=f/\sigma_0,
\ee
\be
L_r^{(0)}=4\pi r^2 F_r^{(0)}, \quad L_r^{(2)}=4\pi r^2 F_r^{(2)}, \quad
F_r^{(0)}=-\lambda^{(0)}{dT^{(0)}/ dr},
\ee
\be
c_1=\frac{\sigma_0^2}{g/r}=\frac{(r/R)^3}{M_r/M}, \quad c_2=\frac{4\pi r^3\rho^{(0)}T^{(0)}c_p}{L_r^{(0)}}\sigma_0, \quad c_3=\frac{4\pi r^3\rho^{(0)}\epsilon^{(0)}}{L_r^{(0)}},
\ee
\be
V=-\frac{d\ln p^{(0)}}{d\ln r}, \quad \nabla=\frac{d\ln T^{(0)}}{d\ln p^{(0)}}, \quad \nabla_{ad}=\left(\frac{\partial\ln T}{\partial\ln p}\right)_{ad},\quad \alpha_T=-\left(\frac{\partial \ln\rho}{\partial\ln T}\right)_p, \quad {\chi_\rho}=\left(\frac{\partial \ln p}{\partial\ln \rho}\right)_T, \quad \chi_T=\left(\frac{\partial\ln p}{\partial\ln T}\right)_\rho,
\ee
\be
\hat\kappa_T=\left(\frac{\partial \ln\kappa}{\partial\ln T}\right)_p, \quad \hat\kappa_p=\left(\frac{\partial \ln\kappa}{\partial\ln P}\right)_T,
 \quad
\hat\epsilon_P=\left(\frac{\partial\ln\epsilon}{\partial\ln p}\right)_T,  \quad \hat\epsilon_T=\left(\frac{\partial\ln\epsilon}{\partial\ln T}\right)_P,
\ee
\be
H^{(2)}=-\overline{\nabla\cdot\left(\rho^\prime\pmb{v}^\prime\right)}, \label{eq:h2}
\ee
\begin{eqnarray}
I^{(2)} &= &  c_3\overline{\frac{\rho^\prime}{\rho^{(0)}}\frac{\epsilon^\prime}{\epsilon^{(0)}}}
-\frac{c_2}{c_p \sigma_0}\left[
\overline{\pmb{v}^\prime\cdot\nabla s^\prime}
+\overline{\left(\frac{T^\prime}{T^{(0)}}+\frac{\rho^\prime}{\rho^{(0)}}\right)
\left(\frac{\partial s^\prime}{\partial t}+\Omega\frac{\partial s^\prime}{\partial\phi}+{v_r^\prime}\frac{\partial s^{(0)}}{\partial r}\right)}\right]\nonumber \\
&& +\frac{1}{V\nabla}\left(\overline{\nabla_H\frac{\lambda^\prime}{\lambda^{(0)}}\cdot\nabla_H\frac{T^\prime}{T^{(0)}}}
+\overline{\frac{\lambda^\prime}{\lambda^{(0)}}\nabla_H^2\frac{T^\prime}{T^{(0)}}}\right)
-c_2\frac{s^{(0)}}{c_p}\frac{\partial}{\partial\tau}Q^{(2)}(s)
+c_3\left[Q^{(2)}(\epsilon)+Q^{(2)}(\rho)
\right],
\label{eq:i2}
\end{eqnarray}
\be
J^{(2)}=Q^{(2)}(\kappa)+Q^{(2)}(\rho)
-D^{(2)}
+\overline{\frac{\lambda^\prime}{\lambda^{(0)}}\left(-\frac{1}{ V\nabla}r\frac{\partial}{\partial r}\frac{T^\prime}{T^{(0)}}
+\frac{T^\prime}{T^{(0)}}\right)},
\ee
\be
D^{(2)}=3\overline{\left(\frac{T^\prime}{T^{(0)}}\right)^2}+
\overline{\left(\frac{\kappa^\prime}{\kappa^{(0)}}\right)^2}
+\overline{\left(\frac{\rho^\prime}{\rho^{(0)}}\right)^2}
-3\overline{\left(\frac{T^\prime}{T^{(0)}}\right)\left(\frac{\kappa^\prime}{\kappa^{(0)}}
+\frac{\rho^\prime}{\rho^{(0)}}\right)}
+\overline{\frac{\kappa^\prime}{\kappa^{(0)}}\frac{\rho^\prime}{\rho^{(0)}}},
\ee
\be
Q^{(2)}(h)=\frac{(p^{(0)})^2}{2}\frac{1}{h}\frac{\partial^2 h}{\partial p^2}\overline{\left(\frac{p^\prime}{p^{(0)}}\right)^2}
+\frac{(T^{(0)})^2}{2}\frac{1}{h}\frac{\partial^2h}{\partial T^2}\overline{\left(\frac{T^\prime}{T^{(0)}}\right)^2}
+{p^{(0)}T^{(0)}}\frac{1}{h}\frac{\partial^2 h}{\partial p\partial T}\overline{\frac{p^\prime}{p^{(0)}}\frac{T^\prime}{T^{(0)}}},
\label{eq:q(2)}
\ee
and $h$ denotes functions that depend on $p$ and $T$, $\pmb{v}_H^{(2)}=v_q^{(2)}\pmb{e}_q+v_{\bar q}^{(2)}\pmb{e}_{\bar q}$, and
\be
\nabla_H=r\left(\pmb{e}_q\frac{\partial}{\partial\bar q}+\pmb{e}_{\bar q}\frac{\partial}{\partial q}\right).
\ee
Note that equations (\ref{eq:qcompmom}) and (\ref{eq:barqcompmom}) will be used to derive
algebraic equations relating the variables $v_q^{(2)}$ and $v_{\bar q}^{(2)}$ to the variables
$v_r^{(2)}$ and $p^{(2)}$.

\subsection{Expansion of non-linear terms in terms of Spin-Weighted Spherical Harmonics}

In this Appendix, we give explicit expressions for various products of linear wave functions, which are given by
series expansion in terms of spin-weighted spherical harmonic functions ${}_sY_l^m$ in the basis coordinates $(r,q,\bar q)$ and
basis { unit} vectors $\pmb{e}_r$, $\pmb{e}_q$ and $\pmb{e}_{\bar q}$.
 
To evaluate the terms $\overline{\pmb{v}^\prime\cdot\nabla\pmb{v}^\prime}$,
we need covariant derivatives of the velocity vector, or the displacement vector ({ see} Schenk et al 2002)
\be
\xi^r_{;r}=\sum_{l\ge|m|}\partial f_0^{lm}{}_0Y_l^m,
\ee
\be
\xi^{\bar q}_{;r}=
\sum_{l\ge|m|}\partial f_{+1}^{lm}{}_{+1}Y_l^m,
\ee
\be
\xi^{q}_{;r}=
\sum_{l\ge|m|}\partial f_{-1}^{lm}{}_{-1}Y_l^m,
\ee
\be
\xi^q_{;q}=
\sum_{l\ge|m|}G_{+0}^{lm}{}_0Y_l^m,
\ee
\be
\xi^{\bar q}_{;\bar q}=
\sum_{l\ge|m|}G_{-0}^{lm}{}_0Y_l^m,
\ee
\be
\xi^{\bar q}_{;q}=
\sum_{l\ge|m|}H_{+2}^{lm}{}_{+2}Y_l^m,
\ee
\be
\xi^{q}_{;\bar q}=
\sum_{l\ge|m|}H_{-2}^{lm}{}_{-2}Y_l^m,
\ee
\be
\xi^r_{;q}=
\sum_{l\ge|m|}F_{+1}^{lm}{}_{+1}Y_l^m,
\ee
\be
\xi^r_{;\bar q}=
\sum_{l\ge|m|}F_{-1}^{lm}{}_{-1}Y_l^m,
\ee
where we have used the notation $\xi^j_{;k}$ for the covariant derivatives $\nabla_k\xi^j$, and 
\be
\partial f_0^{lm}\equiv\frac{\partial f_0^{lm}}{\partial r}
=\sum_{j=1}^{j_{\rm max}}\left(r\frac{\partial}{\partial r}+1\right)S_{l_j}\delta^l_{l_j},
\ee
\be
\partial f_{+1}^{lm} \equiv \frac{\partial f^{lm}_{+1}}{\partial r} =
\sum_{j=1}^{j_{\rm max}} \left(r\frac{\partial}{\partial r}+1\right)
\left(\sqrt{\frac{\Lambda_{l^\prime_j}}{2}} {\rm i}
T_{l^\prime_j}\delta^l_{l^\prime_j}  - \sqrt{\frac{\Lambda_{l_j}}{2}}
H_{l_j}\delta^l_{l_j}\right),
\ee
\be
\partial f_{-1}^{lm}\equiv\frac{\partial f_{-1}^{lm}}{\partial r}=\sum_{j=1}^{j_{\rm max}}\left(r\frac{\partial}{\partial r}+1\right)\left(\sqrt{\frac{\Lambda_{l^\prime_j}}{2}} {\rm i} T_{l^\prime_j}\delta^l_{l^\prime_j}
+\sqrt{\frac{\Lambda_{l_j}}{2}} H_{l_j}\delta^l_{l_j}\right),
\ee
\be
G_{+0}^{lm}\equiv\frac{1}{r}\left(f_0^{lm}-\sqrt{\frac{\Lambda_l}{2}} f_{-1}^{lm}\right)=\sum_{j=1}^{j_{\rm max}}\left[S_{l_j}\delta^l_{l_j} -\frac{1}{ 2}\left(\Lambda_{l_j}H_{l_j}\delta^l_{l_j}+
\Lambda_{l^\prime_j}{\rm i} T_{l^\prime_j}\delta^l_{l^\prime_j}\right)\right],
\ee
\be
G_{-0}^{lm}\equiv\frac{1}{ r}\left(f_0^{lm}+\sqrt{\frac{\Lambda_l}{2}}f_{+1}^{lm}\right)=\sum_{j=1}^{j_{\rm max}}\left[S_{l_j}\delta^l_{l_j}-\frac{1}{ 2}\left(\Lambda_{l_j}H_{l_j}\delta^l_{l_j}-
\Lambda_{l^\prime_j}\rmi T_{l^\prime_j}\delta^l_{l^\prime_j}\right)\right],
\ee
\be
H_{+2}^{lm}\equiv-\sqrt{\frac{\Lambda_l-2}{2}}\frac{f_{+1}^{lm}}{r}=-\frac{1}{2}\sum_{j=1}^{j_{\rm max}}\left(\sqrt{\aleph_{l^\prime_j}}\rmi T_{l^\prime_j}
\delta^l_{l^\prime_j}-\sqrt{\aleph_{l_j}}H_{l_j}\delta^l_{l_j}\right),
\ee
\be
H_{-2}^{lm}\equiv\sqrt{\frac{\Lambda_l-2}{2}}\frac{f_{-1}^{lm}}{r}=\frac{1}{ 2}\sum_{j=1}^{j_{\rm max}}\left(\sqrt{\aleph_{l^\prime_j}}\rmi T_{l^\prime_j}
\delta^l_{l^\prime_j}+\sqrt{\aleph_{l_j}}H_{l_j}\delta^l_{l_j}\right),
\ee
\be
F_{+1}^{lm}\equiv -\frac{1}{r}\left(f_{+1}^{lm}+\sqrt{\frac{\Lambda_l}{2}}f_0^{lm}\right)=\frac{1}{\sqrt{2}}\sum_{j=1}^{j_{\rm max}}\left[\sqrt{\Lambda_{l_j}}\left(H_{l_j}-S_{l_j}\right)\delta^l_{l_j}-\sqrt{\Lambda_{l^\prime_j}}\rmi T_{l^\prime_j}\delta^l_{l^\prime_j}\right],
\ee
\be
F_{-1}^{lm}\equiv -\frac{1}{r}\left(f_{-1}^{lm}-\sqrt{\frac{\Lambda_l}{2}}f_0^{lm}\right)=-\frac{1}{\sqrt{2}}\sum_{j=1}^{j_{\rm max}}\left[\sqrt{\Lambda_{l_j}}\left(H_{l_j}-S_{l_j}\right)\delta^l_{l_j}+\sqrt{\Lambda_{l^\prime_j}}\rmi T_{l^\prime_j}\delta^l_{l^\prime_j}\right],
\ee
\be
f^{lm}_0(r)=r\sum_{j=1}^{j_{\rm max}}S_{l_j}(r)\delta_{l_j}^l,
\ee
\be
f^{lm}_{-1}(r)=\frac{1}{\sqrt{2}}r\sum_{j=1}^{j_{\rm max}}\left(\sqrt{\Lambda_{l'_j}}\rmi T_{l'_j}(r)\delta^l_{l'_j}
+\sqrt{\Lambda_{l_j}}H_{l_j}(r)\delta^l_{l_j}\right),
\ee
\be
f^{lm}_{+1}(r)=\frac{1}{\sqrt{2}}r\sum_{j=1}^{j_{\rm max}}\left(\sqrt{\Lambda_{l'_j}}\rmi T_{l'_j}(r)\delta^l_{l'_j}
-\sqrt{\Lambda_{l_j}}H_{l_j}(r)\delta^l_{l_j}\right),
\ee
and
\be
\aleph_l=\Lambda_l\left(\Lambda_l-2\right).
\ee
Thus, the terms $\overline{\pmb{v}^\prime\cdot\nabla\pmb{v}^\prime}$ in equation (\ref{eq:vecg2}) may be given as

\begin{eqnarray}
\overline{\pmb{v}^\prime\cdot\nabla\pmb{v}^\prime}&=& \frac{1}{ 2}|\omega|^2\Re\left(\pmb{\xi}^*\cdot\nabla\pmb{\xi}\right)\nonumber \\
& =& \frac{1}{ 2}|\omega|^2\sum_{l_1,l_2}(-1)^{m}\Re\biggl\{\left[\left(f_0^{l_1}\right)^*\partial f_0^{l_2}\left({}_0Y_{l_1}^{-m}{}_0Y_{l_2}^m\right)
-\left(f_{-1}^{l_1}\right)^*F_{-1}^{l_2}\left({}_{+1}Y_{l_1}^{-m}{}_{-1}Y_{l_2}^m\right)
-\left(f_{+1}^{l_1}\right)^*F_{+1}^{l_2}\left({}_{-1}Y_{l_1}^{-m}{}_{+1}Y_{l_2}^m\right)\right]\pmb{e}_r
\nonumber \\
&&  +\left[\left(f_0^{l_1}\right)^*\partial f_{-1}^{l_2}\left({}_0Y_{l_1}^{-m}{}_{-1}Y_{l_2}^m\right)
-\left(f_{-1}^{l_1}\right)^*H_{-2}^{l_2}\left({}_{+1}Y_{l_1}^{-m}{}_{-2}Y_{l_2}^m\right)
-\left(f_{+1}^{l_1}\right)^*G_{+0}^{l_2}\left({}_{-1}Y_{l_1}^{-m}{}_{0}Y_{l_2}^m\right)\right]\pmb{e}_q
\nonumber \\
&&  +\left[\left(f_0^{l_1}\right)^*\partial f_{+1}^{l_2}\left({}_0Y_{l_1}^{-m}{}_{+1}Y_{l_2}^m\right)
-\left(f_{+1}^{l_1}\right)^*H_{+2}^{l_2}\left({}_{-1}Y_{l_1}^{-m}{}_{+2}Y_{l_2}^m\right)
-\left(f_{-1}^{l_1}\right)^*G_{-0}^{l_2}\left({}_{+1}Y_{l_1}^{-m}{}_{0}Y_{l_2}^m\right)\right]\pmb{e}_{\bar q}\biggr\}.
\end{eqnarray}

Since the term $\overline{\rho^\prime\nabla p^\prime}$ in equation (\ref{eq:vecg2}) is given by
\be
\overline{\rho^\prime\nabla p^\prime}
=
\frac{1}{2}\sum_{l_1,l_2}(-1)^m \Re\left\{\rho_{l_1}^{\prime *}
\left[\frac{\partial p_{l_2}^\prime}{\partial r} \left({}_0Y_{l_1}^{-m}{}_0Y_{l_2}^m\right)\pmb{e}_r
+\sqrt{\frac{\Lambda_{l_2}}{2}}\frac{p_{l_2}^\prime}{ r}\left({}_0Y_{l_1}^{-m}{}_{-1}Y_{l_2}^m\right)\pmb{e}_q
- \sqrt{\frac{\Lambda_{l_2}}{2}}\frac{p_{l_2}^\prime}{ r}\left({}_0Y_{l_1}^{-m}{}_{+1}Y_{l_2}^m\right)\pmb{e}_{\bar q}\right]\right\},
\ee
we obtain
\begin{eqnarray}
G_r^{(2)} &= & -\frac{1}{2}|\omega|^2\sum_{l_1,l_2}(-1)^m\Re\left[\left(f_0^{l_1}\right)^*\partial f_0^{l_2}\left({}_0Y_{l_1}^{-m}{}_0Y_{l_2}^m\right)
-\left(f_{-1}^{l_1}\right)^*F_{-1}^{l_2}\left({}_{+1}Y_{l_1}^{-m}{}_{-1}Y_{l_2}^m\right)
-\left(f_{+1}^{l_1}\right)^*F_{+1}^{l_2}\left({}_{-1}Y_{l_1}^{-m}{}_{+1}Y_{l_2}^m\right)\right]\nonumber \\
&&+\frac{1}{2}g\sum_{l_1,l_2}(-1)^m\Re\left[\frac{\rho^{\prime *}_{l_1}}{\rho^{(0)}}\frac{\rho^{\prime }_{l_2}}{\rho^{(0)}}
\left({}_0Y_{l_1}^{-m}{}_0Y_{l_2}^m\right)\right]
+\frac{1}{2}g\sum_{l_1,l_2}(-1)^m\Re\left[\frac{\rho^{\prime*}_{l_1}}{\rho^{(0)}}
\frac{1}{\rho^{(0)}g}\frac{\partial p_{l_2}^\prime}{\partial r}\left({}_0Y_{l_1}^{-m}{}_0Y_{l_2}^m\right)\right],
\end{eqnarray}
\begin{eqnarray}
G_q^{(2)} &= & -\frac{1}{2}|\omega|^2\sum_{l_1,l_2}(-1)^m\Re\left[\left(f_0^{l_1}\right)^*\partial f_{-1}^{l_2}\left({}_0Y_{l_1}^{-m}{}_{-1}Y_{l_2}^m\right)
-\left(f_{-1}^{l_1}\right)^*H_{-2}^{l_2}\left({}_{+1}Y_{l_1}^{-m}{}_{-2}Y_{l_2}^m\right)
-\left(f_{+1}^{l_1}\right)^*G_{+0}^{l_2}\left({}_{-1}Y_{l_1}^{-m}{}_{0}Y_{l_2}^m\right)\right]
\nonumber \\
&& +\frac{1}{2}g\sum_{l_1,l_2}(-1)^m\Re\left[\frac{\rho^{\prime*}_{l_1}}{\rho^{(0)}}\sqrt{\frac{\Lambda_{l_2}}{2}}\frac{p_{l_2}^\prime}{gr\rho^{(0)}}
\left({}_0Y_{l_1}^{-m}{}_{-1}Y_{l_2}^m\right)\right],
\end{eqnarray}
\begin{eqnarray}
G_{\bar q}^{(2)} &= & -\frac{1}{2}|\omega|^2\sum_{l_1,l_2}(-1)^m\Re\left[\left(f_0^{l_1}\right)^*\partial f_{+1}^{l_2}\left({}_0Y_{l_1}^{-m}{}_{+1}Y_{l_2}^m\right)
-\left(f_{+1}^{l_1}\right)^*H_{+2}^{l_2}\left({}_{-1}Y_{l_1}^{-m}{}_{+2}Y_{l_2}^m\right)
-\left(f_{-1}^{l_1}\right)^*G_{-0}^{l_2}\left({}_{+1}Y_{l_1}^{-m}{}_{0}Y_{l_2}^m\right)\right]
\nonumber\\
&& -\frac{1}{2}g\sum_{l_1,l_2}(-1)^m\Re\left[\frac{\rho^{\prime*}_{l_1}}{\rho^{(0)}}
\sqrt{\frac{\Lambda_{l_2}}{2}}\frac{p_{l_2}^\prime}{gr\rho^{(0)}}\left({}_0Y_{l_1}^{-m}{}_{+1}Y_{l_2}^m\right)\right].
\end{eqnarray}

The term $\overline{\nabla\cdot\left(\rho^\prime\pmb{v}^\prime\right)}$ in equation (\ref{eq:h2})
may be evaluated as
\begin{eqnarray}
 \overline {\nabla  \cdot \left( {\rho '\pmb{v} '}  \right)} 
  &= & \frac{1}{2}\Re\biggl\{\rmi\omega\sum_{l_1,l_2}\left[\rho_{l_1}^{\prime *}\left(\frac{1}{r^2}\frac{\partial}{\partial r}r^3S_{l_2}  -\Lambda_{l_2}H_{l_2}\right)+S_{l_2}r\frac{\partial}{\partial r}\rho_{l_1}^{\prime *}\right](-1)^m\left({}_0Y_{l_1}^{-m}{}_0Y_{l_2}^m\right)\biggr\}
  \nonumber\\
 && -\frac{1}{2}\frac{1}{2}\Re\biggl\{{\rmi\omega}\biggl[\sum_{l_1,l_2}\rho_{l_1}^{\prime *}H_{l_2}\sqrt{\Lambda_{l_1}\Lambda_{l_2}}  \left(-1)^{m+1}({}_{+1}Y_{l_1}^{-m}{}_{-1}Y_{l_2}^m+{}_{-1}Y_{l_1}^{-m}{}_{+1}Y_{l_2}^m\right)
 \nonumber\\
&& +\sum_{l_1,l^\prime_2}\rho_{l_1}^{\prime *}iT_{l^\prime_2}\sqrt{\Lambda_{l_1}\Lambda_{l^\prime_2}} 
(-1)^{m+1} \left({}_{+1}Y_{l_1}^{-m}{}_{-1}Y_{l^\prime_2}^m-{}_{-1}Y_{l_1}^{-m}{}_{+1}Y_{l^\prime_2}^m\right)\biggr]\biggr\}.
\end{eqnarray}
The terms $\overline{\nabla_H\lambda^\prime\cdot\nabla_HT^\prime}$ and $\overline{\pmb{\xi}\cdot\nabla s^\prime}$
in equation (\ref{eq:i2}) may be given by
\be
\overline{\nabla_H\frac{\lambda^\prime}{\lambda^{(0)}}\cdot\nabla_H\frac{T^\prime}{T^{(0)}}}
=-\frac{1}{2}\frac{1}{2}\sum_{l_1,l_2}\Re\left[\sqrt{\Lambda_{l_1}\Lambda_{l_2}}\frac{\lambda^{\prime*}_{l_1}}{\lambda^{(0)}}
\frac{T^\prime_{l_2}}{T^{(0)}}(-1)^{m+1}\left({}_{+1}Y_{l_1}^{-m}{}_{-1}Y_{l_2}^m+{}_{-1}Y_{l_1}^{-m}{}_{+1}Y_{l_2}^m\right)
\right],
\ee
and
\begin{eqnarray}
\overline{\pmb{\xi}\cdot\nabla s^\prime}& = 
& \frac{1}{2}\sum_{l_1,l_2}\Re\left[S_{l_2}r\frac{\partial s^{\prime*}_{l_1}}{\partial r}(-1)^m\left({}_0Y_{l_1}^{-m}{}_0Y_{l_2}^m\right)\right]\nonumber\\
&& -\frac{1}{2}\frac{1}{ 2}\sum_{l_1,l_2}\Re\biggl[\sqrt{\Lambda_{l_1}\Lambda_{l_2}}H_{l_2}s^{\prime*}_{l_1}(-1)^{m+1}\left({}_{-1}Y_{l_1}^{-m}{}_{+1}Y_{l_2}^m+{}_{+1}Y_{l_1}^{-m}{}_{-1}Y_{l_2}^m\right)\nonumber\\
&&+\sqrt{\Lambda_{l_1}\Lambda_{l_2^\prime}}iT_{l_2^\prime}s^{\prime*}_{l_1}
(-1)^{m+1}\left({}_{+1}Y_{l_1}^{-m}{}_{-1}Y_{l_2^\prime}^m-{}_{-1}Y_{l_1}^{-m}{}_{+1}Y_{l_2^\prime}^m\right)\biggr].
\end{eqnarray}

We also note that the Coriolis term and the term $\nabla p^{(2)}$ in equation (\ref{eq:secondmomcon}) are given as
\begin{eqnarray}
\pmb{v}^{(2)}\cdot\nabla\pmb{v}^{(0)}+\pmb{v}^{(0)}\cdot\nabla\pmb{v}^{(2)}
&= &-\rmi\sqrt{2}f\left[\sum_{l}\sqrt{\Lambda_l}\hat v_{H,l}^{(2)}\left({}_1Y_1^0{}_{-1}Y_l^0-{}_{-1}Y_1^0{}_1Y_l^0\right)
+\sum_{l'}\sqrt{\Lambda_{l^\prime}}\rmi\hat v_{T,l^\prime}^{(2)}\left({}_1Y_1^0{}_{-1}Y_{l^\prime}^0
+{}_{-1}Y_1^0{}_1Y_{l'}^0\right)\right]\pmb{e}_r
\nonumber\\
&& -\rmi\sqrt{2}f\left[\sum_{l}\sqrt{\Lambda_l}\hat v_{H,l}^{(2)}\left({}_0Y_1^0{}_{-1}Y_l^0\right)
+\sum_{l^\prime}\sqrt{\Lambda_{l^\prime}}\rmi\hat v_{T,l^\prime}^{(2)}\left({}_0Y_1^0{}_{-1}Y_{l^\prime}^0\right)-\sqrt{2}\sum_l
\hat v_{S,l}\left({}_{-1}Y_1^0{}_0Y_l^0\right)\right]\pmb{e}_q
\nonumber\\
&& -\rmi\sqrt{2}f\left[\sum_{l}\sqrt{\Lambda_l}\hat v_{H,l}^{(2)}\left({}_0Y_1^0{}_{1}Y_l^0\right)
-\sum_{l'}\sqrt{\Lambda_{l^\prime}}\rmi\hat v_{T,l^\prime}^{(2)}\left({}_0Y_1^0{}_{1}Y_{l^\prime}^0\right)-\sqrt{2}\sum_l
\hat v_{S,l}\left({}_{1}Y_1^0{}_0Y_l^0\right)\right]\pmb{e}_{\bar q},
\label{eq:corioli}
\end{eqnarray}
and
\be
\nabla p^{(2)}
=\sum_l \left(\frac{\partial p^{(2)}_l}{\partial r}{}_0Y_l^0 \pmb{e}_r
+\frac{p^{(2)}_l}{r}\sqrt{\frac{\Lambda_l}{2}}{}_{-1}Y_l^0 \pmb{e}_q
-\frac{p^{(2)}_l}{r}\sqrt{\frac{\Lambda_l}{2}}{}_{+1}Y_l^0 \pmb{e}_{\bar q}\right).
\ee

\subsection{Projection of second-order equations onto Spin-Weighted Spherical Harmonics ${}_sY_l^m$}

Multiplying equations (\ref{eq:dvr2dr}) to (\ref{eq:dt2dr}) by ${}_0Y_k^0$, and carrying out angular integration over spherical surface, we obtain
\be
r\frac{\partial}{\partial r}\frac{\hat v^{(2)}_{S,k}}{ r\sigma_0}=-\frac{\partial}{\partial\tau}\frac{\rho^{(2)}_k}{\rho^{(0)}}
+\frac{\Lambda_k \hat v^{(2)}_{H,k}}{r\sigma_0}-\left(3+\frac{d\ln\rho^{(0)}}{d\ln r}\right)\frac{\hat v^{(2)}_{S,k}}{r\sigma_0}
+\frac{H_k^{(2)}}{\rho^{(0)}\sigma_0},
\label{eq:z1}
\ee
\be
r\frac{\partial}{\partial r}\frac{p^{(2)}_k}{\rho^{(0)} gr}=-
c_1\frac{\partial}{\partial \tau}\frac{\hat v^{(2)}_{S,k}}{r\sigma_0}
+2\rmi c_1 \bar f \int {}_0Y_k^0\left(\frac{ v_q^{(2)}}{r\sigma_0}{}_1Y_1^0
+\frac{ v_{\bar q}^{(2)}}{r\sigma_0}{}_{-1}Y_1^0\right)d\Omega
-\frac{d\ln\rho^{(0)} gr}{d\ln r}\frac{p^{(2)}_k}{\rho^{(0)} gr}+\frac{\rho^{(2)}_k}{\rho^{(0)}}+\frac{G_{r,k}^{(2)}}{g},
\label{eq:z2}
\ee
\begin{eqnarray}
r\frac{\partial}{\partial r}\frac{L_{r,k}^{(2)}}{L_r^{(0)}} &= &  -c_2\left[\frac{\partial}{\partial \tau}\frac{T^{(2)}_k}{T^{(0)}}
-\nabla_{ad}\frac{\partial}{\partial \tau}\frac{p^{(2)}_k}{p^{(0)}}+\frac{\hat v_{S,k}^{(2)}}{r\sigma_0}V\left(\nabla_{ad}-\nabla\right)\right]
\nonumber\\
&& +c_3\left[\left(\hat\epsilon_P+\frac{1}{\chi_\rho}\right)\frac{p_k^{(2)}}{ p^{(0)}}+
\left(\hat\epsilon_T-\alpha_T\right)\frac{T^{(2)}_k}{T^{(0)}}\right]
-\frac{d\ln L_r^{(0)}}{d\ln r}\frac{L_{r,k}^{(2)}}{L_r^{(0)}}-\frac{\Lambda_k}{V\nabla}\frac{T^{(2)}_k}{ T^{(0)}}+I_k^{(2)},
\label{eq:z3}
\end{eqnarray}
\be
r\frac{\partial}{\partial r}\frac{T^{(2)}_k}{ T^{(0)}}=V\nabla\left(4-\hat\kappa_T+\alpha_T
\right)\frac{T^{(2)}_k}{T^{(0)}}
-V\nabla\left(\hat\kappa_p+\frac{1}{\chi_\rho}\right)\frac{p^{(2)}_k}{p^{(0)}}-V\nabla\frac{L_{r,k}^{(2)}}{L_r^{(0)}}-V\nabla J_k^{(2)},
\label{eq:z4}
\ee
where
\be
G_{r,k}^{(2)}\equiv \int {}_0Y_k^0 G_q^{(2)}d\Omega, \quad {H_k^{(2)}}\equiv\int{}_0Y_k^0{H^{(2)}}d\Omega, \quad 
I_k^{(2)}\equiv\int{}_0Y_k^0I^{(2)}d\Omega, \quad J_k^{(2)}\equiv \int {}_0Y_k^0J^{(2)}d\Omega,
\ee
and
\be
\int {}_0Y_k^0\left({v_q^{(2)} }{}_1Y_1^0
+{v_{\bar q}^{(2)}}{}_{-1}Y_1^0\right)d\Omega
=-\frac{1}{\sqrt{2}}
\left[\sum_l\sqrt{\Lambda_l}\hat v_{H,l}^{(2)}\left(C_{0(-1)1}^{kl1}-C_{01(-1)}^{kl1}\right)
+\sum_{l'}\sqrt{\Lambda_{l'}}\rmi\hat v_{T,{l'}}^{(2)}\left(C_{0(-1)1}^{kl'1}+C_{01(-1)}^{kl'1}\right)\right],
\ee
with 
\be
C_{abc}^{ijk}=\left[\matrix{i & j & k\cr 0 & 0 & 0 \cr a & b & c \cr}\right].
\label{eq:csquare}
\ee
The symbol $[\cdots]$ in equation (\ref{eq:csquare}) has been defined in Appendix A.

Multiplying equations (\ref{eq:qcompmom}) and (\ref{eq:barqcompmom}) 
by  
${}_1Y_k^0$ and ${}_{-1}Y_k^0$, and
carrying out angular integration over spherical surface, we obtain
\be
-\sum_l\sqrt{\frac{\Lambda_l}{2}}\frac{\partial \hat v_{H,l}^{(2)}}{\partial t}\delta_{kl}
-\sum_{l'}\sqrt{\frac{\Lambda_{l'}}{2}}\frac{\partial \rmi\hat v_{T,{l'}}^{(2)}}{ \partial t}\delta_{kl'}
+2\rmi f\int{}_1Y_k^0\left(v_r^{(2)}{}_{-1}Y_1^0-v_q^{(2)}{}_0Y_1^0\right)d\Omega
-\sum_l\sqrt{\frac{\Lambda_l}{2}}\frac{p_l^{(2)}}{r\rho^{(0)}}\delta_{kl}
=  G_{q,k}^{(2)},
\label{eq:qmom}
\ee
\be
 \sum_l\sqrt{\frac{\Lambda_l}{2}}\frac{\partial \hat v_{H,l}^{(2)}}{\partial t}\delta_{kl}
 -\sum_{l'}\sqrt{\frac{\Lambda_{l'}}{2}}\frac{\partial \rmi\hat v_{T,{l'}}^{(2)}}{\partial t}\delta_{kl'}
 +2\rmi f\int{}_{-1}Y_k^0\left(v_r^{(2)}{}_1Y_1^0+v_{\bar q}^{(2)}{}_0Y_1^0\right)d\Omega
  +\sum_l\sqrt{\frac{\Lambda_l}{2}}\frac{p_l^{(2)}}{r\rho^{(0)}}\delta_{kl}=G_{\bar q,k}^{(2)},
  \label{eq:barqmom}
\ee
where
\be
G_{q,k}^{(2)}\equiv \int {}_1Y_k^0 G_q^{(2)}d\Omega, \quad G_{\bar q,k}^{(2)}\equiv \int{}_{-1}Y_k^0G_{\bar q}^{(2)}d\Omega,
\ee
and
\be
\int{}_1Y_k^0\left(v_r^{(2)}{}_{-1}Y_1^0-v_q^{(2)}{}_0Y_1^0\right)d\Omega=-\frac{1}{\sqrt{2}}\left(\sum_l\sqrt{\Lambda_l}\hat v_{H,l}^{(2)}C_{1(-1)0}^{kl1}+\sum_{l'}\sqrt{\Lambda_{l'}}\rmi\hat v_{T,{l'}}^{(2)}C_{1(-1)0}^{kl'1}-\sqrt{2}\sum_l\hat v_{S,l}^{(2)}C_{10(-1)}^{kl1}\right),
\ee
\be
\int{}_{-1}Y_k^0\left(v_r^{(2)}{}_1Y_1^0+v_{\bar q}^{(2)}{}_0Y_1^0\right)d\Omega=-\frac{1}{\sqrt{2}}\left(\sum_l\sqrt{\Lambda_l}\hat v_{H,l}^{(2)}C_{(-1)10}^{kl1}
-\sum_{l'}\sqrt{\Lambda_{l'}}\rmi\hat v_{T,{l'}}^{(2)}C_{(-1)10}^{kl'1}-\sqrt{2}\sum_l\hat v_{S,l}^{(2)}C_{(-1)01}^{kl1}\right).
\ee
We may obtain a set of algebraic equations using equations (\ref{eq:qmom}) and (\ref{eq:barqmom}).
The sum (\ref{eq:qmom}) + (\ref{eq:barqmom}) gives
\be
-\sum_{l^\prime}2\sqrt{\frac{\Lambda_{l^\prime}}{2}}\frac{\partial \rmi\hat v_{T,{l^\prime}}^{(2)}}{\partial t}\delta_{kl^\prime}
-\rmi\sqrt{2}f\left[\sum_l\left(C_{1(-1)0}^{kl1}+C_{(-1)10}^{kl1}\right)\sqrt{\Lambda_l}\hat v_{H,l}^{(2)}
-\sqrt{2}\sum_l\left(C_{10(-1)}^{kl1}+C_{(-1)01}^{kl1}\right)\hat v_{S,l}^{(2)}\right]=G_{q,k}^{(2)}+G_{\bar q,k}^{(2)},
\label{eq:vt}
\ee
and the difference (\ref{eq:qmom}) $-$ (\ref{eq:barqmom}) gives
\be
-\sum_l2\sqrt{\frac{\Lambda_l}{2}}\frac{\partial \hat v_{H,l}^{(2)}}{\partial t}\delta_{kl}
-\rmi\sqrt{2}f\sum_{l^\prime}\left( C_{1(-1)0}^{kl^\prime 1}+C_{(-1)10}^{kl^\prime 1}\right)\sqrt{\Lambda_{l^\prime}}\rmi\hat v_{T,{l^\prime}}^{(2)}-\sum_l2\sqrt{\frac{\Lambda_l}{2}}\frac{p_l^{(2)}}{ r\rho^{(0)}}\delta_{kl}
=G_{q,k}^{(2)}-G_{\bar q,k}^{(2)},
\label{eq:vh}
\ee
where we have used
\be
C_{1(-1)0}^{kl1}-C_{(-1)10}^{kl1}=0, \quad C_{10(-1)}^{kl1}-C_{(-1)01}^{kl1}=0, \quad C_{0(-1)1}^{kl1}-C_{01(-1)}^{kl1}=0.
\ee

\subsection{Explicit spectral coefficients of non-linear terms in the basis of Spin-Weighted Spherical Harmonics}

Here we give explicit expressions for the quantities $G_{r,k}^{(2)}$, $G_{q,k}^{(2)}$, $G_{\bar q,k}^{(2)}$, $H_k^{(2)}$, $I^{(2)}_k$, $J^{(2)}_k$, which are projections onto spin-weighted spherical harmonic functions:
\begin{eqnarray}
{G_{r,k}^{(2)}} &= & - \frac{\left| \omega  \right|^2}{2} \sum_{l_1,l_2} {\Re\left[ 
{\left(f_0^{l_1 }\right)^* \partial f_0^{l_2 } B_{000}^{kl_1 l_2 }  
+ \left(f_{ - 1}^{l_1 }\right)^* F_{ - 1}^{l_2 } B_{01\left( { - 1} \right)}^{kl_1 l_2 }  
+ \left(f_{ + 1}^{l_1}\right)^* F_{ + 1}^{l_2} B_{0\left( { - 1} \right)1}^{kl_1 l_2 } } \right]}\nonumber\\
&& + \frac{1}{2}\sum_{l_1,l_2} {\Re\left( g{\frac{{\rho ^{\prime *}_{l_1 } }}{{\rho ^{\left( 0 \right)} }}\frac{{\rho '_{l_2 } }}{{\rho ^{\left( 0 \right)} }}  
+ \frac{{\rho ^{\prime *} _{l_1 } }}{{\rho ^{\left( 0 \right)} }}\frac{1}{{\rho ^{\left( 0 \right)} }}\frac{{\partial p'_{l_2 } }}{{\partial r}}} \right)} B_{000}^{kl_1 l_2 },
\end{eqnarray}
\begin{eqnarray}
G_{q,k}^{(2)} &= & - \frac{\left| \omega  \right|^2}{2} \sum\limits_{l_2 l_1 } {\Re\left[ 
{\left(f_0^{l_1 }\right)^* \partial f_{ - 1}^{l_2 } B_{10(-1)}^{kl_1 l_2 }  
+ \left(f_{ - 1}^{l_1 }\right)^* H_{-2}^{l_2 } B_{11(-2)}^{kl_1 l_2 }  
+ \left(f_{ + 1}^{l_1 }\right)^* G_{+0}^{l_2 } B_{1(-1)0}^{kl_1 l_2 } } \right]}
\nonumber\\
&& + \frac{1}{2}\sum\limits_{l_2 l_1 } \Re\left({\frac{{\rho^{\prime *}_{l_1 } }}{{\rho ^{\left( 0 \right)} }}\sqrt {\frac{{\Lambda _{l_2 } }}{2}} \frac{{p'_{l_2 } }}{{r\rho ^{\left( 0 \right)} }} } \right)B_{10\left( { - 1} \right)}^{kl_1 l_2 },
\end{eqnarray}
\begin{eqnarray}
G_{\bar q,k}^{(2)}  &= &- \frac{\left| \omega  \right|^2}{2} \sum_{l_1,l_2} {\Re\left[ 
{\left(f_0^{l_1 }\right)^* \partial f_{ + 1}^{l_2 } B_{\left( { - 1} \right)01}^{kl_1 l_2 }  
+ \left(f_{ + 1}^{l_1 }\right)^* H_{ + 2}^{l_2 } B_{\left( { - 1} \right)\left( { - 1} \right)2}^{kl_1 l_2 }  
+ \left(f_{ - 1}^{l_1 }\right)^* G_{ - 0}^{l_2 } B_{\left( { - 1} \right)10}^{kl_1 l_2 } } \right]}
\nonumber\\
&&- \frac{1}{2}\sum\limits_{l_1 ,l_2 } \Re\left({\frac{{\rho^{\prime *}_{l_1 } }}{{\rho ^{\left( 0 \right)} }}\sqrt {\frac{{\Lambda _{l_2 } }}{2}} \frac{{p'_{l_2 } }}{{r\rho ^{\left( 0 \right)} }} }\right)B_{\left( { - 1} \right)01}^{kl_1 l_2 },
\end{eqnarray}
\begin{eqnarray}
{H_k^{(2)}}  &=&  -\frac{1}{ 2}\Re\biggl\{\rmi\omega\biggl[\sum_{l_1,l_2}\rho_{l_1}^{\prime *}\left(\frac{1}{ r^2}\frac{\partial}{\partial r}r^3 S_{l_2}-\Lambda_{l_2}H_{l_2}\right)B_{000}^{kl_1l_2}
+\sum_{l_1l_2}S_{l_2}r\frac{\partial\rho^{\prime *}_{l_1}}{\partial r}B_{000}^{kl_1l_2}
\nonumber\\
&& -\frac{1}{2}\sum_{l_1l_2}\rho_{l_1}^{\prime*}H_{l_2}\sqrt{\Lambda_{l_1}\Lambda_{l_2}} B_B^{kl_1l_2}
-\frac{1}{2}\sum_{l_1l_2^\prime}\rho_{l_1}^{\prime *}iT_{l_2^\prime}\sqrt{\Lambda_{l_1}\Lambda_{l_2^\prime}}
B_A^{kl_1l_2^\prime}
\biggr]\biggr\},
\end{eqnarray}
\begin{eqnarray}
I_k^{(2)} 
&=& \frac{1}{2}\sum_{l_1l_2}\Re\left\{c_3\frac{\rho_{l_1}^{\prime *}}{\rho^{(0)}}\frac{\epsilon^\prime_{l_2}}{\epsilon^{(0)}}-{\rmi\bar\omega c_2}\left[{S_{l_2}}r\frac{\partial}{\partial r}\frac{s^{\prime *}_{l_1}}{c_p}
+\frac{d\ln c_p}{d\ln r}{S_{l_2}}\frac{s^{\prime *}_{l_1}}{c_p}
+\left(\frac{T^{\prime *}_{l_1}}{T^{(0)}}
+\frac{\rho^{\prime *}_{l_1}}{\rho^{(0)}}\right)\left(\frac{s^\prime_{l_2}}{ c_p}+{S_{l_2}}V(\nabla_{ad}-\nabla)\right)\right]\right\}B_{000}^{kl_1l_2}
\nonumber\\
&& +\frac{1}{2}\frac{1}{2}\sum_{l_1,l_2}\Re\left[{\rmi\bar\omega c_2 }\left(\sqrt{\Lambda_{l_1}\Lambda_{l_2}}H_{l_2}
\frac{s^{\prime *}_{l_1}}{c_p}B_B^{kl_1l_2}
+\sqrt{\Lambda_{l_1}\Lambda_{l^\prime_2}}iT_{l^\prime_2}\frac{s^{\prime *}_{l_1}}{ c_p}
B_A^{kl_1 l_2^\prime}
\right)\right]
\nonumber\\
&&  -\frac{1}{2}\sum_{l_1l_2}\frac{\Lambda_{l_2}}{V\nabla}\Re\left(\frac{\lambda_{l_1}^{\prime *}}{\lambda^{(0)}}\frac{T_{l_2}^{\prime }}{T^{(0)}}\right)B_{000}^{kl_1l_2}
-\frac{1}{2}\frac{1}{2}\sum_{l_1,l_2}\frac{\sqrt{\Lambda_{l_1}\Lambda_{l_2}}}{ V\nabla}\Re\left(\frac{\lambda^{\prime*}_{l_1}}{\lambda^{(0)}}
\frac{T^\prime_{l_2}}{T^{(0)}}\right)B_B^{kl_1l_2}
\nonumber\\
&&  -\frac{1}{2}c_2\frac{s}{c_p}\sum_{l_1l_2}\frac{\partial}{\partial\tau}\Re\left[Q^{(2)}_{l_1,l_2}(s)
\right]B_{000}^{kl_1l_2}
+\frac{1}{2}c_3\sum_{l_1l_2}\Re\left[Q^{(2)}_{l_1,l_2}(\epsilon)+Q^{(2)}_{l_1,l_2}(\rho)
\right]B_{000}^{kl_1l_2},
\end{eqnarray}
\be
J_k^{(2)}=\frac{1}{2}\sum_{l_1l_2}
\Re\left[Q^{(2)}_{l_1,l_2}(\kappa)+Q^{(2)}_{l_1,l_2}(\rho)
-D^{(2)}_{l_1l_2}
+\frac{\lambda_{l_1}^{\prime *}}{\lambda^{(0)}}\left(-
\frac{1}{ V\nabla}r\frac{\partial}{\partial r}\frac{T^\prime_{l_2}}{ T^{(0)}}+\frac{T^\prime_{l_2}}{T^{(0)}}\right)\right]B_{000}^{kl_1l_2},
\ee
where
\be
B_A^{kl_1l_2}=B_{01(-1)}^{kl_1l_2}-B_{0(-1)1}^{kl_1l_2}, \quad
B_B^{kl_1l_2}=B_{01(-1)}^{kl_1l_2}+B_{0(-1)1}^{kl_1l_2},
\ee
\be
D^{(2)}_{l_1,l_2}=\frac{1}{2}\sum_{l_1l_2}\left[3\frac{T^{\prime*}_{l_1}}{ T^{(0)}}\frac{T^{\prime}_{l_2}}{T^{(0)}}
+\frac{\kappa^{\prime*}_{l_1}}{\kappa^{(0)}}\frac{\kappa^{\prime}_{l_2}}{ \kappa^{(0)}}
+\frac{\rho^{\prime*}_{l_1}}{\rho^{(0)}}\frac{\rho^{\prime}_{l_2}}{\rho^{(0)}}
+\frac{\kappa^{\prime*}_{l_1}}{\kappa^{(0)}}\frac{\rho^{\prime}_{l_2}}{ \rho^{(0)}}
-3\frac{T^{\prime*}_{l_1}}{ T^{(0)}}\left(\frac{\kappa^\prime_{l_2}}{\kappa^{(0)}}+\frac{\rho^{\prime*}_{l_2}}{\rho^{(0)}}\right)\right],
\ee
\be
Q^{(2)}_{l_1,l_2}(h)=\frac{(p^{(0)})^2}{2}
\frac{1}{h}
\frac{\partial^2 h}{\partial p^2}{\frac{p^{\prime *}_{l_1}}{ p^{(0)}}\frac{p^\prime_{l_2}}{p^{(0)}}}
+\frac{(T^{(0)})^2}{2}\frac{1}{h}\frac{\partial^2h}{\partial T^2}{\frac{T^{\prime *}_{l_1}}{T^{(0)}}
\frac{T^\prime_{l_2}}{T^{(0)}}}
+{p^{(0)}T^{(0)}}\frac{1}{h}\frac{\partial^2 h}{\partial p\partial T}{\frac{p^{\prime *}_{l_1}}{p^{(0)}}\frac{T^\prime_{l_2}}{T^{(0)}}},
\ee
and
\be
B_{abc}^{kl_1l_2}=(-1)^{b+m}\left[\matrix{k & l_1 & l_2\cr 0 & -m & m \cr a & b & c \cr}\right].
\ee

To evaluate $Q^{(2)}$, we need to calculate second derivatives of thermal quantities.
For example, for the specific entropy $s$
\be
\frac{\partial^2 s}{\partial p^2}=-\frac{1}{ p^2}\frac{c_p\nabla_{ad}}{\chi_\rho}\left[\left(\frac{\partial\ln\alpha_T}{\partial\ln\rho}\right)_T-1\right],
\ee
\be
\frac{\partial^2 s}{\partial T^2}=\frac{c_p}{T^2}\left[\left(\frac{\partial\ln c_p}{\partial\ln T}\right)_\rho
-\alpha_T\left(\frac{\partial\ln c_p}{\partial\ln\rho}\right)_T-1\right],
\ee
\be
\frac{\partial ^2 s}{\partial T\partial p}=\frac{1}{pT}\frac{c_p}{\chi_\rho}\left(\frac{\partial \ln c_p}{\partial \ln \rho}\right)_T.
\ee 
For the density $\rho$, we have
\be
\frac{\partial ^2\rho}{\partial p^2}=\frac{\rho}{ p^2}\frac{1}{\chi_\rho^2}\left[1-\chi_\rho-\left(\frac{\partial\ln\chi_\rho}{\partial\ln\rho}\right)_T\right],
\ee
\be
\frac{\partial^2\rho}{\partial T^2}=\frac{\rho}{T^2}\alpha_T\left[\alpha_T\left(\frac{\partial\ln\alpha_T}{\partial\ln\rho}\right)_T
-\left(\frac{\partial\ln\alpha_T}{\partial\ln T}\right)_\rho+\alpha_T+1\right],
\ee
\be
\frac{\partial^2\rho}{\partial p\partial T}=-\frac{\rho}{pT}\frac{\alpha_T}{\chi_\rho}\left[\left(\frac{\partial\ln\alpha_T}{\partial\ln\rho}\right)_T+1\right].
\ee
For the opacity, which is given as a function of $\rho$ and $T$, we have
\be
\frac{\partial^2\kappa}{\partial p^2}=\frac{\kappa}{ p^2}\frac{\kappa_\rho}{\chi_\rho^2}\left[\kappa_\rho-\chi_\rho
-\left(\frac{\partial\ln\chi_\rho}{\partial\ln\rho}\right)_T+\left(\frac{\partial\ln\kappa_\rho}{\partial\ln\rho}\right)_T\right]
\ee

\be
\frac{\partial^2\kappa}{\partial T^2}=\frac{\kappa}{ T^2}\left[\beta\left(\beta-1\right)
+\left(\frac{\partial\beta}{\partial\ln T}\right)_\rho-\alpha_T\left(\frac{\partial\beta}{\partial\ln\rho}\right)_T\right]
\ee
\be
\frac{\partial^2\kappa}{\partial p\partial T}=\frac{\kappa}{p T}\frac{1}{\chi_\rho}\left[\kappa_\rho\left(\kappa_T-\alpha_T\kappa_\rho\right)
+\kappa_T\left(\frac{\partial\ln\kappa_T}{\partial\ln\rho}\right)_T-\alpha_T\kappa_\rho\left(\frac{\partial\ln\alpha_T\kappa_\rho}{\partial\ln\rho}\right)_T\right]=\frac{\kappa}{pT}\frac{1}{\chi_\rho}\left[\beta\kappa_\rho+\left(\frac{\partial\beta}{\partial\ln\rho}\right)_T\right],
\ee
where
\be
\beta=\kappa_T-\alpha_T\kappa_\rho=\hat\kappa_T, \quad 
\kappa_\rho=\left(\frac{\partial\ln\kappa}{\partial\ln\rho}\right)_T, \quad 
\kappa_T=\left(\frac{\partial\ln\kappa}{\partial\ln T}\right)_\rho.
\ee

\end{appendix}


\bsp	
\label{lastpage}
\end{document}